%
%
\input harvmac
\newcount\yearltd\yearltd=\year\advance\yearltd by 0

\noblackbox

\input epsf

\def\tilde{\widetilde}
\def\hat{\widehat}
\newcount\figno
\figno=0
\def\fig#1#2#3{
\par\begingroup\parindent=0pt\leftskip=1cm\rightskip=1cm\parindent=0pt
\baselineskip=11pt \global\advance\figno by 1 \midinsert
\epsfxsize=#3 \centerline{\epsfbox{#2}} \vskip 12pt {\bf Figure\
\the\figno: } #1\par
\endinsert\endgroup\par
}
\def\figlabel#1{\xdef#1{\the\figno}}
\def\encadremath#1{\vbox{\hrule\hbox{\vrule\kern8pt\vbox{\kern8pt
\hbox{$\displaystyle #1$}\kern8pt}
\kern8pt\vrule}\hrule}}

\def\half{{\textstyle{1\over2}}}

\def\half{{1\over 2}}

 \def\d{{\delta}}
 
 \def\t{{\theta}}
 \def\a{{\alpha}}
 
 \def\frac#1#2{{#1\over #2}}
 \def\l{{\lambda}}

 \def\g{{\gamma}}
 \def\s{{\sigma}}
 
 \def\b{{\beta}}
 \def\t{{\tau}}

 \def\CO{{\cal O}}

 \def\CD{{\cal D}}
 \def\p{\partial}

 \def\r{\rightarrow}
\def\bal{{\bar{\alpha}}}
\def\bb{{\bar{\beta}}}
\def\bg{{\bar{\gamma}}}
\def\bd{{\bar{\delta}}}
\def\bl{{\bar{\lambda}}}
\def\br{{\bar{\rho}}}
\def\e{{\epsilon}}
\def\r{{\rho}}
\def\bs{{\bar{\sigma}}}
\def\bt{{\bar{\tau}}}
\def\IR{\relax{\rm I\kern-.18em R}}
\def\tdzero{D_{\tau}}


\lref\thooftFH{ G.~'t Hooft, ``Renormalization of massless
Yang-Mills fields,'' Nucl.\ Phys.\ B {\bf 33}, 173 (1971).
}

\lref\grosswit{ D.~J.~Gross and E.~Witten, ``Possible third order
phase transition in the large N lattice gauge theory,'' Phys.\
Rev.\ D {\bf 21}, 446 (1980).
}

\lref\creutz{ M.~Creutz, ``On invariant integration Over
$SU(N)$,'' J.\ Math.\ Phys.\  {\bf 19}, 2043 (1978).
}

\lref\first{ O.~Aharony, J.~Marsano, S.~Minwalla, K.~Papadodimas
and M.~Van Raamsdonk, ``The Hagedorn / deconfinement phase
transition in weakly coupled large $N$ gauge theories,''
arXiv:hep-th/0310285, to appear in ATMP.
}

\lref\toruspaper{O.~Aharony, J.~Marsano, S.~Minwalla, K.~Papadodimas,
M.~Van Raamsdonk and T.~Wiseman, to appear.}

\lref\sundborg{ B.~Sundborg, ``The Hagedorn transition,
deconfinement and $N = 4$ SYM theory,'' Nucl.\ Phys.\ B {\bf 573},
349 (2000) [arXiv:hep-th/9908001].
}

\lref\GrossBR{ D.~J.~Gross, R.~D.~Pisarski and L.~G.~Yaffe, ``QCD
and instantons at finite temperature,'' Rev.\ Mod.\ Phys.\  {\bf
53}, 43 (1981).
}

\lref\splitdimone{ G.~Leibbrandt and J.~Williams, ``Split
dimensional regularization for the Coulomb Gauge,'' Nucl.\ Phys.\
B {\bf 475}, 469 (1996) [arXiv:hep-th/9601046].
}

\lref\splitdimtwo{ Y.~H.~Chen, R.~J.~Hsieh and C.~l.~Lin, ``Split
dimensional regularization for the temporal gauge,''
arXiv:hep-th/9610165.
}

\lref\splitdimthree{ G.~Leibbrandt, ``The three-point function in
split dimensional regularization in the  Coulomb gauge,'' Nucl.\
Phys.\ B {\bf 521}, 383 (1998) [arXiv:hep-th/9804109].
}

\lref\splitdimfour{ G.~Heinrich and G.~Leibbrandt, ``Split
dimensional regularization for the Coulomb gauge at two loops,''
Nucl.\ Phys.\ B {\bf 575}, 359 (2000) [arXiv:hep-th/9911211].
}

\lref\ArnoldPS{ 
J. Kapusta, ``Finite temperature field theory'' (Cambridge University
Press, 1989) and references therein.
}

\lref\Cutkosky{ R.~E.~Cutkosky, ``Harmonic functions and matrix
elements for hyperspherical quantum field models,'' J.\ Math.\
Phys.\  {\bf 25}, 939 (1984).
}

\lref\AdamsJB{
A.~Adams and E.~Silverstein,
``Closed string tachyons, AdS/CFT, and large N QCD,''
Phys.\ Rev.\ D {\bf 64}, 086001 (2001)
[arXiv:hep-th/0103220].
}

\lref\AharonyPA{
O.~Aharony, M.~Berkooz and E.~Silverstein,
``Multiple-trace operators and non-local string theories,''
JHEP {\bf 0108}, 006 (2001)
[arXiv:hep-th/0105309].
}

\lref\HadizadehBF{
S.~Hadizadeh, B.~Ramadanovic, G.~W.~Semenoff and D.~Young,
``Free energy and phase transition of the matrix model on a plane-wave,''
arXiv:hep-th/0409318.
}

\lref\LuciniKU{
B.~Lucini, M.~Teper and U.~Wenger,
``The deconfinement transition in SU(N) gauge theories,''
Phys.\ Lett.\ B {\bf 545}, 197 (2002)
[arXiv:hep-lat/0206029].
}

\lref\LuciniZR{
B.~Lucini, M.~Teper and U.~Wenger,
``The high temperature phase transition in SU(N) gauge theories,''
JHEP {\bf 0401}, 061 (2004)
[arXiv:hep-lat/0307017].
}

\lref\GrossHE{
D.~J.~Gross and E.~Witten,
``Possible third order phase transition in the large N lattice gauge theory,''
Phys.\ Rev.\ D {\bf 21}, 446 (1980); ~~~
S.~Wadia,
``A study of $U(N)$ lattice gauge theory in two-dimensions,''
preprint EFI-79/44-CHICAGO; ~~~
S.~R.~Wadia,
``$N = \infty$ phase transition in a class of exactly soluble model
lattice gauge theories,''
Phys.\ Lett.\ B {\bf 93}, 403 (1980).
}

\lref\HawkingDH{
S.~W.~Hawking and D.~N.~Page,
``Thermodynamics of black holes in anti-de Sitter space,''
Commun.\ Math.\ Phys.\  {\bf 87}, 577 (1983).
}

\lref\WittenZW{
E.~Witten,
``Anti-de Sitter space, thermal phase transition, and confinement in
gauge theories,''
Adv.\ Theor.\ Math.\ Phys.\  {\bf 2}, 505 (1998)
[arXiv:hep-th/9803131].
}

\lref\tHooftJZ{
G.~'t Hooft,
``A planar diagram theory for strong interactions,''
Nucl.\ Phys.\ B {\bf 72}, 461 (1974).
}

\lref\mathematica{This file is available at
\tt{http://www.fas.harvard.edu/\~{}papadod/3loop/3loop.html}\rm{}.}
%

\lref\gfive{ K.~Kajantie, M.~Laine, K.~Rummukainen and Y.~Schroder,
``The pressure of hot QCD up to $g^6 \ln(1/g)$,''
Phys.\ Rev.\ D {\bf 67}, 105008 (2003) [arXiv:hep-ph/0211321].
}

\def\my_Title#1#2{\nopagenumbers\abstractfont\hsize=\hstitle\rightline{#1}%
\vskip .5in\centerline{\titlefont
#2}\abstractfont\vskip.5in\pageno=0}

\my_Title {\vbox{\baselineskip12pt \hbox{WIS/03/05-JAN-DPP}
\hbox{\tt hep-th/0502149}}} {\vbox{\centerline{A First Order
Deconfinement Transition in} \vskip 5pt \centerline{Large $N$
Yang-Mills Theory on a Small $S^3$}}}

\centerline{Ofer Aharony$^{a}$, Joseph Marsano$^{b,c}$, Shiraz Minwalla$^{c,b}$,}
\centerline{Kyriakos Papadodimas$^{b,c}$ and Mark Van Raamsdonk$^{d}$}

\medskip

\centerline{\sl $^{a}$Department of Particle Physics, Weizmann Institute of
Science, Rehovot 76100, Israel}
\centerline{\sl $^{b}$Jefferson Physical Laboratory, Harvard University,
Cambridge, MA 02138, USA}
\centerline{\sl $^{c}$Department of Theoretical Physics, Tata Institute of Fundamental Research,}
\centerline{\sl Homi Bhabha Rd, Mumbai 400005, India}
\centerline{\sl $^{d}$Department of Physics and Astronomy,
University of British Columbia,}
\centerline{\sl Vancouver, BC, V6T 1Z1, Canada}
\medskip


\medskip

\noindent

We give an analytic demonstration that the $3+1$ dimensional large $N$
$SU(N)$ pure Yang-Mills theory, compactified on a small $S^3$ so that
the coupling constant at the compactification scale is very small, has
a first order deconfinement transition as a function of temperature.
We do this by explicitly computing the relevant terms in the canonical
partition function up to 3-loop order; this is necessary because the
leading (1-loop) result for the phase transition is precisely on the
borderline between a first order and a second order transition. Since
numerical work strongly suggests that the infinite volume large $N$
theory also has a first order deconfinement transition, we conjecture
that the phase structure is independent of the size of the $S^3$. To deal with divergences in our calculations, we are led to introduce a novel method of regularization useful for nonabelian gauge theory on $S^3$. 

\vskip 0.5cm \Date{February 2005}



\newsec{Introduction}

It is widely believed that $3+1$-dimensional $SU(N)$ Yang Mills
theory on $\IR^3$ confines at low temperatures, but is deconfined
at high temperatures. Compelling numerical evidence indicates
that in the absence of quarks, when all fields
are in the adjoint representation, there is a sharp phase
transition separating the confined and the deconfined phases, which
occurs at a temperature $T \sim \Lambda_{QCD}$. Since the Yang Mills
theory is strongly coupled at the transition temperature, the
deconfinement phase transition is rather poorly understood. In
particular, using the currently available analytic techniques it
is not possible to determine even the order of the transition for
$N \neq 3$; lattice simulations suggest that the transition
is of second order for $N=2$ and of first order for $N\geq 3$
(see \refs{\LuciniKU,\LuciniZR} for the latest results for $N>3$).

The intractability of the thermal behaviour of Yang Mills theory
on $\IR^3$ is related to the absence of a dimensionless coupling
constant. It is thus interesting to note that Yang Mills theory
compactified on an $S^3$ of radius $R$ has an effective
dimensionless coupling constant, $R \Lambda_{QCD}$. Indeed, when
$R \Lambda_{QCD} \ll 1$, the Yang Mills coupling constant is weak
even at the lowest energy scale in the theory, $E \sim 1 / R$. As
a consequence, at small values of $\Lambda_{QCD} R$, the thermal
behaviour of this system is completely tractable\foot{More
precisely, it is tractable for temperatures smaller than an upper
bound that scales to infinity in the limit $\Lambda_{QCD} R \to
0$, see section 5. This would not be true if the gauge field had zero
modes on the compact space, which is why we chose a sphere rather
than, say, a torus.}. Unfortunately, the most interesting feature
of infinite volume thermodynamics -- the sharp deconfinement phase
transition -- is smoothed out into crossover behaviour at any
finite $R$, assuming that $N$ is also finite.

However, in the `thermodynamic' $N \to \infty$ 't Hooft limit (with
fixed $\lambda \equiv g_{YM}^2 N$ \tHooftJZ), this
deconfinement phase transition remains sharp even at finite $R$. In
this limit it is possible to study the dynamics of the deconfinement
transition as a function of the effective coupling constant $R
\Lambda_{QCD}$. When $R\Lambda_{QCD} \gg 1$, this system approaches
the theory on $\IR^3$. On the other hand, in the opposite limit $R
\Lambda_{QCD} \to 0$ the theory is weakly coupled, and may be solved exactly
\refs{\sundborg, \first}; quite remarkably it turns out that this
`free' gauge theory undergoes a confinement-deconfinement phase
transition at a temperature of order $1/R$.\foot{As discussed in
\first, the $\lambda = 0$ theory must still obey a Gauss Law
constraint which requires physical states to be gauge invariant. This
constraint leads to nontrivial thermodynamics even at zero coupling.}

At strictly zero coupling, the transition is first order, but lies
precisely at the border between first and second order behaviour, as
reviewed below. Consequently, to understand the nature of the
transition at weak nonzero coupling, the leading effects of
interaction terms must be taken into account via a perturbative
calculation. This calculation is the goal of the present paper.

Before describing the calculation and our result, we recall the
essential details of the story in the $\lambda=0$ limit. It was
demonstrated in \refs{\sundborg, \first} that in the limit $R
\Lambda_{QCD} \rightarrow 0$ in which the theory becomes free, the
thermal partition function of Yang Mills theory on a 3-sphere of
radius $R$ reduces (up to an overall constant) to an integral over a
single unitary $SU(N)$ matrix\foot{We will generally ignore the
distinction between $SU(N)$ and $U(N)$ groups in this paper; the only
difference between their partition functions is an overall
coupling-independent factor coming from the free $U(1)$ photons.}
\eqn\eqform{
Z(\beta)= \int [dU] \exp[-S_{{\rm eff}}(U)],
}
where
\eqn\eqformt{
S_{{\rm eff}}(U)= -\sum_{n=1}^{\infty}
{z_V(e^{-n\beta/R}) \over n} \Tr (U^n) \Tr (U^{-n}); ~~~~~~~z_V(x)=
{6x^2 - 2x^3 \over (1-x)^3},
}
and $\beta\equiv 1/T$. The matrix $U$ is the holonomy of the gauge
field around the thermal circle (more precisely, $U=e^{i \beta
\alpha}$ with $\alpha$ the zero mode of $A_0$ on $S^3 \times S^1$
\first). All other modes of the theory are massive, and the effective
action is obtained by integrating them out.

As usual, in the large $N$ limit, it is
convenient to replace the integral over the unitary matrix $U$ by
an integral over the eigenvalue distribution
$\rho(\theta)\equiv{1\over N} \sum_i \delta(\theta - \theta_i)$
(where $e^{i \theta_i}$, for $i=1,\cdots,N$, are the eigenvalues
of $U$). Let $u_n$ denote the $n^{th}$ Fourier mode of the
eigenvalue distribution, $u_n\equiv \int e^{i n \theta}
\rho(\theta) d\theta =\Tr (U^n) / N$. In the large $N$ limit,
\eqform\ may be rewritten as \first\
\eqn\eqformm{Z(\beta)=\int
du_n d{\bar u}_n \exp\left[ -N^2 \sum_{n=1}^{\infty}
{\left(1-z_V(e^{-n\beta/R})\right) \over n} |u_n|^2 \right],}
with a complicated integration boundary for the $u_n$'s coming from the non-negativity of $\rho(\theta)$. Equation \eqformm\ may be
evaluated in the saddle point approximation in the large $N$
limit. This system has one obvious saddle point located at $u_n=0$
(for all $n \geq 1$). This saddle point is stable and dominates
\eqformm\ whenever $(1-z_V(e^{-n\beta/R}))$ is positive for all $n$,
which is the case for $T<T_c={1\over {R \ln(2+\sqrt{3})}}
\approx{0.759326 \over R}$.

At $T=T_c$, $z_V(e^{-\beta/R})=1$ and the potential for $u_1$ in
\eqformm\ is exactly flat (all other $u_n$ remain massive). For
$T>T_c$, the variable $u_1$ becomes tachyonic about the saddle
point described above. At these temperatures, the system is
dominated by a new saddle point, one in which $u_1$ has an
expectation value of order unity (see \first). Thus, free Yang
Mills theory on $S^3$ undergoes a phase transition at $T=T_c$; the
order parameter for this transition is the expectation value of
the Polyakov loop $u_1$ (more precisely, since $u_1$ has an
arbitrary phase, the order parameter is actually $|u_1|^2$
\first). Since the saddle point changes discontinuously at
$T=T_c$, this phase transition is of first order. However, this
phase transition is extremely finely tuned\foot{As one indication,
the order parameter $u_1$ is massless at the phase transition
point, a feature usually associated with second order
transitions.}, in a sense we will now explain.

As described above, the exact Wilsonian effective action for the order
parameter $u_1$, expanded about the low temperature saddle point in
the limit of zero gauge coupling, is $S_{eff}=N^2 \left(1
-z_V(e^{-\beta/R})\right)|u_1|^2$. This effective action is corrected at
nonzero gauge coupling; it was argued in \first\ that at nonzero
coupling it takes the form\foot{To obtain the explicit effective
action, we first integrate out all massive modes on the sphere, and
subsequently all of the other $u_n$ modes for $n>1$.}
\eqn\sefff{{S_{eff} \over
N^2}= \left[ \left(1 -z_V(e^{-\beta/R})\right) +\CO(\lambda)\right]
|u_1|^2 + \lambda^2 b(\beta)|u_1|^4 + \CO(\lambda^4),}
where $\lambda\equiv g_{YM}^2 N$ is the 't Hooft coupling
evaluated at the energy scale $1/R$, and $b(\beta)$ is a
perturbatively computable function of the temperature.

The action \sefff\ describes a system that undergoes a phase
transition at $T=T_c+\CO(\lambda)$. The nature of this phase
transition depends crucially on the sign of $b=b(\beta_c)$, where
$\beta_c=1/ T_c$. If $b$ is negative, the phase transition is of
first order, as in the free theory. However, unlike in the free
theory, this transition occurs at a lower temperature than the
temperature at which $u_1$ in \sefff\ becomes massless (the latter
temperature was identified in \refs{\sundborg,\first} with the
Hagedorn temperature of the large $N$ Yang-Mills theory).

On the other hand, if $b$ is positive, the phase transition
continues to occur precisely at the temperature at which $u_1$
becomes massless; however, it is now of second order and is
followed, at a slightly higher temperature, by another phase
transition of third order, similar to that of \GrossHE\ (see
section 6 of \first\ for more details).

The leading perturbative contribution to the value of $b$ at the
phase transition temperature is determined by a set of 2-loop and
3-loop vacuum diagrams in the Yang Mills theory on a sphere
\first. In this paper we evaluate $b$ by computing the relevant
Feynman diagrams.

Our main result is that $b(\beta_c)\simeq -5.7\times 10^{-4}$.
Note, in particular, that $b$ is negative; consequently, the
deconfinement transition for large $N$ $SU(N)$ Yang-Mills theory
on a 3-sphere with small but nonzero $R \Lambda_{QCD}$  is of
first order. As noted above, lattice simulations suggest that the
large $N$ deconfinement phase transition is also of first order in
the opposite infinite-volume limit $R \Lambda_{QCD} \to \infty$.
Thus, it is tempting to conjecture that the phase diagram for the
large $N$ $SU(N)$ Yang Mills theory on $S^3$ takes the form shown
in figure 1.

\fig{The simplest possible phase diagram for large $N$ Yang Mills
theory on $S^3$.}{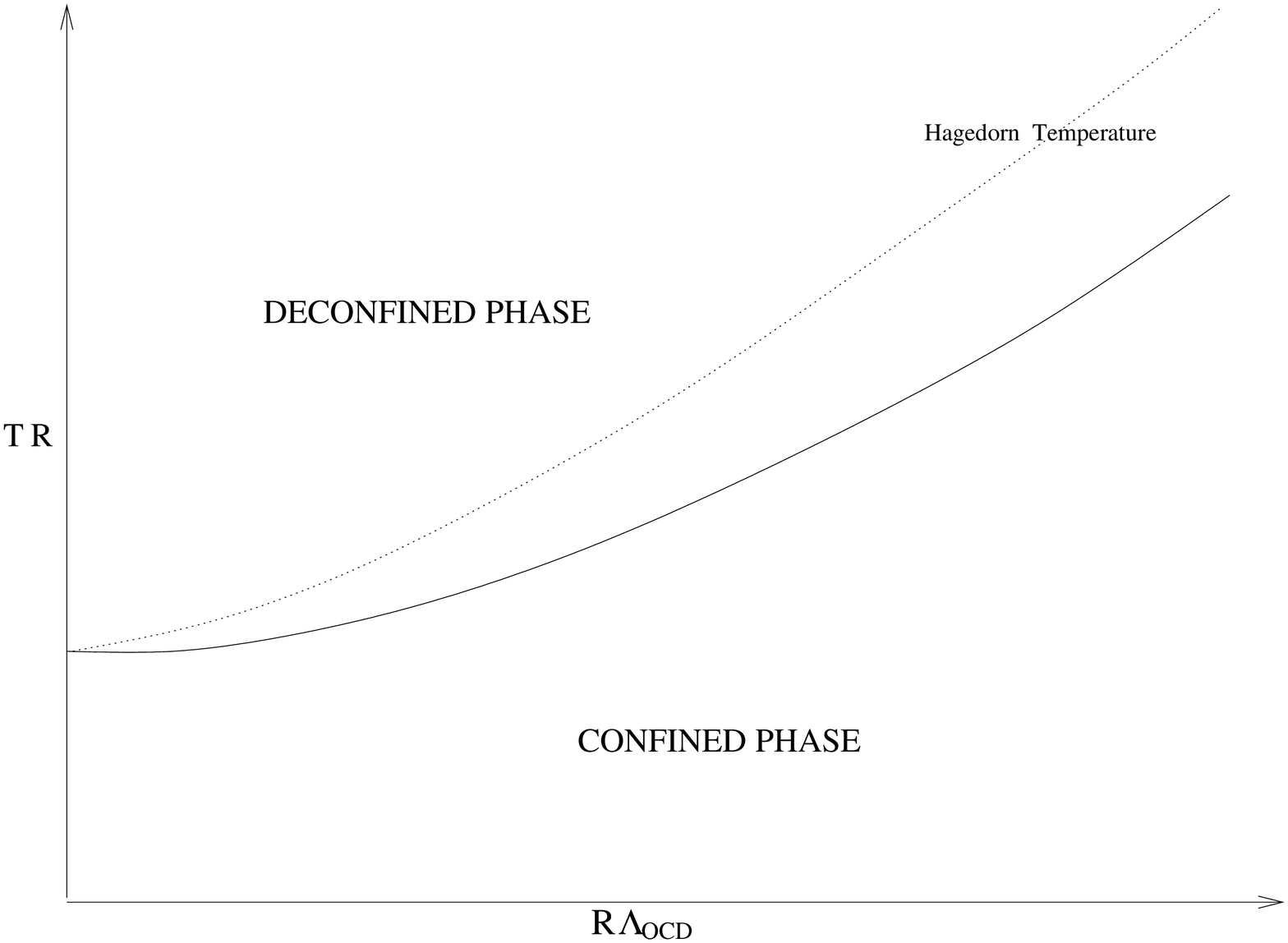}{4.0truein} \figlabel{\posbmicro}

The solid line in figure \posbmicro\ denotes the phase transition.
The dotted line describes the boundary of stability (the locus at
which the coefficient of $|u_1|^2$ is zero) of the low temperature
phase, which would be interpreted by a low temperature observer as
a limiting or Hagedorn temperature.

Of course, the conjectured phase diagram in figure \posbmicro\
merely represents the simplest phase diagram consistent with our
knowledge of the behaviour of all order parameters at weak and strong 
coupling. It
is possible that the true phase diagram is more complicated; for
instance, the confined phase at small $R \Lambda_{QCD}$ could be
separated from the confined phase at large $R\Lambda_{QCD}$ by a
phase transition. 

This paper is organized as follows. In section 2, we set up
our calculation, and describe, in general terms, the procedure we
employ to compute $b$. We then proceed in section 3 to enumerate and evaluate
all diagrams that contribute to $b$.

The lengthy calculation we present is straightforward in principle but
involves some subtleties.  First, it is necessary to deal with UV
divergences which show up in the perturbative computation. Due to
technical difficulties associated with implementing dimensional
regularization on $S^3$, we have found it convenient to regularize the
Feynman diagrams in a rather unusual fashion: we use a
(non-gauge-invariant) momentum cut-off, and simultaneously add in a
set of compensating non-gauge-invariant counterterms, such that the
full theory is gauge-invariant when the cutoff is removed. This
regularization procedure is described briefly in section 2.4, with a
more detailed discussion, including checks, examples, and the explicit
evaluation of the counterterms needed for our calculation presented in
section 4 and in appendix A.

In section 5, we discuss potential problems related to the infra-red
divergences associated with finite-temperature field theory. Naively
there should be no trouble at finite volume, where all modes are
massive. However, it turns out that there is still a breakdown of
perturbation theory at sufficiently high temperatures, when the
dynamically generated mass scales exceed the Kaluza-Klein scale
$1/R$. Fortunately, as we argue in section 5, these effects are not
important at the transition temperature and do not affect our
determination of $b$.

In section 6, we present some conclusions and discussion.

Finally, in order to compute sums over $S^3$ spherical harmonics,
which replace the loop integrals of the flat space theory, we needed
to derive various spherical harmonic identities. These identities,
together with some basic properties of the spherical harmonics for
$S^3$ are collected in appendix B.

\newsec{The Setup for the Perturbative Calculation}

As we have described, the order of the deconfinement phase transition
for pure Yang-Mills theory at small volume is determined by the sign
of the quartic coefficient $b$ in the effective action \sefff\ at the
$\lambda = 0$ transition temperature. In this section, we set up the
calculation of this coefficient, which may be determined from leading
order perturbative corrections to the matrix model action \eqformt\
for the $SU(N)$ pure Yang-Mills theory on $S^3$. For simplicity, we
will generally take the radius of the $S^3$ to be one; it can always be
reinstated by dimensional analysis. The actual
computation is presented in the next section.

\subsec{Basic objective}

The basic set-up for the computation was described in section 4 of
\first. We consider pure $SU(N)$ Yang-Mills theory on $S^3$, at
finite temperature. The thermal partition function may be computed by
evaluating the Euclidean path integral with Euclidean time
compactified on a circle of radius $\beta=1/T$. The Euclidean action
is given as usual by
\eqn\pureymlag{
{\cal L} = {1\over 4} \int_0^{\beta} dt \int d^3x \, \tr(F_{\mu \nu}
F^{\mu \nu}).  }
For calculations on $S^3$, it is convenient to work in the gauge
\eqn\gf{\del_i A^i=0,}
where $i=1,2,3$ runs over the sphere coordinates, and $\del_i$ are
(space-time) covariant derivatives. Equation \gf\ fixes the gauge
only partially; it leaves spatially independent (but time
dependent) gauge transformations unfixed. We fix this residual
gauge invariance by setting the constant mode of $A_0$ to be constant
in time,
\eqn\gff{\p_t \int_{S^3} A_0=0.}
It will be convenient to give this (time independent) zero mode a name; we define
\eqn\defalpha{\alpha = {g_{YM}\over \omega_3} \int_{S^3} A_0,}
where $\omega_3$ is the volume of the 3-sphere. $\alpha$ will play
a special role in what follows, because it turns out that it is
the only zero mode (mode whose action vanishes at quadratic
order) in the decomposition of Yang Mills theory into Kaluza-Klein
modes on $S^3 \times S^1$.

As $\alpha$ is a zero mode, it cannot
be integrated out in perturbation theory (roughly speaking,
$\alpha$ fluctuations are always strongly coupled in the bare
action). In order to perturbatively evaluate the free energy we
will therefore adopt a two step procedure. In the first step we integrate
out all nonzero modes and generate an effective action for
$\alpha$. As described in \first, this action will be non-trivial
even at zero coupling, and it is corrected in perturbation theory
in $\lambda$. In the second step, we analyze the remaining
integral over $\alpha$.

On general grounds described in \first, the finite temperature
effective action for $\alpha$ can be written completely in terms of the unitary
matrix $U \equiv e^{i\beta \alpha}$ in the form :
\eqn\seff{\eqalign{S_{eff}=&\sum_{m} C_{m,-m}(x) \tr(U^m)
\tr(U^{-m}) + \lambda \beta \sum_{m, n} C_{m,n,-m-n}(x) \tr(U^m)
\tr(U^{n}) \tr(U^{-m-n}) / N\cr & + \lambda^2  \beta \sum_{m, n, p}
C_{m,n, p,-m-n-p}(x) \tr(U^m) \tr(U^{n}) \tr(U^{p}) \tr(U^{-m-n
-p}) / N^2 + \cdots,}}
where $x\equiv e^{-\beta}$. Here, the coefficients obey appropriate
constraints such that the action is real, and in general are corrected
at higher orders in $\lambda$ and $1/N$.  As described in \first, the
free energy $F$ of the Yang Mills theory is then given by the matrix
integral \eqn\intaft{e^{-\beta F}=
\int [dU] e^{-S_{eff}(U)},} where the Fadeev-Popov determinant
corresponding to the gauge-fixing \gff\ transforms the integral
over $\alpha$ into an integral over the gauge group with Haar measure $[dU]$.

At large $N$, the unitary integral may be evaluated using saddle point techniques.
Defining $u_n$ = $\tr(U^n)/N$, we write the effective action in
the form
\eqn\actfort{ Z = \int [d u_i]e^{- N^2 S'_{eff}(u_n)}. }
where $S'_{eff}$ includes contributions both from $S_{eff}$ and from the
Vandermonde determinant obtained in changing to the variables $u_n$. The
order $N^2$ contribution to the free energy is then given by the
minimum value of $S'_{eff}(u_n)$, and the deconfinement phase
transition occurs where this minimum is no longer at $|u_n|=0$.

As described in detail in section 6 of \first, in order to compute
the order of this phase transition, we have to look at $S'_{
eff}(u_n)$ near the phase transition point $x_c=2-\sqrt{3}$ where
(as shown in \refs{\sundborg,\first} and reviewed in the
introduction) the mass term of $u_1$ changes sign, and compute the
leading corrections to the potential for $u_1$. The relevant terms
in the action, to leading order in $\lambda$ and in $x-x_c$, take the form
\eqn\leadingf{\eqalign{ S'_{{\rm eff}}(u_n) =& \mu_1 (x_c-x)|u_1|^2
+ \mu_2 |u_2|^2 + \dots \cr &+ \lambda \beta [C_{1,1,-2} (u_1^2
\bar{u}_2 + \bar{u}_1^2 u_2) + \cdots] \cr &+ \lambda^2 \beta
[C_{1,1,-1,-1}|u_1|^4 + \cdots]. }}
At large $N$, the effective action \sefff\ for $u_1$ is obtained from
this by classically minimizing over $u_n$ for fixed $u_1$. In
particular, the variable $u_2$ may now be integrated out in \leadingf\
by setting it to its classical value
\eqn\forttwo{ u_2 = - \lambda {\beta_c C_{1,1,-2} \over \mu_2}
u_1^2 + {\cal O} (\lambda^2). }
This yields the following effective action for $u_1$ :
\eqn\forfeff{ S'_{eff}(u_1) = \mu_1(x_c-x) |u_1|^2 + b\lambda^2
|u_1|^4 + \cdots, }
where
\eqn\bdefdag{b =\beta_c C_{1,1,-1,-1}-\frac{\beta_c^2
C_{1,1,-2}^2}{\mu_2}.}
Our goal will be to compute the coefficients appearing here at the
leading order in $\lambda$, and thus determine the sign of $b$, and
the order of the transition, at weak coupling. Since an $n$-loop
diagram has at most $n+1$ index lines, a term in $S_{eff}$ with $m$
traces gets its lowest order contributions at $m-1$ loops. Thus, the
term $\mu_2$ arises at one-loop order, $C_{1,1,-2}$ requires a
two-loop computation, while $C_{1,1,-1,-1}$ requires a three-loop
calculation.

\subsec{Gauge-fixed action}

We now set up the perturbative computation that will determine
$S_{eff}(U)$, the effective action for $U$ (which we treat as a
background field). The Fadeev-Popov determinant corresponding to the gauge
fixing condition \gf\ is
\eqn\fpd{ \det{\del_i D^i } = \int \CD c
\CD {\bar c} e^{- \tr({\bar c} \del_i D^i c)},}
where $D^i$ denotes a gauge covariant derivative
\eqn\gaugecovariant{D_i c=\del_i-i g_{YM} [A_i,c],}
and $c$ and ${\bar c}$ are complex
ghosts in the adjoint representation of the gauge group. The
quadratic terms in the gauge-fixed Yang-Mills action \pureymlag\ take the form
\eqn\fymact{ -\int d^4x \tr \left( \half A_i (\tdzero^2
+\del^2) A^i + \half A_0 \del^2 A_0 + {\bar c} \del^2 c\right),}
where $\del^2$ is the Laplacian on the sphere and
\eqn\deft{\tdzero X \equiv \p_0 X -i[\alpha, X].}
The interaction terms in \pureymlag\ are given by
\eqn\ymint{\eqalign{ \int d^4x \tr (  & ig_{YM} \tdzero A^i
[A_i,A_0] -ig_{YM}[A^i,A^0] \del_i A_0 -i g_{YM}\del_i A_j [A^i,
A^j] + \cr & {g_{YM}^2\over 4} [A_i, A_j][A^j, A^i]
-{g_{YM}^2\over 2} [A_0,A_i][A^0,A^i] -i g_{YM}\del_i {\bar c}
[A_i, c]). \cr}}

%

\subsec{The spherical harmonic expansion}

On $S^3$, integrals over spatial momenta are replaced by sums over the
quantum numbers of $SO(4)$ angular momenta.  It will thus be useful to
expand the fields explicitly in terms of an orthonormal basis of
functions on $S^3$ which are angular momentum eigenstates, and write
the action explicitly in terms of a standard set of integrals over
these spherical harmonic functions. We denote the scalar and vector
spherical harmonics on $S^3$ by $S^\alpha(\Omega)$ and
$V_i^\beta(\Omega)$, where $\alpha = (j_\alpha ,m_\alpha ,m'_\alpha)$
and $\beta = (j_\beta, m_\beta, m'_\beta,
\epsilon_\beta)$ are the angular momentum (and parity for the
vector field) quantum numbers for the various modes. The properties of
these functions are reviewed in appendix B.

We expand the modes of the fields in terms of these spherical harmonics as
follows :
\eqn\expand{\eqalign{ A_0(t,\theta) & = \sum_{\alpha}
a^{\alpha}(t) S^{\alpha}(\theta); \cr A_i(t,\theta) &=
\sum_{\beta} A^{\beta}(t) V_i^{\beta}(\theta); \cr c(t,\theta) &=
\sum_{\alpha} c^{\alpha}(t) S^{\alpha}(\theta). \cr}}
Note that general vector functions also include modes proportional to
$\nabla S^\alpha$, but these are eliminated by our gauge
choice. Below, it will be useful to denote the complex conjugates of
$S^{\alpha}$ and $V_i^{\beta}$ by $S^{\bar \alpha}$ and $V_i^{\bar
\beta}$.

In terms of these spherical harmonics, we define
\eqn\sphconv{\eqalign{C^{\alpha \beta \gamma} &=
\int_{S^3} S^\alpha \vec{V}^\beta \cdot \vec{\nabla} S^\gamma, \cr
D^{\alpha \beta \gamma} &= \int_{S^3} \vec{V}^\alpha \cdot
\vec{V}^\beta S^\gamma, \cr E^{\alpha \beta \gamma} &= \int_{S^3}
\vec{V}^\alpha \cdot (\vec{V}^\beta \times \vec{V}^\gamma), \cr}}
where explicit expressions for $C$, $D$ and $E$ may be found
in \Cutkosky\ and are collected in appendix B. Note that
$C$ is
antisymmetric in $\alpha$ and $\gamma$, $D$ is symmetric in
$\alpha$ and $\beta$, and $E$ is totally antisymmetric.

Using the spherical harmonic expansions, we may now write the action
for gauge-fixed pure Yang-Mills theory on $S^3$ explicitly in terms of
modes. The quadratic action becomes
\eqn\quadratic{\eqalign{ S_2 = \int dt \tr( {1 \over 2} A^{\bar{\alpha}}
( - \tdzero^2 + (j_\alpha+1)^2) A^\alpha +
{1 \over 2} a^{\bar{\alpha}} j_\alpha (j_\alpha + 2) a^\alpha +
\bar{c}^{\bar{\alpha}} j_\alpha (j_\alpha + 2) c^\alpha ).
}}
In addition, we have cubic interactions
\eqn\cubic{\eqalign{ S_3 = g_{YM} \int dt \, \tr( & i
\bar{c}^{\bar \alpha} [A^\gamma, c^\beta] C^{\bar{\alpha} \gamma
\beta} +2i a^\alpha A^\gamma a^\beta C^{\alpha \gamma \beta} \cr
&-i[A^\alpha, D_\tau A^\beta]a^\gamma D^{\alpha \beta \gamma} + i
 A^\alpha A^\beta A^\gamma \epsilon_\alpha (j_\alpha + 1)
E^{\alpha \beta \gamma}), }}
and quartic interactions
\eqn\quartic{\eqalign{ S_4 = g_{YM}^2 \int dt \, \tr( & -{1 \over
2} [a^\alpha, A^\beta][a^\gamma, A^\delta]\left(D^{\beta
\bar{\lambda} \alpha} D^{\delta \lambda \gamma} + {1 \over
j_\lambda (j_\lambda+2)} C^{\alpha \beta \bar{\lambda}} C^{\gamma
\delta \lambda} \right) \cr & - {1 \over 2} A^\alpha A^\beta
A^\gamma A^\delta \left( D^{\alpha \gamma \bar{\lambda}} D^{\beta
\delta \lambda} - D^{\alpha \delta \bar{\lambda}} D^{\beta \gamma
\lambda} \right)).}}
The propagators of the various fields follow from \quadratic\ and are
given by
\eqn\propc{ \langle \bar{c}^{\bar \alpha}_{ab}(t)
c_{cd}^\beta(t') \rangle = {1 \over j_\alpha (j_\alpha +2)}
\delta^{\alpha \beta} \delta(t - t') \delta_{ad} \delta_{cb}, }
\eqn\propa{ \langle a_{ab}^\alpha(t) a_{cd}^\beta(t') \rangle = {1
\over j_\alpha (j_\alpha +2)} \delta^{\alpha \bar{\beta}} \delta(t
- t') \delta_{ad} \delta_{cb} , }
\eqn\propbiga{
\langle
A^\alpha_{ab}(t) A^\beta_{cd}(t') \rangle = \delta^{\alpha
\bar{\beta}} \Delta^{ad,cb}_{j_\alpha}(t-t'). }
Here, $\Delta$ is defined to be a periodic function of time satisfying
\eqn\propdef{
(-\tdzero^2 +(j+1)^2 )\Delta_j(t)= \delta(t)
}
where we have suppressed matrix indices. For $0 \le t \le \beta$, the
explicit solution is given by
\eqn\defdelta{ \Delta_{j}(t) \equiv {e^{i \alpha t} \over 2(j+1)}
\left( {e^{-(j+1)t} \over 1 - e^{i \alpha \beta} e^{-(j+1) \beta}}
- {e^{(j+1)t} \over 1 - e^{i \alpha \beta} e^{(j+1) \beta}}
\right) }
with the value for other values of $t$ defined by the
periodicity. Here, $\alpha$ is shorthand for $\alpha \otimes 1 - 1
\otimes \alpha$, and a term $\alpha^n \otimes \alpha^m$ in the
expansion of $\Delta$ should be understood to carry indices
$(\alpha^n)^{ad}(\alpha^m)^{cb}$ in \propbiga.

For our calculations, the following correlators are also useful
\eqn\propdta{
\langle D_\tau A^\alpha_{ab}(t) A^\beta_{cd}(t') \rangle = -
\langle A^\alpha_{ab}(t) D_\tau A^\beta_{cd} (t') \rangle =
\delta^{\alpha \bar{\beta}} D_\tau
\Delta^{ad,cb}_{j_\alpha}(t-t'), } \eqn\dtdt{ \langle D_\tau
A^\alpha_{ab}(t) D_\tau A^\beta_{cd}(t') \rangle = \delta^{\alpha
\bar{\beta}} \delta(t - t') \delta_{ad} \delta_{cb} -
\delta^{\alpha \bar{\beta}} (j_\alpha + 1)^2
\Delta^{ad,cb}_{j_\alpha}(t-t'). }

\subsec{Regularization and counterterms}

As usual with four dimensional gauge theories, certain perturbative
calculations lead naively to ultraviolet divergences. In our
calculation of the coefficient $b$ in \sefff, we will find that all
divergences cancel, but only after summing a collection of
logarithmically divergent diagrams. It is therefore necessary to
introduce a regularization scheme, and in order to obtain the correct
finite result, this must respect gauge invariance.

In principle, there is no obstacle in taking the usual approach of
applying dimensional regularization. In our case, this amounts to
considering gauge theory on $S^3 \times S^1 \times \IR^{d-4}$.\foot{As
we describe in section 4, in order that all momentum
sums/integrals are rendered finite when employing dimensional
regularization in Coulomb gauge, it is necessary to analytically
continue both the number of dimensions which participate in the
Coulomb gauge condition, and the number of dimensions which do not, a procedure referred to as split dimensional regularization.}
While it is straightforward to write down the appropriate expressions
for Feynman diagrams in this dimensionally regulated theory, we have
found it difficult to evaluate them because of the combination of
momentum sums and integrals that appear. Thus, we have found it
helpful to apply a more unusual approach, which nevertheless gives
precisely the answers that would have been obtained through
dimensional regularization.

In practice, we apply a sharp momentum cutoff to the total angular
momentum quantum number for modes on the sphere. This does not respect
gauge invariance. However, as explained originally by 't Hooft
\thooftFH\ in his series of classic papers on the renormalizability of
Yang Mills theory, a non gauge invariant regularization yields gauge
invariant results when employed with a bare Yang Mills action that
includes an appropriate set of non-gauge-invariant counterterms. As we
describe in section 4, these counterterms may be determined by
demanding that simple Green's functions evaluated using the cutoff and
counterterms agree with the same Green's functions evaluated using
dimensional regularization. Apart from curvature-dependent
counterterms (which do not contribute to our calculation), this
comparison may be carried out in flat space, since all counterterms
must be local.

In section 4, we present the calculation to determine the precise
coefficients for all counterterms necessary in our calculation,
together with a more complete discussion of the regularization scheme
and a variety of consistency checks. In the end, we should expect
non-zero coefficients for all counterterms with dimension less than or
equal to four which respect $SO(3)$ invariance.

\newsec{The Perturbative Computation}

In this section, we proceed to calculate the coefficients in
\leadingf\ necessary to determine the order of the deconfinement
transition at weak coupling. We have
\eqn\general{\eqalign{
e^{-S_{eff}(U)} &= \int [da] [dA] [dc] e^{-S(\alpha,a,A,c)} \cr
&= e^{-S_{eff}^{1 \; loop}} \langle e^{-S_3-S_4} \rangle,
}}
where the expectation value in the last line is evaluated in the free
theory with propagators given in section 2. The required leading order
contributions to $\mu_2$, $C_{1,1,-2}$, and $C_{1,1,-1,-1}$ appear at
one, two, and three loops respectively.

\subsec{Simplifying the action by integrating out $A_0$ and $c$}

Since the action is quadratic in $a$ and $c$, these may be integrated
out explicitly to yield additional interaction vertices for the
$A$'s. The first contribution arises from loops of $a$ or $c$.  For
the calculation up to three loops that we are interested in here, the
relevant vertices in the resulting effective action (combining $a$ and
$c$ loops) are a quadratic vertex
\eqn\defatwo{ A_2 = g_{YM}^2 N \delta(0)
{D^{\gamma_1 \alpha \beta} D^{\gamma_2 \bar{\alpha} \bar{\beta}}
\over j_\beta (j_\beta + 2) } \tr(A^{\gamma_1} A^{\gamma_2}), }
a cubic vertex
\eqn\defathree{ A_3 = - 2 i g_{YM}^3 N \delta(0)
{C^{\bar{\alpha} \gamma_1 \bar{\beta}} D^{\gamma_2 \delta \beta}
D^{\gamma_3 \bar{\delta} \alpha} \over j_\alpha (j_\alpha + 2)
j_\beta (j_\beta +2)} \tr(A^{\gamma_1} A^{\gamma_2} A^{\gamma_3}),
}
and a quartic vertex
\eqn\quartvert{\eqalign{ A_4 = -g_{YM}^4 N
\delta(0) & \left( 3 {C^{\alpha \gamma_1 \bar{\beta}} C^{\beta
\gamma_2 \bar{\rho}} D^{\gamma_3 \lambda \rho} D^{\gamma_4
\bar{\lambda} \bar{\alpha}} \over j_\alpha (j_\alpha + 2) j_\beta
(j_\beta + 2) j_\rho (j_\rho + 2)}  + {1 \over 2} { D^{\gamma_1
\lambda \alpha} D^{\gamma_2 \bar{\lambda} \beta} D^{\gamma_3  \rho
\bar{\beta}} D^{\gamma_4 \bar{\rho} \bar{\alpha}} \over j_\alpha
(j_\alpha + 2) j_\beta (j_\beta + 2)} \right) \cr & \times \left(
\tr(A^{\gamma_1} A^{\gamma_2} A^{\gamma_3} A^{\gamma_4}) + {1
\over N} \tr(A^{\gamma_1} A^{\gamma_2}) \tr(A^{\gamma_3}
A^{\gamma_4}) \right. \cr & \qquad \qquad \left. + {1 \over N}
\tr(A^{\gamma_1} A^{\gamma_3}) \tr(A^{\gamma_2} A^{\gamma_4})+ {1
\over N} \tr(A^{\gamma_1} A^{\gamma_4}) \tr(A^{\gamma_2}
A^{\gamma_3}) \right). }}
We have included here only terms that can contribute to planar
diagrams. Note that all of these are proportional to a divergent
factor $\delta(0)$, and so any diagrams containing these vertices must
eventually cancel.\foot{The divergence associated with $\delta(0)$
terms may be regulated using a momentum cutoff in the $S^1$
direction. We have checked that in the properly regulated theory, the
naive cancellations of $\delta(0)$ terms described below persist
without introducing any new finite contributions.} Note also that
since these vertices arise from loops, they have additional factors of
$g_{YM}^2 N$ compared to
\cubic, \quartic. Therefore, inserting any one of these vertices into
a diagram counts as an additional loop.

In addition to these, we have vertices arising from open strings
of $a$'s containing two vertices linear in $a$ and some number of
vertices quadratic in $a$. These start at quartic order, and for
our three loop calculation, we will need the quartic, quintic, and
sextic vertices. These are
\eqn\Bverts{\eqalign{ B_4 &= {g_{YM}^2
\over 2} {D^{\alpha_1 \beta_1 \gamma} D^{\alpha_2 \beta_2
\bar{\gamma}} \over j_\gamma (j_\gamma + 2)} \tr([A^{\alpha_1},
D_\tau A^{\beta_1}] [A^{\alpha_2}, D_\tau A^{\beta_2}]), \cr B_5
&= -i g_{YM}^3 {D^{\alpha_1 \beta_1 \lambda} C^{\bar{\lambda}
\gamma \bar{\sigma}} D^{\alpha_2 \beta_2 \sigma} \over j_\lambda
(j_\lambda + 2) j_\sigma (j_\sigma + 2)} \tr([A^{\alpha_1}, D_\tau
A^{\beta_1}][A^\gamma, [A^{\alpha_2}, D_\tau A^{\beta_2}]]), \cr
B_6 &= {g_{YM}^4 \over 2} \left( 3 {D^{\alpha_1 \beta_1 \sigma}
C^{\bar{\sigma} \gamma_1 \tau} C^{\bar{\tau} \gamma_2 \lambda}
D^{\alpha_2 \beta_2 \bar{\lambda}} \over j_\lambda (j_\lambda + 2)
j_\sigma (j_\sigma + 2) j_\tau (j_\tau + 2) } + {D^{\alpha_1
\beta_1 \sigma} D^{\gamma_1 \lambda \bar{\sigma}} D^{\gamma_2
\bar{\lambda} \bar{\tau}} D^{\alpha_2 \beta_2 \tau} \over j_\sigma
(j_\sigma + 2) j_\tau (j_\tau + 2) } \right) \cr & \qquad
\tr([[A^{\alpha_1}, D_\tau A^{\beta_1}],A^{\gamma_1}]
[[A^{\alpha_2}, D_\tau A^{\beta_2}],A^{\gamma_2}]). \cr }}
Using these effective vertices, it is straightforward to check
that all divergent contributions proportional to $\delta(0)$
cancel. Any diagram with a vertex $A_2$ will cancel the
$\delta(0)$ part of a similar diagram with $A_2$ replaced with a
$B_4$ having its two covariant derivative legs contracted.
Similarly, any diagram with a vertex $A_3$ will cancel the
$\delta(0)$ part of a similar diagram with $A_3$ replaced with a
$B_5$ having its two covariant derivative legs contracted.
Finally, any diagram containing $A_4$ will cancel against a
combination of two diagrams: one with $A_4$ replaced by $B_6$ with
its two covariant derivative legs contracted, and one with $A_4$
replaced by two $B_4$'s with the four covariant derivative legs
contracted into a loop.

Thus, all $\delta(0)$ terms coming from the $A_n$ vertices cancel
out, and it is easy to check that there are no additional
$\delta(0)$ terms coming from diagrams with $B_n$ vertices apart
from those needed to cancel the $A_n$ vertex diagrams. As a
result, we may proceed with the calculation by keeping only the
transverse photons $A^{\alpha}$, and evaluating all diagrams built
from the original cubic and quartic vertices in the transverse
photons plus the additional vertices $B_4$, $B_5$, and $B_6$,
ignoring any terms proportional to $\delta(0)$.

\fig{The diagrams contributing to the free energy up to 3-loop
order. In this figure we present a particular planar form for each
diagram, but in some cases the same diagram may also be drawn in
the plane in different ways. There is also an additional
counter-term diagram that will be discussed
later.}{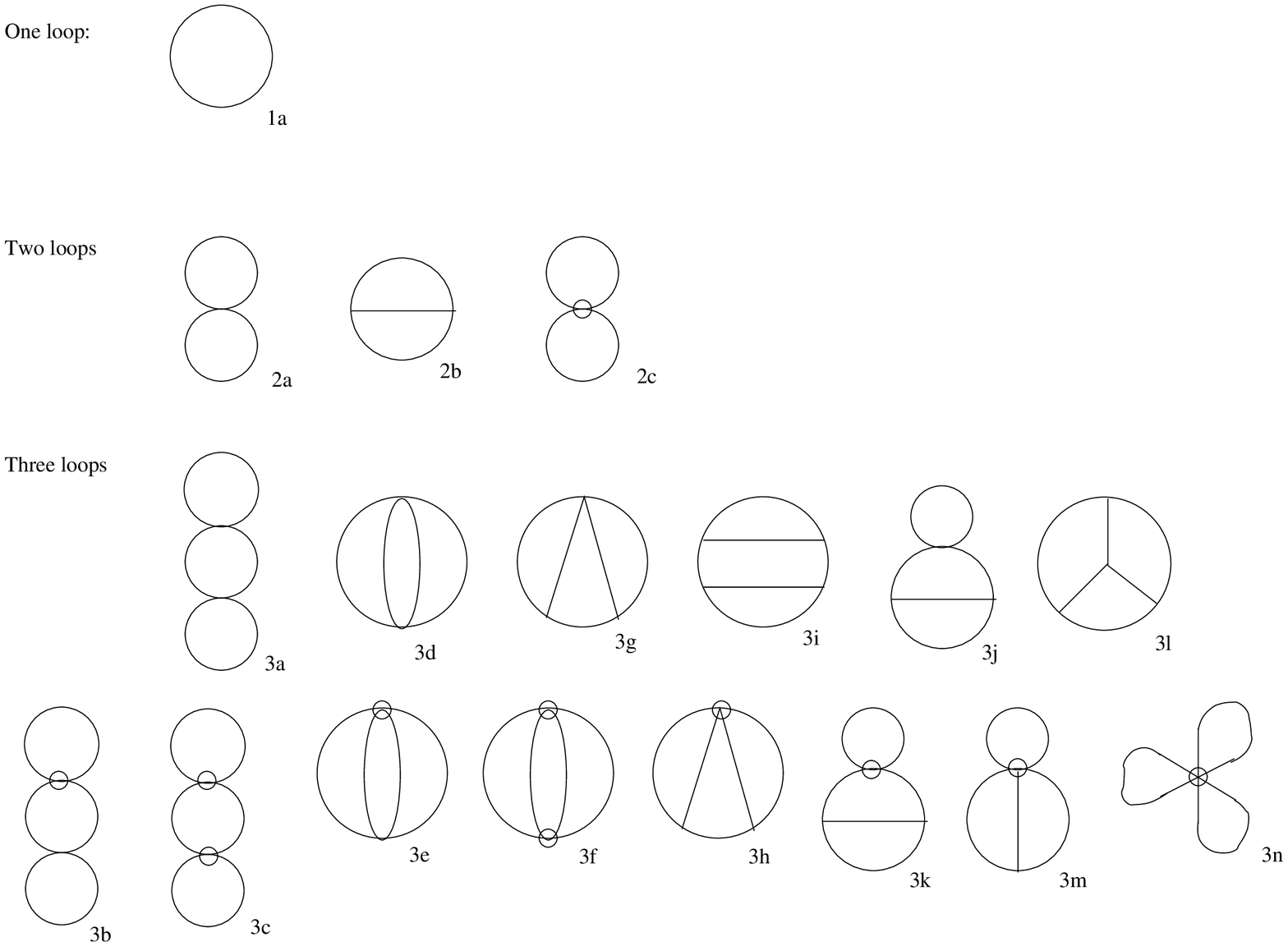}{7truein} \figlabel{\diagrams}
The diagrams contributing to the free energy at one, two and three
loop orders, after having integrated out $A_0$ and $c$, are shown
in figure \diagrams. The $B$-type vertices are denoted by circles.

\subsec{1-Loop}

The one-loop computation of the path integral was described in
\first. The result (writing only the leading terms in the large
$N$ limit) is
\eqn\effact{S_{eff}(U) = {11\beta N^2\over 120} -
\sum_{n=1}^{\infty} \frac{z_V(x^n)}{n} \tr(U^n) \tr(U^{-n}), }
where $z_V(x)$ is the single-particle partition function for a
free vector field on $S^3$, given by
\eqn\forzv{z_V(x) = {{6x^2 - 2x^3}\over {(1-x)^3}}.}
Changing variables to $u_n = \tr(U^n)/N$ and including the additional
Vandermonde determinant from the measure, we have
\eqn\effacttwo{S'_{eff}(u_n) = {11\beta \over 120} +
\sum_{n=1}^{\infty} {1 \over n} (1 - {z_V(x^n)}) |u_n|^2, }
and thus
\eqn\mutwo{
\mu_2(x_c) = {1 \over 2}(1 - z_V(x_c^2)) \approx 0.481125 \; .
}

\subsec{Two Loops}

In this section we compute the coefficient $C_{1,1,-2}$ in \leadingf\
by evaluating the three two-loop diagrams of figure \diagrams.  Note
that since the propagators depend only on the relative time between
vertices, it is always possible to change variables so that the
integrand is independent of one of the time variables. The integral
over this variable then gives an overall factor of $\beta$ so it is
convenient to define $F(U) = S_{eff}(U)/\beta$. In terms of the
propagators and spherical harmonic integrals defined above, we find
that the three two-loop contributions to $F$ are\foot{Recall that in
diagram 2c, we ignore the part proportional to $\delta(0)$.} (with
summation over the spherical harmonic indices
$\alpha$, $\beta$ and $\gamma$ implied)
\eqn\stwo{\eqalign{
F_{2a} &= - {g_{YM}^2 \over 2} (D^{\alpha
\beta  \gamma} D^{\bar{\alpha} \bar{\beta} \bar{\gamma}} -
D^{\alpha \bar{\alpha} \gamma} D^{\beta \bar{\beta} \bar{\gamma}})
\Delta_{j_\alpha}(0, \alpha_{ab})
 \Delta_{j_\beta}(0, \alpha_{ac}),  \cr
F_{2b} &= - {g_{YM}^2 \over 6}
\hat{E}^{\alpha \beta \gamma} 
\hat{E}^{\bar{\alpha} \bar{\beta}\bar{\gamma}}
\int dt \Delta_{j_\alpha}(t, \alpha_{ab})
\Delta_{j_\beta}(t, \alpha_{bc}) \Delta_{j_\gamma}(t, \alpha_{ca}),
\cr
F_{2c} &= g_{YM}^2 {D^{\alpha \beta \gamma}
D^{\bar{\alpha} \bar{\beta} \bar{\gamma}} \over j_\gamma
(j_\gamma+2)} (D_\tau \Delta_{j_\alpha}(0, \alpha_{ac}) D_\tau
\Delta_{j_\beta}(0, \alpha_{ab}) + (j_\beta + 1)^2
\Delta_{j_\alpha}(0, \alpha_{bc})
\Delta_{j_\beta}(0, \alpha_{ab}) ) ,
}}
where we have defined
\eqn\ehat{ \hat{E}^{\a \b \g} \equiv E^{\a \b \g} (\e_\a (j_\a +
1) + \e_\b (j_\b + 1) + \e_\g (j_\g + 1)) \; . }
In the expressions above, each of the propagators contributes factors
of $\alpha$ to two of the three index loops, which we label by
$a$,$b$, and $c$. The notation $\alpha_{ab}$ indicates that for the
tensor products $(\alpha \otimes 1 - 1 \otimes \alpha)$ appearing in
the propagator, the first and second elements of the tensor product
appear in the traces associated with index loops $a$ and $b$
respectively.

The expressions \stwo\ involve sums over products of spherical harmonic
integrals. In all cases here and below, the sums over quantum numbers
$m$, $m'$ and $\epsilon$ may be carried out explicitly using the
formulae in appendix B. To express the results, it is useful,
following
\Cutkosky, to define some functions which appear in integrals of
products of spherical harmonics
\eqn\rtwoo{R_2(x,y,z)={
(-1)^{\sigma'}\over\pi}\left[\frac{(x+1)(z+1)(\sigma'-x)(\sigma'-y)(\sigma'-
z)(\sigma'+1)}{(y+1)}\right]^{1/2},}
\eqn\Rcombo{\eqalign{ R_{3 \e_x \e_z}(x,y,z) =& {(-1)^{\s +
(\e_x+\e_z)/2} \over \pi}\left({(y+1) \over 32
(x+1)(z+1)}\right)^{1 \over 2} \cr & \cdot \left( (\e_x(x+1) +
\e_z(z+1) + y+2)(\e_x(x+1) + \e_z(z+1)
 + y) \right. \cr
& \quad \left. (\e_x (x+1) + \e_z (z+1)  - y)(\e_x (x+1) +
\e_z(z+1) -
 y - 2) \right)^{1 \over 2}, \cr
R_{4 \e_x \e_y \e_z}(x,y,z) =& {(-1)^{\s'+1} \over \pi} {\rm
sign}(\e_x+\e_y+\e_z) \left({(\s'+1)(\s'-x)(\s'-y) (\s'-z) \over
4(x+1)(y+1)(z+1)}\right)^{1 \over 2} \cr &
\!\!\!\!\!\!\!\!\!\!\!\!\!\!\!\!\!\!\!\!\!\!\!\!\!\!\!\!\!\!\!\! 
\cdot \left( (\e_x(x+1) +
\e_y(y+1) + \e_z(z+1) +2 ) (\e_x(x+1) + \e_y(y+1) + \e_z (z+1)-2)
\right)^{1 \over 2}, \cr \hat{R}_{4 \e_x \e_y \e_z}(x,y,z) =& R_{4
\e_x \e_y \e_z}(x,y,z) (\e_x (x+1) + \e_y (y+1) + \e_z (z+1)), }}
where the right-hand sides of the equations are defined to be
non-zero only if the triangle inequality $|x-z| \le y \le x+z$
holds, and if $\s\equiv (x+y+z)/2$ (in $R_3$) and $\s' \equiv
(x+y+z+1)/2$ (in $R_2$ and $R_4$) are integers. We also define
$R_{3+}\equiv R_{3++}=R_{3--}$, $R_{3-}\equiv R_{3+-}=R_{3-+}$,
$R_{4+}\equiv R_{4+++}$ and $R_{4-}\equiv R_{4++-}$.

With these definitions,
we find after performing the sums over $m$, $m'$ and $\epsilon$, that
\eqn\twoa{\eqalign{ F_{2a}
&={2 g_{YM}^2 \over 3 \pi^2}  j_\alpha (j_\alpha + 2)
 j_\beta (j_\beta + 2)  \Delta_{j_\alpha}(0, \alpha_{ab})
 \Delta_{j_\beta}(0, \alpha_{ac}), \cr
F_{2b} &= -g_{YM}^2 ({1 \over 3} (j_\alpha + j_\beta + j_\gamma +
3)^2 R_{4+}^2(j_\alpha,j_\beta,j_\gamma) + (j_\alpha + j_\beta -
j_\gamma + 1)^2 R_{4-}^2(j_\alpha,j_\beta,j_\gamma)) \cr & \qquad
\int dt \Delta_{j_\alpha}(t, \alpha_{ab}) \Delta_{j_\beta}(t,
\alpha_{bc}) \Delta_{j_\gamma}(t, \alpha_{ca}), \cr
F_{2c} &={2 g_{YM}^2 \over j_\gamma (j_\gamma+2)}
(R_{3+}^2(j_\alpha,j_\gamma,j_\beta) +
R_{3-}^2(j_\alpha,j_\gamma,j_\beta)) \cr & \qquad \qquad (D_\tau
\Delta_{j_\beta} (0, \alpha_{ab}) D_\tau \Delta_{j_\alpha} (0,
\alpha_{ac}) + (j_\beta + 1)^2 \Delta_{j_\alpha} (0, \alpha_{bc})
\Delta_{j_\beta}(0, \alpha_{ab})). }}
These expressions are all to be summed over the $j$'s, with the
sums unconstrained in $F_{2a}$, and constrained in $F_{2b}$ and
$F_{2c}$ by the rules given above.

As described above, in order to analyze the phase transition we
need to compute the specific term in $F_{2a}+F_{2b}+F_{2c}$ of the
form
\eqn\intterm{ C_{1,1,-2} g_{YM}^2 (\tr(U) \tr(U)
\tr((U^\dagger)^2) + \tr(U^2) \tr(U^\dagger) \tr(U^\dagger)), }
and to determine the coefficient $C_{1,1,-2}$ at the deconfinement
temperature of the free Yang-Mills theory, $x_c=2-\sqrt{3}$. For each
diagram, we therefore expand the product of propagators in powers of
$U$, and sum the coefficients of all terms of the form \intterm.

For $F_{2a}$, all sums may be done explicitly, and we find
\eqn\fthreesummed{\eqalign{ F_{2a} = {g_{YM}^2 \over 24 \pi^2}
\sum_{n,m \ge 0} \tilde{F}(x^n) \tilde{F}(x^m) (& \tr(U^m) \tr(U^n)
\tr(U^{-n-m}) + \{m \to -m \} +\cr & \{n \to -n\} + \{m,n \to
-m,-n\}) \cr +{ g_{YM}^2 \over 3 \pi^2} (\sum {j(j+2)\over (j+1)})
\sum_n & \tilde{F}(x^n) N \tr(U^n) \tr(U^{-n}) + {g_{YM}^2 \over 6
\pi^2}(\sum {j(j+2)\over (j+1)})^2 N^3, }}
where we have left the divergent sums (which will not be relevant
for the computation we are doing in this paper) explicit. The
function $\tilde{F}$ is related to the single particle partition
function $z_V$ by
\eqn\deftildef{ \tilde{F}(e^{-\beta}) =
\int_\beta^\infty da z_V(e^{-a}), }
or explicitly
\eqn\formtildef{
\tilde{F}(x) = 2\ln(1-x) + {2x \over (1-x)^2}. }
Thus, the contribution to $C_{1,1,-2}$ from $F_{2a}$ is
\eqn\ctwoa{
C_{2a} = {1 \over 24 \pi^2} ( \tilde{F}(x) \tilde{F}(x) +
2\tilde{F}(x) \tilde{F}(x^2)).
}
For the other two cases we could not compute the sums explicitly, but
it is not difficult to numerically evaluate the desired coefficient at
the transition temperature $x_c=2-\sqrt{3}$.  We find that the
contributions of the three diagrams to $C_{1,1,-2}$ are given by
\eqn\Ctwo{\eqalign{ C_{2a} &= 6.53536 \times 10^{-4},
\cr C_{2b} &= -22.87088 \times 10^{-4}, \cr C_{2c} &= 9.16396
\times 10^{-4}, }}
so that the total coefficient is
\eqn\twoloop{\eqalign{ C_{1,1,-2} = -7.1716 \times 10^{-4} \; . }}

\subsec{Three Loops}

The leading contribution to the coefficient $C_{1,1,-1,-1}$ in
\leadingf\ comes from the fourteen three-loop diagrams of figure
\diagrams. We first give the expressions for each diagram in terms of
propagators and spherical harmonic integrals. For diagram 3a, we find
\eqn\Fthreea{\eqalign{
F_{3a} &= -{g_{YM}^4 \over 2} (D^{\a \g \l} D^{\bal \d \bl} D^{\b
\bd \t} D^{\bb \bg \bt} - 2 D^{\a \bal \l} D^{\g \d \bl} D^{\b \bg
\t} D^{\bb \bd \bt} + D^{\a \bal \l} D^{\g \d \bl} D^{\b \bb \t}
D^{\bg \bd \bt}) \cr & \qquad \int dt \; \Delta_{j_\a} (0,
\alpha_{ab})  \Delta_{j_\g} (t, \alpha_{ca})  \Delta_{j_\d} (t,
\alpha_{ac}) ( \Delta_{j_\b} (0, \alpha_{ad})+ \Delta_{j_\b} (0,
\alpha_{dc}) ). }}
For diagram 3b, we find
\eqn\Fthreeb{\eqalign{ F_{3b} =& g_{YM}^4 (D^{\a \g \l} D^{\bal \d
\bl} - D^{\a \bal \l} D^{\g \d \bl}) {D^{\bg \b \r} D^{\bd \bb
\br} \over j_\r (j_\r + 2)}  \cr & \qquad \left\{ \Delta_{j_\a}(0,
\alpha_{ab})(D_\tau \Delta_{j_\b}(0, \alpha_{ad}) + D_\tau
\Delta_{j_\b}(0, \alpha_{dc})) \right. \cr & \qquad \qquad \qquad
\int dt \; (D_\tau \Delta_{j_\g}(t, \alpha_{ac}) \Delta_{j_\d}(t,
\alpha_{ca}) - \Delta_{j_\g}(t, \alpha_{ac}) D_\tau
\Delta_{j_\d}(t, \alpha_{ca})) \cr & \qquad + \Delta_{j_\a}(0,
\alpha_{ab})(\Delta_{j_\b}(0, \alpha_{ad}) + \Delta_{j_\b}(0,
\alpha_{dc})) \cr & \left. \qquad \qquad \qquad \int dt \; ((j_\b
+1)^2 \Delta_{j_\g}(t, \alpha_{ac}) \Delta_{j_\d}(t, \alpha_{ca})
- D_\tau \Delta_{j_\g}(t, \alpha_{ac}) D_\tau \Delta_{j_\d}(t,
\alpha_{ca})) \right\}. }}
For diagram 3c, the result is
\eqn\Fthreec{\eqalign{ F_{3c} =& -{g_{YM}^4 \over 2} {D^{\a \g \l}
D^{\bal \d \bl} D^{\bg \b \r} D^{\bd \bb \br} \over j_\lambda
(j_\lambda+2) j_\r (j_\r + 2)} \cr & \int dt (\Delta_{j_\b}(0,
\alpha_{ad})+ \Delta_{j_\b}(0, \alpha_{dc})) \cr & \qquad \left\{
((j_\a+1)^2 (j_\b+1)^2+(j_\g+1)^2 (j_\d+1)^2) \Delta_{j_\a}(0,
\alpha_{ab}) \Delta_{j_\d}(t, \alpha_{ac}) \Delta_{j_\g}(t,
\alpha_{ca})  \right. \cr & \qquad - (j_\beta + 1)^2 D_\tau
\Delta_{j_\g}(t, \alpha_{ca})(4 D_\t \Delta_{j_\a}(0, \alpha_{ab})
\Delta_{j_\d}(t, \alpha_{ac}) + 2 \Delta_{j_\a}(0, \alpha_{ab})
D_\tau \Delta_{j_\d}(t, \alpha_{ac})) \cr & \qquad \left. - 2
(j_\d + 1)^2 \Delta_{j_\a}(0, \alpha_{ab}) \Delta_{j_\d}(0,
\alpha_{ac}) \right\} \cr & + \int dt (D_\tau \Delta_{j_\b}(0,
\alpha_{ad}) + D_\tau \Delta_{j_\b}(0, \alpha_{dc})) \cr & \qquad
\left\{(j_\g+1)^2 \Delta_{j_\g}(t, \alpha_{ca})( 4 D_\tau
\Delta_{j_\d}(t, \alpha_{ac}) \Delta_{j_\a}(0, \alpha_{ab}) + 2
\Delta_{j_\d}(t, \alpha_{ac}) D_\tau \Delta_{j_\a}(0,
\alpha_{ab})) \right. \cr & \qquad - 2 D_\tau \Delta_{j_\a}(0,
\alpha_{ab}) D_\t \Delta_{j_\d}(t, \alpha_{ac}) D_\tau
\Delta_{j_\g}(t, \alpha_{ca}) \cr & \qquad \left. - 2 (D_\tau
\Delta_{j_\a}(0, \alpha_{ab}) \Delta_{j_\d}(0, \alpha_{ac}) + 2
\Delta_{j_\a}(0, \alpha_{ab}) D_\tau \Delta_{j_\d}(0,
\alpha_{ac})) \right\}. }}
For diagram 3d, we find
\eqn\Fthreed{\eqalign{
F_{3d} =& -{g_{YM}^4 \over 4} D^{\a \g \l} D^{\b \d \bl} D^{\bal
\bg \r} D^{\bb \bd \br} \cr & \qquad  \int dt \Delta_{j_\alpha}
(t, \alpha_{ab}) \Delta_{j_\beta}(t, \alpha_{bc}) (2
\Delta_{j_\g}(t, \alpha_{cd}) \Delta_{j_\delta}(t, \alpha_{da}) +
\Delta_{j_\d}(t, \alpha_{cd}) \Delta_{j_\g}(t, \alpha_{da}) ) \cr
&-{g_{YM}^4 \over 4} D^{\a \b \l} D^{\g \d \bl} D^{\bg \bb \r}
D^{\bal \bd \br} \cr & \qquad \int dt \Delta_{j_\alpha} (t,
\alpha_{ab}) \Delta_{j_\beta}(t, \alpha_{bc}) (\Delta_{j_\g}(t,
\alpha_{cd}) \Delta_{j_\delta}(t, \alpha_{da}) - 4
\Delta_{j_\d}(t, \alpha_{cd}) \Delta_{j_\g}(t, \alpha_{da}) ) . }}
Evaluating diagram 3e, we obtain
\eqn\Fthreee{\eqalign{ F_{3e} =& g_{YM}^4 (2D^{\a \d \l} D^{\b \g
\bl} - D^{\a \b \l} D^{\g \d \bl} - D^{\a \g \l} D^{\b \d \bl})
{D^{\bg \bd \r} D^{\bal \bb \br} \over j_\r (j_\r + 2)}  \cr &
\qquad \int dt \left\{ D_\tau \Delta_{j_\a}(t, \alpha_{ba})
\Delta_{j_\b}(t, \alpha_{ad}) \Delta_{j_\gamma}(t, \alpha_{cb})
D_\tau \Delta_{j_\d}(t, \alpha_{dc})) \right. \cr & \qquad \qquad
\left. - \Delta_{j_\a}(t, \alpha_{ba}) D_\tau \Delta_{j_\b}(t,
\alpha_{ad}) \Delta_{j_\gamma}(t, \alpha_{cb}) D_\tau
\Delta_{j_\d}(t, \alpha_{dc})) \right\}. }}
Diagram 3f gives
\eqn\Fthreef{\eqalign{ F_{3f} &=- {g_{YM}^4 \over 2} {D^{\a \g \l}
D^{\b \d \bl} D^{\bal \bg \r} D^{\bb \bd \br} \over j_\l (j_\l +2)
j_\r (j_\r +2)} \cr &\qquad \left[4 \Delta_{j_\a} (0, \alpha_{ab})
D_\tau \Delta_{j_\d} (0, \alpha_{cd}) D_\tau \Delta_{j_\b} (0,
\alpha_{da}) \right. \cr &\qquad \qquad -2 ((j_\b +1)^2 + (j_\d +
1)^2) \Delta_{j_\a} (0, \alpha_{ab}) \Delta_{j_\d} (0,
\alpha_{cd}) \Delta_{j_\b} (0, \alpha_{da}) \cr &\qquad +\int dt
\left\{ \Delta_{j_\a} (t, \alpha_{ab}) \Delta_{j_\g} (t,
\alpha_{bc}) \Delta_{j_\d} (t, \alpha_{cd}) \Delta_{j_\b} (t,
\alpha_{da}) \right. \cr &\qquad \qquad \qquad \qquad
(j_\gamma+1)^2((j_\d + 1)^2 + (j_\b +1)^2) \cr & \qquad \qquad +2
D_\tau \Delta_{j_\a} (t, \alpha_{ab}) D_\tau \Delta_{j_\g} (t,
\alpha_{bc})D_\tau \Delta_{j_\d} (t, \alpha_{cd}) D_\tau
\Delta_{j_\b} (t, \alpha_{da}) \cr & \left. \left. \qquad \qquad
-4 (j_\gamma+1)^2 \Delta_{j_\a} (t, \alpha_{ab}) \Delta_{j_\g} (t,
\alpha_{bc}) D_\tau \Delta_{j_\d} (t, \alpha_{cd}) D_\tau
\Delta_{j_\b} (t, \alpha_{da}) \right\} \right] \cr &- {g_{YM}^4
\over 2} {D^{\a \b \l} D^{\g \d \bl} D^{\bal \bg \r} D^{\bb \bd
\br} \over j_\l (j_\l +2) j_\r (j_\r +2)} \cr &\qquad \left[4
D_\tau \Delta_{j_\a} (0, \alpha_{ab}) D_\tau \Delta_{j_\g} (0,
\alpha_{bc}) \Delta_{j_\b} (0, \alpha_{da}) \right. \cr &\qquad
\qquad -2 (j_\a +1)^2 \Delta_{j_\a} (0, \alpha_{ab}) \Delta_{j_\g}
(0, \alpha_{bc}) \Delta_{j_\b} (0, \alpha_{da}) \cr &\qquad \qquad
-2 \Delta_{j_\a} (0, \alpha_{ab}) D_\tau \Delta_{j_\g} (0,
\alpha_{bc}) D_\tau \Delta_{j_\b} (0, \alpha_{da}) \cr &\qquad
+\int dt \left\{ D_\tau \Delta_{j_\a} (t, \alpha_{ab}) D_\tau
\Delta_{j_\g} (t, \alpha_{bc}) D_\tau  \Delta_{j_\d} (t,
\alpha_{cd}) D_\tau \Delta_{j_\b} (t, \alpha_{da}) \right. \cr &
\qquad \qquad +(j_\a +1)^2 (j_\d +1)^2  \Delta_{j_\a} (t,
\alpha_{ab}) \Delta_{j_\g} (t, \alpha_{bc}) \Delta_{j_\d} (t,
\alpha_{cd}) \Delta_{j_\b} (t, \alpha_{da}) \cr & \qquad \qquad +2
(j_\d +1)^2  \Delta_{j_\a} (t, \alpha_{ab})D_\tau \Delta_{j_\g}
(t, \alpha_{bc}) \Delta_{j_\d} (t, \alpha_{cd}) D_\tau
\Delta_{j_\b} (t, \alpha_{da}) \cr & \left. \left. \qquad \qquad
-4 (j_\d+1)^2 D_\tau \Delta_{j_\a} (t, \alpha_{ab}) D_\tau
\Delta_{j_\g} (t, \alpha_{bc}) \Delta_{j_\d} (t, \alpha_{cd})
\Delta_{j_\b} (t, \alpha_{da}) \right\} \right]. }}
For diagram 3g, we find
\eqn\Fthreeg{\eqalign{ F_{3g} =& g_{YM}^4 \hat{E}^{\a \d \r}
\hat{E}^{\g \b \br}(D^{\bal \bg \l} D^{\bb \bd \bl} - {1 \over 2}
D^{\bal \bb \l} D^{\bg \bd \bl} - {1 \over 2} D^{\bal \bd \l}
D^{\bb \bg \bl}) \cr & \qquad \int dt dt' \Delta_{j_\b}(t',
\alpha_{da}) \Delta_{j_\g}(t', \alpha_{cd}) \Delta_{j_\d}(t,
\alpha_{bc}) \Delta_{j_\a}(t, \alpha_{ab}) \Delta_{j_\r}(t'-t,
\alpha_{ac}), }}
where $\hat{E}$ was defined in the previous subsection. Diagram 3h evaluates to
\eqn\Fthreeh{\eqalign{ F_{3h} =& g_{YM}^4 {1 \over j_\l (j_\l +
2)} D^{\a \g \l} D^{\b \d \bl} \hat{E}^{\bal \bg \r} \hat{E}^{\bd
\bb \br} \cr & \qquad \int dt_1 dt_2 \; D_\tau \Delta_{j_\a} (t_1,
\alpha_{ab}) \Delta_{j_\g}(t_1, \alpha_{bc})
\Delta_{j_\r}(t_1-t_2, \alpha_{ca}) \cr & \qquad \qquad \qquad
(\Delta_{j_\d}(t_2, \alpha_{cd}) D_\tau \Delta_{j_\b}(t_2,
\alpha_{da}) - D_\tau \Delta_{j_\d}(t_2, \alpha_{cd})
\Delta_{j_\b}(t_2, \alpha_{da}))\cr & + g_{YM}^4 {1 \over j_\l
(j_\l + 2)} D^{\a \b \l} D^{\g \d \bl} \hat{E}^{\bal \bg \r}
\hat{E}^{\bd \bb \br} \cr & \qquad \int dt_1 dt_2 \; D_\tau
\Delta_{j_\a} (t_1, \alpha_{ab}) \Delta_{j_\b}(t_2, \alpha_{da})
\Delta_{j_\r}(t_1-t_2, \alpha_{ca}) \cr & \qquad \qquad \qquad
(\Delta_{j_\d}(t_2, \alpha_{cd}) D_\tau \Delta_{j_\g}(t_1,
\alpha_{bc}) - D_\tau \Delta_{j_\d}(t_2, \alpha_{cd})
\Delta_{j_\g}(t_1, \alpha_{bc})). }}
Diagram 3i gives
\eqn\Fthreei{\eqalign{ F_{3i} =& -{g_{YM}^4 \over 4} \hat{E}^{\a
\b \r} \hat{E}^{\bal \s \bb} \hat{E}^{\bs \d \g} \hat{E}^{\bd \br
\bg} \cr & \qquad \int dt_1 dt_2 dt_3 \Delta_{j_\a}(t_1-t_2,
\alpha_{ab}) \Delta_{j_\b}(t_1-t_2, \alpha_{bc})\Delta_{j_\g}(t_3,
\alpha_{cd}) \cr & \qquad \qquad \qquad \qquad \Delta_{j_\d}(t_3,
\alpha_{da}) \Delta_{j_\r}(t_1-t_3, \alpha_{ca})\Delta_{j_\s}(t_2,
\alpha_{ac}). }}
For diagram 3j, we find
\eqn\Fthreej{\eqalign{ F_{3j} =& g_{YM}^4(D^{\a \r \l} D^{\b \br
\bl} - D^{\r \br \l} D^{\a \b \bl}) \hat{E}^{\bal \t \s}
\hat{E}^{\bb \bs \bt} \cr & \qquad \int dt dt' \Delta_{j_\r}(0,
\alpha_{ab}) \Delta_{j_\b}(t-t', \alpha_{ac}) \Delta_{j_\s}(t',
\alpha_{ad}) \Delta_{j_\a}(t, \alpha_{ca}) \Delta_{j_\t}(t',
\alpha_{dc}). }}
Diagram 3k yields
\eqn\Fthreek{\eqalign{ F_{3k} =& g_{YM}^4 {1 \over j_\lambda
(j_\lambda + 2)} D^{\alpha \beta \lambda} D^{\bal \g \bl}
\hat{E}^{\bb \r \s} \hat{E}^{\bg \bs \br} \cr & \qquad \int dt_1
dt_2 \; \Delta_{j_\r}(t_1 - t_2, \alpha_{cd}) \Delta_{j_\s}(t_1 -
t_2, \alpha_{da}) \cr & \qquad \qquad \qquad \left\{ 2 D_\tau
\Delta_{j_\alpha}(0, \alpha_{ab}) \Delta_{j_\b}(t_1, \alpha_{ac})
D_\tau \Delta_{j_\gamma}(t_2, \alpha_{ca}) \right. \cr & \qquad
\qquad \qquad + \Delta_{j_\alpha}(0, \alpha_{ab}) D_\tau
\Delta_{j_\b}(t_1, \alpha_{ac}) D_\tau \Delta_{j_\gamma} (t_2,
\alpha_{ca}) \cr & \qquad \qquad \qquad \left. -(j_\alpha +1)^2
\Delta_{j_\alpha}(0, \alpha_{ab}) \Delta_{j_\b}(t_1, \alpha_{ac})
\Delta_{j_\gamma}(t_2, \alpha_{ca}) \right\}. }}
For diagram 3l, we obtain
\eqn\Fthreel{\eqalign{ F_{3l} =& -{g_{YM}^4 \over 12} \hat{E}^{\a
\b \t} \hat{E}^{\bb \g \r} \hat{E}^{\bg \bal \s} \hat{E}^{\br \bs
\bt} \cr & \qquad \int dt_1 dt_2 dt_3 \Delta_{j_\a}(t_2-t_3,
\alpha_{ab}) \Delta_{j_\b}(t_3-t_1,
\alpha_{ac})\Delta_{j_\g}(t_1-t_2, \alpha_{ad}) \cr & \qquad
\qquad \qquad \qquad \Delta_{j_\r}(t_1, \alpha_{dc})
\Delta_{j_\s}(t_2, \alpha_{bd})\Delta_{j_\t}(t_3, \alpha_{cb}). }}
For diagram 3m, we find
\eqn\Fthreem{\eqalign{ F_{3m} &= 4g_{YM}^4 {D^{\a \b \l} C^{\bl
\bal \r} D^{\g \d \br} \hat{E}^{\bb \bg \bd} \over j_\l (j_\l +2)
j_\r (j_\r + 2)} \cr & \qquad \int dt ( D_\tau \Delta_{j_\a} (0,
\alpha_{ab}) \Delta_{j_\b} (t, \alpha_{ca}) - \Delta_{j_\a} (0,
\alpha_{ab}) D_\tau \Delta_{j_\b} (t, \alpha_{ca}) ) \cr &\qquad
\qquad \qquad ( \Delta_{j_\g} (t, \alpha_{dc}) D_\tau
\Delta_{j_\d} (t, \alpha_{ad}) - D_\tau \Delta_{j_\g} (t,
\alpha_{dc}) \Delta_{j_\d} (t, \alpha_{ad})) \cr &+2 g_{YM}^4
{D^{\a \d \l} C^{\bl \b \r} D^{\g \bd \br} \hat{E}^{\bal \bg \bb}
\over j_\l (j_\l +2) j_\r (j_\r + 2)} \cr & \qquad \int dt \left\{
D_\tau \Delta_{j_\a} (t, \alpha_{ab}) \Delta_{j_\b} (t,
\alpha_{bd}) D_\tau \Delta_{j_\g} (t, \alpha_{da}) \Delta_{j_\d}
(0, \alpha_{ca}) \right. \cr &\qquad \qquad + 2 D_\tau
\Delta_{j_\a} (t, \alpha_{ab}) \Delta_{j_\b} (t, \alpha_{bd})
\Delta_{j_\g} (t, \alpha_{da}) D_\tau \Delta_{j_\d} (0,
\alpha_{ca}) \cr &\qquad \qquad \left. -(j_\delta + 1)^2
\Delta_{j_\a} (t, \alpha_{ab}) \Delta_{j_\b} (t, \alpha_{bd})
\Delta_{j_\g} (t, \alpha_{da}) \Delta_{j_\d} (0, \alpha_{ca})
\right\}. }}
Finally, diagram 3n gives the result
\eqn\Fthreen{\eqalign{ F_{3n} &= g_{YM}^4 {D^{\a \g \r} D^{\b \bg
\s} \over j_\r (j_\r +2) j_\s (j_\s + 2)} \left( 3 {C^{\br \bal
\l} C^{\bl \bb \bs} \over j_\l (j_\l + 2)} + 3 {C^{\br \bb \l}
C^{\bl \bal \bs} \over j_\l (j_\l + 2)} + D^{\bal \bl \br} D^{\bb
\l \bs} + D^{\bal \bl \bs} D^{\bb \l \br} \right) \cr & \qquad
\left\{ D_\tau \Delta_{j_\a} (0, \alpha_{cb}) \Delta_{j_\g} (0,
\alpha_{ac}) D_\tau \Delta_{j_\b} (0, \alpha_{ad}) \right. \cr &
\qquad \qquad +2 D_\tau \Delta_{j_\a} (0, \alpha_{cb}) D_\tau
\Delta_{j_\g} (0, \alpha_{ac}) \Delta_{j_\b} (0, \alpha_{ad}) \cr
& \left. \qquad \qquad - (j_\g + 1)^2 \Delta_{j_\a} (0,
\alpha_{cb}) \Delta_{j_\g} (0, \alpha_{ac}) \Delta_{j_\b} (0,
\alpha_{ad}) \right\} \cr &- g_{YM}^4 {D^{\a \g \r} D^{\b \bg \s}
\over j_\r (j_\r + 2) j_\s (j_\s + 2)} \left( 3 {C^{\br \bal \l}
C^{\bl \bb \bs} \over j_\l (j_\l + 2)} + D^{\bal \bl \br} D^{\bb \l
\bs} \right) \cr & \qquad \left\{ D_\tau \Delta_{j_\a} (0,
\alpha_{ab}) \Delta_{j_\g} (0, \alpha_{ac}) D_\tau \Delta_{j_\b}
(0, \alpha_{ad}) \right. \cr & \qquad \qquad +2 D_\tau
\Delta_{j_\a} (0, \alpha_{ab}) D_\tau \Delta_{j_\g} (0,
\alpha_{ac}) \Delta_{j_\b} (0, \alpha_{ad}) \cr & \left. \qquad
\qquad + (j_\g + 1)^2 \Delta_{j_\a} (0, \alpha_{ab}) \Delta_{j_\g}
(0, \alpha_{ac}) \Delta_{j_\b} (0, \alpha_{ad}) \right\} \cr &-
g_{YM}^4 {D^{\a \g \r} D^{\bal \bg \s} \over j_\r (j_\r + 2) j_\s
(j_\s + 2)} \left( 3 {C^{\br \b \l} C^{\bl \bb \bs} \over j_\l
(j_\l + 2)} + D^{\l \b \br} D^{\bl \bb \bs} \right) \cr & \qquad
\left\{ 2 D_\tau \Delta_{j_\a} (0, \alpha_{ab}) D_\tau
\Delta_{j_\g} (0, \alpha_{ad}) \Delta_{j_\b} (0, \alpha_{bc})
\right. \cr & \left. \qquad \qquad + ((j_\a + 1)^2 +(j_\g + 1)^2)
\Delta_{j_\a} (0, \alpha_{ab}) \Delta_{j_\g} (0, \alpha_{ad})
\Delta_{j_\b} (0, \alpha_{bc}) \right\}. \cr }}

\subsec{Evaluation of three-loop diagrams}

We now evaluate the coefficient $C_{1,1,-1,-1}$ of
$\beta g_{YM}^4 |\tr(U)|^4$ in \seff\ arising from the three loop
diagrams computed above. The expressions for the diagrams are of
the general form:
\eqn\dformone{F=\sum_{j,m,\epsilon}S'(j,m,\epsilon) I'(j),}
where $S'$ is the term containing all spherical harmonic factors
$(C,D,E)$ and $I'$ is the term involving the propagators and the
integrals over $t$. $S'$ depends on the $j$'s, $m$'s and
$\epsilon$'s of the spherical harmonics but $I'$ only on the
$j$'s.  Using the identities of appendix B we can perform the sum
over the $m$'s and $\epsilon$'s analytically and find
$S(j)=\sum_{m,\epsilon} S'(j,m,\epsilon)$.
Then, the diagram can be written as a sum
over $j$'s
\eqn\dformtwo{F=\sum_{j}S(j)I'(j).}
We then expand $I'(j)$ in
powers of $\tr(U^n)$ and determine $I(j)$, the coefficient of
$|\tr(U)|^4$ in this expansion.  The contribution to $b$ from this
diagram may then be written in the form

\eqn\dformthree{C_{1,1,-1,-1}=\sum_{j}S(j)I(j).}
We can find $S$ for any diagram by using the identities given in
appendix B, and the corresponding integral $I$ can be found in the
Mathematica file \mathematica. The sum \dformthree\ over $j$'s is the only
part of the calculation that we have generally had to perform
numerically.

Diagram 3a is the simplest and we can actually calculate it
analytically. The expression \Fthreea\ above may be simplified
using the angular momentum sums of appendix B.
We find
\eqn\fthreea{\eqalign{ C_{1,1,-1,-1}^{3a} &= -{1 \over 18 \pi^4}
\sum_{a,b,c} {a(a+2)b(b+2)c(c+2) \over (a+1)(b+1)(c+1)^3}x^{a+b+2}
\cr & \qquad \qquad \qquad \left\{(x^{c+1}+1)(x^{c+1}+2) - (c+1)
\ln(x) (2x^{2c+2} + 3 x^{c+1}) \right\}. }}
Note that this sum contains a logarithmic divergence in the sum over
$c$ when taking the $x$-independent term (equal to $2$) in the curly
brackets.

Such logarithmic divergences appear in the contribution to
$C_{1,1,-1,-1}$ from most of our 3-loop diagrams, since there are
terms where only two of the three unconstrained sums over loop momenta
have exponential damping factors. These divergences arise from {\it
non-planar} 1-loop subdiagrams, as we explain in the next section when
we discuss the corresponding counterterms.

In practice, when doing the computation we will place a cutoff $J$ on the
angular momentum sums. We can then write the result with a cutoff
in the form \eqn\gendiv{ C_{1,1,-1,-1}(J) = \alpha \sum_{c=1}^J{1
\over c+1} + C_{1,1,-1,-1}^{finite} + {\cal O} (1/J), } where
$\alpha$ is the coefficient of the logarithmic divergence, which
we can define by \eqn\defalpha{ \alpha = \lim_{J \to \infty}
(J+1)(C_{1,1,-1,-1}(J) - C_{1,1,-1,-1}(J-1)), }
and\foot{To improve numerical convergence we use Aitken's method :
if $\lim_{n\to\infty}a_n = r$ then $\lim_{n\to\infty}{a_n
a_{n+2}-a_{n+1}^2 \over a_n + a_{n+2}-2 a_{n+1}}=r$, but the
convergence of the second sequence is faster.}
\eqn\deffinite{ C_{1,1,-1,-1}^{finite} \equiv  \lim_{J \to \infty}
(C_{1,1,-1,-1}(J) - \alpha \sum_{c=1}^J{1 \over c+1} ). }
In particular, from the expression above for the 3a sum \fthreea\
we find
\eqn\sumthreea{ \alpha_{3a} = -{1\over {9\pi^4}} \left({x\over
(1-x)^2} + \log(1-x) \right)^2 = -4.0356 \times 10^{-5}, \qquad
C_{1,1,-1,-1}^{3a,finite} = -3.666 \times 10^{-7},}
where we have evaluated $\alpha_{3a}$ and
$C_{1,1,-1,-1}^{3a,finite}$ at the transition point
$x_c=2-\sqrt{3}$ \first.

For the 13 remaining diagrams, we were unable to evaluate the sum
over $j$ in \dformthree\ analytically.  However, we have performed
the sum numerically (at the phase transition temperature), and we
will present the results in section 3.7.

\subsec{Counterterm diagrams}

We have seen that the contributions to $C_{1,1,-1,-1}$ arising from
individual 3-loop diagrams contain logarithmic divergences. It turns
out that all of these must cancel in the sum over diagrams. To see
this, note first that no single-trace counterterm can contribute to
the coefficient of $|\tr(U)|^4$, since such a contribution requires
four index loops, while planar counterterm diagrams at order
$\lambda^2$ are two loop diagrams with only three index loops. In
fact, the only possible counterterm contribution comes from
double-trace counterterms\foot{See, for instance, \refs{\AdamsJB,
\AharonyPA} for recent
discussions of how double-trace terms can contribute at leading
order in the large $N$ limit.} of the form $\tr(AA)\tr(AA)$, which give
rise to diagrams of the form shown in figure 3. But no such
counterterm is gauge invariant, so in any gauge invariant
regularization scheme (such as dimensional regularization), there are no
counterterm contributions at all, and all divergences must cancel in
the sum over diagrams. Our regularization scheme does not respect
gauge invariance, but the coefficients of logarithmic divergences are
insensitive to the regularization scheme, so we must still find that
all divergences cancel.\foot{In comparing with dimensional
regularization, the coefficient of $1/\epsilon$ poles will be
proportional to the coefficient of the logarithmic divergence in a
cutoff scheme.}

The preceding argument does not mean that we can ignore counterterms
altogether. Indeed, the finite contribution resulting from the sum
over diagrams does depend on the regularization scheme, and it is
crucial to include the contributions of finite counterterms in order
to obtain the correct result. The counterterms that can contribute
take the form
\eqn\countertermform{L_{CT}=\left({\l \over N}\right)^2{1\over
\pi^2} \left(c_1 \tr(A_i A_i) \tr(A_j A_j)+2 c_2 \tr (A_i A_j)
\tr (A_i A_j)\right),}
where $c_1$ and $c_2$ will be finite numerical coefficients chosen so
that the results of calculations in our scheme match with results
using dimensional regularization.

{\fig{Counterterm contribution to the free energy at order
$\lambda^2$.}{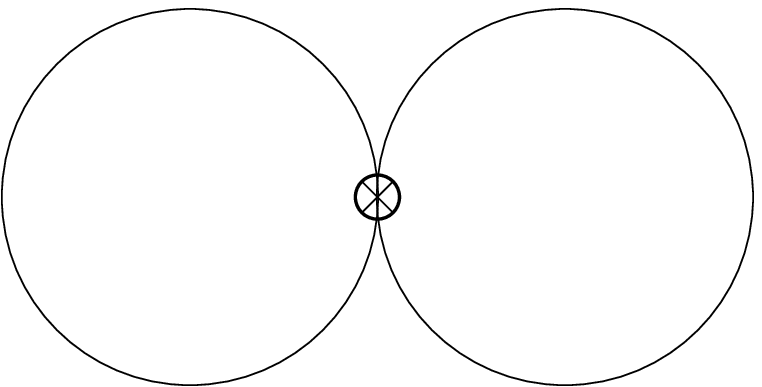}{2.0truein}\figlabel\Ctermdiag}

At order $\lambda^2$, this vertex contributes to the free energy via
counterterm graphs of the form depicted in figure
\Ctermdiag. Following the same procedure as in the previous
subsections, it is not difficult to isolate the coefficient $C_c$ of
$g^4_{YM}|\tr(U)^4|$ in the contribution of this diagram to the
free energy :
\eqn\countercontrib{C_c= {1\over
\pi^2}\sum_{\a ,\g,\kappa}\left(c_1 D^{\a \bar{\a} \kappa}D^{\g
\bar{\g} \bar{\kappa}}+2 c_2 D^{\a\g\kappa}
D^{\bar{\a}\bar{\g}\bar{\kappa}}\right){1\over j_\a +1} {1\over
j_\g +1} x^{j_\a+j_\g +2}.}
Using spherical harmonic identities from appendix B and performing the sum we
find:
\eqn\countercontribtwo{C_c= {2\over \pi^4}\left(c_1
+{2\over 3} c_2\right) \left[ {x\over (1-x)^2}
+\ln(1-x)\right]^2,}
which may be conveniently written as
\eqn\countercottribc{C_c= -(18 c_1 + 12 c_2)\a_{3\a},}
where $\a_{3\a}$ is as in \sumthreea.

In order to determine the counterterm coefficients $c_1$ and $c_2$, it
is enough to look at the simplest correlator to which the counterterms
\countertermform\ contribute, namely the nonplanar one-loop four point
function on $\IR^4$ with two external legs attached to each of the index
loops. As we describe in detail in section 4.4 and appendix A.5 below,
the coefficients are determined by demanding that the combination of
logarithmically divergent one-loop diagrams (shown in figure 6) with
the counterterm diagrams reproduces the result for the same correlator
evaluated in dimensional regularization. For our regularization scheme
with a sharp cutoff, we find
\eqn\countertermres{
c_1 = c_2 = {1 \over 60},
}
so that
\eqn\counterresult{
C_c = - {1 \over 2} \alpha_{3a}.
}

This net contribution from the counterterm diagrams is all we need to
complete our calculation. However, it turns out that a very useful
check of our three loop results arises from splitting up the
counterterm contribution into pieces associated with the individual
one loop diagrams of figure 6. Note that each of these is
logarithmically divergent, and appears as a nonplanar subdiagram in
exactly one of the three loop vacuum diagrams of figure 2 (obtained by
joining up the four free vector lines of the one loop diagram in pairs
in such a way that we obtain a planar three-loop diagram, and
replacing any $a$ or $c$ lines with the corresponding effective
vertex). Thus, the logarithmic divergences in the contribution to $b$
from three loop vacuum diagrams are directly related to logarithmic
divergences in a specific set of non-planar one-loop four point
diagrams.\foot{The coefficient of $|\tr(U)|^4$ has no additional
divergences from planar one-loop subdiagrams of the three-loop
diagrams, since we have seen that there are no single-trace
counterterms that could cancel them (or remove the regulator
dependence of the finite results if any such divergences cancelled
upon summing diagrams).} If we denote by $L_{CT}^X$ the counterterm
Lagrangian density necessary to make this particular set of one-loop
diagrams agree with dimensional regularization, then the combination
of the three loop diagram $X$ and the counterterm diagrams associated
with $L_{CT}^X$ should be finite, providing a check on the divergent
part of each individual 3-loop diagram.

In addition to the overall result \counterresult\ we therefore define
partial contributions $C_c^X$ arising from the counterterm diagrams
associated with a given three loop diagram $X$. These will generally
have both a logarithmically divergent piece, and a finite piece, and
take the form\foot{The constants
${\cal{A}}_1,{\cal{A}}_2,{\cal{A}}_3,{\cal{A}}_4$ appearing in
Appendix A all evaluate to the same value with the choice of damping
function (essentially a step function) used here, while the UV cutoff
$M$ is simply $J/R$.}
\eqn\alphacCfin{C_c^X=\alpha^X_c\ln\left(\frac{J}{R\mu}\right)+C_c^{X \; {\rm
finite}}.}
The values of $\alpha_c$ and $C_c^{{\rm finite}}$ for each diagram are
tabulated in the next section together with our results for the three
loop diagrams.

\subsec{Results}

In this section, we tabulate our numerical results for the
contributions to $C_{1,1,-1,-1}$. For each three-loop diagram, we give
the coefficient $\alpha$ of the logarithmic divergence and the
remaining finite piece, as defined in \defalpha\ and \deffinite. We
also give the coefficients $\alpha_c$ and $C_c^{finite}$ for the
logarithmically divergent and finite parts of the associated
counterterm diagrams, as defined in \alphacCfin. We find:
\medskip
{
\offinterlineskip
\tabskip=0pt
\halign{
\vrule height2.75ex depth1.25ex width 0.6pt #\tabskip=1em &
 \hfil # \hfil &\vrule  # &
 \hfil $#$ \hfil &\vrule  # &
 \hfil $#$ \hfil &\vrule  # &
 \hfil $#$ \hfil &\vrule  # &
 \hfil $#$ \hfil & #\vrule \tabskip=0pt\cr
\noalign{\hrule height 0.6pt}
& Diagram && \a/\a_{3\a} &&  C_{1,1,-1,-1}^{{\rm finite}} &&
 \alpha_c/ \a_{3\a} && C_c^{{\rm finite}}/ \a_{3\a} & \cr
\noalign{\hrule} & 3a &&   1  && -3.666\times 10^{-7} && -1 &&-107/120
&\cr & 3b &&  -1  && -4.805\times 10^{-5}  && 1 && 13/60 &\cr & 3c &&
-3/4 && 2.224\times 10^{-5} && 3/4 && 1/20 &\cr & 3d &&  5/4 &&
2.21\times 10^{-5} && -5/4 && -47/60 &\cr & 3e && 1/2  && -5.85\times 10^{-5}
&& -1/2 && -7/30 &\cr & 3f && -9/4 && -2.69\times 10^{-4} && 9/4 && 3/5
&\cr & 3g && -3   && 1.15\times 10^{-4} && 3 && 1/5 &\cr & 3h && 0    &&
1.91\times 10^{-4} && 0 && -1/10 &\cr & 3i && 5/2  && -6.6\times 10^{-4} &&
-5/2 && 1/12 &\cr & 3j && -3   && 2.414\times 10^{-4} && 3 && 11/10 &\cr
& 3k && 1/2  && 1.42\times 10^{-4} && -1/2 && 7/60 &\cr & 3l && 5/4  &&
-1.\times 10^{-4} && -5/4 && 1/24 &\cr & 3m && 2    && 4.6\times 10^{-5} && -2
&& 1/15 &\cr & 3n && 1    && -9.2\times 10^{-5} && -1 && -29/30 &\cr
\noalign{\hrule height 0.6pt} & Sum && 0 && -4.5\times  10^{-4} && 0 &&
-1/2 & \cr \noalign{\hrule height 0.6pt} }}
\medskip
The numerical results in the middle column of the table are all
accurate at least within the number of digits appearing in the
table. We estimate the maximal total numerical error of our result
for the sum to be less than $3\times 10^{-5}$. This accuracy could
be improved with additional computer time, but we found no reason to
do this since we are only interested in the sign of $b$. Note that
the sum of all divergent contributions vanishes (without including
counterterms).

Note that, in the table above, $\alpha^X+\alpha_c^X=0$ for every
diagram $X$; that is, the sum of any three-loop vacuum graph and its
associated counterterm graphs is finite. This sum depends on the
renormalization scale $\mu$ (see \deffinite\ and \alphacCfin) but
should be independent of the regulating function $R(q)$ \foot{In
order to verify this independence -- and as a check on the logic of
our regularization scheme and our numerics --  we have recomputed
$\a/\a_{3a},  C_{1,1,-1,-1}^{{\rm finite}},
 \alpha_c/ \a_{3a}$  and  $C_c^{{\rm finite}}/ \a_{3a}$ for diagrams
 3a, 3b, 3c, 3g and 3n with a different regulating function (we took
 $R(q/M)$ to be a double step function, $R(x)=1$ for $x<1$, $R(x)=\half$
 for $1<x<2$ and $R(x)=0$ for $x>2$). We evaluated these quantities
 analytically for diagram 3a and numerically for all the other diagrams.
 As expected, in every case the coefficient of the logarithmic divergence
$\a/\a_{3a}$ and $\alpha_c/ \a_{3a}$ was unchanged in this new
regulating scheme. Also, as expected, while finite parts of each
diagram and its associated counterterm yielded different values in
this new regulating scheme, the sum of every diagram with its
associated counterterm graphs was unmodified.}. 
This $\mu$-dependence vanishes upon summing all
diagrams{\foot{This follows because $\sum_{\{{\rm
diagrams}\ X\}}\alpha^X=\sum_{\{{\rm diagrams}\ X\}}\alpha_c^X=0$.}}, 
leaving us with the scheme-independent answer
\eqn\finansCs{C_{1,1,-1,-1}=-4.5\times 10^{-4}-\alpha_{3a}/2=-4.3\times 10^{-4}.}

\subsec{Effective potential}

We can now put together our results for the terms of interest in
our effective potential \leadingf\ for the eigenvalues, evaluated
at the deconfinement temperature:
\eqn\comreslts{\eqalign{\mu_2&=4.8112\times 10^{-1},\cr \beta_c
C_{1,1,-2}&=-9.4447\times
10^{-4},\cr \beta_c C_{1,1,-1,-1}&=-5.7\times 10^{-4}. }}
From \bdefdag\ we find that the coefficient $b$ in the effective
potential \forfeff\ for $u_1$ is
\eqn\bdefdag{\eqalign{b&=\beta_c C_{1,1,-1,-1}-\frac{\beta_c^2
C_{1,1,-2}^2}{\mu_2}\cr &\simeq -5.7\times 10^{-4}-1.854\times
10^{-6}\cr &\approx-5.7\times 10^{-4}.}}

Note that $b$ is the sum of two terms, one of which is manifestly
negative. In \bdefdag, however, the dominant contribution came
from the first term which, in principle, could have been either
positive or negative.  This suggests that there could exist also
theories for which $b>0$; we will discuss this further in the
final section.

Since we have, after a laborious calculation, determined that
$b<0$, we may conclude that the large $N$ deconfinement transition
of pure Yang-Mills theory on a small $S^3$ is of first order.

\newsec{A Gauge Invariant Regularization Scheme on $S^3$}

In the previous section we have computed a set of two and three
loop Feynman diagrams to determine a particular term ($b$ in
equation \sefff) in the Wilsonian effective action for finite
temperature Yang Mills theory on $S^3$. We found that while
$b$ is finite, individual diagrams that contribute to $b$ diverge
logarithmically. In order to obtain the finite physical value of
$b$ we needed to sum contributions from the various diagrams, at
which point the logarithmic divergence cancels and we are left
with the finite result of interest. The process of isolating a
finite piece from the difference of divergent sub pieces is
delicate, and will yield the correct answer only if the
regularization procedure respects gauge invariance. In this
section we will expand on the discussion of section 2.4 to
describe in more detail the regularization procedure that we
employ in our computation.

As described above, computations of Yang Mills theory on $S^3$ are
most simply performed in the Coulomb gauge $\partial_i A^i=0$,
where the index $i$ runs over the three spatial indices of the
$S^3$. Further, we found it most convenient to regularize all
diagrams by truncating the spherical harmonic sums at spherical
harmonic number $n$ (in flat space this corresponds to imposing a
hard momentum cut off at momentum ${E(n) / R}$ where $E(n) \sim n$
is the energy of the $n^{th}$ spherical harmonic mode). This regularization
scheme is not gauge invariant, but should yield gauge invariant results
when employed with a bare Yang Mills action that includes an
appropriate set of non-gauge-invariant counterterms. The
appropriate counterterms may, in principle, be uniquely determined
(up to the usual ambiguity in the definition of the Yang Mills
coupling constant) by demanding that correlation functions
computed by this theory obey the Ward identities that follow from
gauge invariance, together with local Lorentz invariance.

In this section we will explicitly determine some of the counterterms that
will render our non-gauge-invariant regularization scheme
effectively gauge-invariant. These counterterms fall into two
classes; counterterms that would be needed even in flat space, and
counterterms that are proportional to the spacetime curvature. It
will turn out that no counterterm of the second type (those
proportional to spacetime curvature) contributes to the computation
of $b$, so we will content ourselves with determining only those
counterterms that appear even in flat space. These counterterms
may be determined rather simply by choosing them to ensure that
certain Green's functions (following 't Hooft we use the $A_\mu(p)
A_\nu(-p)$ two-point function, as well as a four-point function of
gauge fields) agree with the same Green's functions evaluated
using dimensional regularization\foot{More precisely, we compare
with a form of dimensional regularization that is tailored to deal
with Yang Mills theory in the Coulomb gauge. This so-called split
dimensional regularization scheme
\refs{\splitdimone,\splitdimtwo,\splitdimthree,\splitdimfour} 
separately extends the number of dimensions participating in the
Coulomb gauge condition (from $3$ to $3-\epsilon$) and the number of
other dimensions (from $1$ to $1-\epsilon'$). One may worry that this
regularization procedure is not Lorentz invariant (since integrals
over temporal and spatial momenta end up being regulated differently),
however the breaking of Lorentz invariance really comes from our
choice of gauge rather than the regularization scheme. In practice, we
will apply split dimensional regularization in the limit where
$\epsilon' \to 0$.}.

In \S4.1 we explain our regularization method in detail and give a
simple example of how it works. Sections 4.2 and 4.3 are devoted to
tests of the validity of our regularization scheme. In \S4.2 we
determine all quadratic counterterms of the first type (those that
appear in flat space) to order $\lambda$.  Even though these
counterterms do not actually contribute to our main computation in
this paper, we use them to test the validity of our regularization
procedure. First, we verify in \S4.3 that our results are consistent
with the Slavnov-Taylor identity and with Lorentz invariance. As
another test, in appendix A.3 we verify that our results lead to the
correct free energy at infinite volume to order $\lambda$.  Finally,
in \S4.4 we proceed to use the same methods to compute the
counterterms that we actually need for the computation of $b$; these
are a set of double-trace counterterms at order $\lambda^2$.

\subsec{General discussion and a simple example}

The regularization we will analyze in this section is a slightly
more general regularization scheme than the sharp cutoff which was
used in the computation of the previous section. We include damping
functions $R(\sqrt{q^2}/M)$ and $\tilde{R}(q_0/\Lambda)$ for the
momentum of each internal $A_i$ line of a given diagram, and damping
functions $\bar{R}(\sqrt{q^2}/\bar{M})$ and
$\bar{\tilde{R}}(q_0/\bar{\Lambda})$ ($q^2\equiv q_i q_i$,
$i=1,2,3$, and we take $\bar{\Lambda} \gg \Lambda$, ${\bar M} \gg M$,
$\Lambda \gg M$ and ${\bar \Lambda} \gg {\bar M}$ for
convenience) for the momentum of each internal $A_0$ or ghost line.
These functions are chosen so that
$R(0)=\bar{R}(0)=\tilde{R}(0)=\bar{\tilde{R}}(0)=1$,
$R'(0)=\bar{R}'(0)=\tilde{R}'(0)=\bar{\tilde{R}}'(0)=0$, and
$R(x\rightarrow\infty)=\bar{R}(x\rightarrow\infty)=
\tilde{R}(x\rightarrow\infty)=\bar{\tilde{R}}(x\rightarrow\infty)=0$.
We choose to treat $A_0$ and ghost lines differently from $A_i$
lines because, in our calculation of $b$, it was convenient to
integrate out $A_0$ and the ghosts directly in the action.  This can
be done with no regularization subtleties in diagrams with both
$A_0$/ghost lines and $A_i$ lines, provided that we take the scale
$\bar{M}\gg M$ and $\bar{\Lambda} \gg \Lambda$. 
We will sometimes be lazy with our notation and
write $R(\sqrt{q^2}/M)$ ($\bar{R}(\sqrt{q^2}/\bar{M})$) as $R(q/M)$
($\bar{R}(q/\bar{M})$).

In perturbation theory, correlation functions are obtained by
evaluating all contributing Feynman diagrams, each of which may be
written as an integral over internal momenta. In general these
integrals diverge and must be regulated; in this section we will
explain how one may convert the simple minded regularization scheme
described in the previous paragraph into dimensional regularization
by an appropriate choice of non gauge invariant counterterms. In the
rest of this subsection we will demonstrate our method on a `toy'
regularized integral\foot{This integral is slightly different in
form from those that will appear in our actual expressions below; in
particular the integrand contains a single propagator but two copies
of the regulator function. Thus, it should merely be thought of as a
simple divergent integral that illustrates all the complications
that arise in the actual process of regularization.}
\eqn\exint{I=\int \frac{d^3q d q_0}{(2\pi)^4}\frac{R({q\over M})
R({p-q \over M})}{q^2+q_0^2},}
and its counterpart in split dimensional regularization (SDR)
\eqn\exinto{I=\int_{SDR} \frac{d^3q d
q_0}{(2\pi)^4}\frac{1}{q^2+q_0^2}.}
The $q_0$ integral in \exint, \exinto\ is finite (which is why we
ignored the $\tilde{R}$ regulators in \exint) and may easily be
done to yield (from now on we suppress explicit reference to the
regulator in intermediate steps)
\eqn\exintqfour{I=\frac{1}{16\pi^3}\int\frac{d^3q}{q}.}
In split dimensional regularization \exintqfour\ evaluates to
zero, whereas in the damping function regularization scheme
\exint\ evaluates to\foot{The finite piece arises because
$\int_0^{\infty}\,dq\,R(q)\,R'(q)=-\frac{1}{2}$.}
\eqn\exintregdep{\alpha(p)=M^2C_2+\frac{p^2}{6}F_2-\frac{p^2}{24\pi^2},}
where we have defined
\eqn\twomomsdef{\eqalign{C_2&=\frac{1}{4\pi^2}\int_0^{\infty}\,dq\,q
R(q)^2,\cr F_2&=\frac{1}{4\pi^2}\int_0^{\infty}\,dq\,q
R(q)\,R^{\prime\prime}(q).\cr}}
Thus, in order that our damping scheme agrees with dimensional
regularization, we should introduce a counterterm that contributes
$-\alpha(p)$.

Not every integral we encounter will be simple enough to
explicitly evaluate in split dimensional regularization (as $I$ of
\exint\ was). However, it will always be possible to decompose the
integrals of interest into the sum of a complicated but convergent
integral and an easily evaluated divergent integral; this will be
sufficient to determine the corresponding counterterms. As an
illustration, we reevaluate the counterterm corresponding to the
integral $I$ in a perversely convoluted manner. Under the change
of variables $q\rightarrow p-q$ (an allowed variable change in
both regularization schemes), $I$ becomes
\eqn\exintt{I=\int\frac{d^4q}{(2\pi)^4}\frac{1}{(p-q)^2+(p_0-q_0)^2}.}
Performing the $q_0$ integration as above yields
\eqn\exinttqfour{\frac{1}{16\pi^3}\int\frac{d^3q}{\sqrt{(p-q)^2}}=
\frac{1}{16\pi^3}\int\frac{d^3q}{q}\left(1+\frac{q\cdot
p}{q^2}-\frac{p^2}{2q^2}+\frac{3(q\cdot
p)^2}{2q^4}+\ldots\right).}
We may now rewrite \exintt\ as a sum over a manifestly convergent
piece and an easily evaluated divergent piece as\foot{We introduce
a parameter $a$ in order to avoid artificially introducing IR
singularities.}
\eqn\exinttsplitone{\eqalign{I=&\frac{1}{16\pi^3}\int\frac{d^3q}
{\sqrt{(p-q)^2}} \cr
=&\frac{1}{16\pi^3}\int\,d^3q
\left(\frac{1}{\sqrt{(p-q)^2}}-\frac{1}{\sqrt{q^2}}\left[1+\frac{q\cdot
p}{q^2}-\frac{p^2}{2(q^2+a^2)}+\frac{3(q\cdot
p)^2}{2q^2(q^2+a^2)}\right]\right)\cr
&+\frac{1}{16\pi^3}\int\frac{d^3q}{\sqrt{q^2}}\left(1+\frac{q\cdot
p}{q^2}-\frac{p^2}{2(q^2+a^2)}+\frac{3(q\cdot
p)^2}{2q^2(q^2+a^2)}\right)}}
In (split) dimensional regularization with $d=3-\epsilon$, the second
line evaluates to
\eqn\painful{\eqalign{\frac{1}{32\pi^3}\int\frac{d^3q}{\sqrt{q^2}(q^2+a^2)}\left[{{3(q\cdot
p)^2}\over q^2}-p^2\right]&=\lim_{d\rightarrow
3}\frac{p^2}{32\pi^3}\left(\frac{3}{d}-1\right)\int\frac{d^dq}{\sqrt{q^2}(q^2+a^2)}\cr
&=\lim_{\epsilon\rightarrow
0}\frac{p^2}{32\pi^3}\left(\frac{\epsilon}{3-\epsilon}\right)
\left(\frac{4\pi}{\epsilon}+{\cal{O}}(\epsilon^0)\right)\cr
&=\frac{p^2}{24\pi^2}.}}
On the other hand, in the cut off scheme it evaluates to
\eqn\exinttregdepone{M^2C_2+\frac{p^2}{6}F_2,}
and \exinttregdepone\ is equal to \painful\ upon the addition of
the counterterm $-\alpha(p)$.

In this particular example, the convergent part of
\exinttsplitone\ is easy to evaluate (it is equal to $-p^2 \over
24 \pi^2$), but this is not true in more complicated examples, and
was not needed in order to evaluate the counterterm.

Of course, the separation of \exinttsplitone\ into a divergent and a
convergent part is ambiguous.  In our work ahead we will find it
convenient to fix this ambiguity by demanding that the divergent
piece (which we call the regulator dependent piece below) should
evaluate to zero in split dimensional regularization\foot{With this
convention, the regulator dependent piece in the example of the
previous paragraphs is the second term in \exinttsplitone\ minus
${p^2 \over 24 \pi^2}$.}.  With this convention, the counterterm
associated with any diagram is simply minus the regulator dependent
integral evaluated in the cutoff regulator scheme.

\subsec{Single-trace quadratic flat-space counterterms at order $\lambda$}

As an explicit example, we can now proceed to compute the
regulator-dependent piece of the gauge boson self-energy
$\Pi_{\mu\nu}=-\frac{1}{2}\langle A_{\mu}A_{\nu}\rangle$ .  To set our
conventions, we write the Yang-Mills action as
\eqn\ymaction{S=\frac{1}{4}\int\,d^4x\,\tr(F_{\mu\nu}F^{\mu\nu}).}
The momentum space Coulomb gauge ($\del_i A_i=0$) propagators take
the form
\eqn\props{\langle A^{ab}_iA^{cd}_j\rangle=\delta^{ad}\delta^{bc}
\left(\frac{q^2g_{ij}-q_iq_j}{q^2(q^2+q_0^2)}\right),\qquad\qquad
\langle A^{ab}_0A^{cd}_0\rangle=\delta^{ad}\delta^{bc}
\left(\frac{1}{q^2}\right).}
In addition, the gauge fixing procedure introduces a set of
complex adjoint ghosts $c,\bar{c}$ with Lagrangian
\eqn\ghostaction{{\cal{L}}_{ghost}=-\tr(\bar{c}\partial^iD_ic),}
where $D_i=\partial_i-ig_{YM}[A_i,\ast]$ is the gauge covariant
derivative. The propagator for the ghosts is identical to that of
the $A_0$ fields,
\eqn\ghostprop{\langle\bar{c}^{ab}c^{cd}\rangle=\delta^{ad}\delta^{bc}
\left(\frac{1}{q^2}\right).}
In the computations of this section we do not explicitly integrate
out $A_0$ and $c$ as we did in the previous section; of course
this does not affect the results.

In Appendix A.1 we define several regulator-dependent constants
and functions of external momentum that arise in our calculation.
In Appendix A.2 we depict the diagrams contributing to the gauge
boson self-energy, and list our results for their
regulator-dependent contributions (defined above). Adding these
together yields the following result for the regulator-dependent
contribution to the gauge boson self-energy (see Appendix A.1 for
notation) :
\eqn\PiikRD{\eqalign{\frac{1}{\lambda}\Pi_{ik}^{(RD)}&=\frac{1}{8\pi^2}
\ln\left(\frac{M}{\mu}\right)
\left[p_ip_k-(p^2+p_0^2)g_{ik}\right]-\frac{1}{15}\left(p^2g_{ik}+4p_ip_k\right)F_2\cr
&+\frac{7}{48\pi^2}\left(p_ip_k-p^2g_{ik}\right)-\frac{p_0^2}{48\pi^2}g_{ik}+
M^2g_{ik}\left(2C_1-\frac{2}{3}C_2\right)\cr
&+\frac{1}{24\pi^2}p_0^2g_{ik}\ln\left(\frac{{\cal{A}}_2}{{\cal{A}}_1^4}\right)+
\frac{1}{40\pi^2}p^2g_{ik}\ln\left(\frac{{\cal{A}}_1^4}{{\cal{A}}_2^9}\right)+
\frac{1}{120\pi^2}p_ip_k\ln\left(\frac{{\cal{A}}_2^{31}}{{\cal{A}}_1^{16}}\right)\cr
&+2 {\bar \Lambda} {\bar M}{\bar B}_1 {\bar C}_1 g_{ik}
+ \Lambda M B_1 \left[\delta^m_i\delta^n_k-g^{mn}g_{ik}\right](H_1)_{mn}\cr
&-\bar{\Lambda}\bar{M}\bar{B}_2\left(\bar{H}_2\right)_{ik}, }}
%
\eqn\PiifourRD{\frac{1}{\lambda}\Pi_{i0}^{(RD)}=\frac{7}{24\pi^2}p_0p_i
\ln\left(\frac{M}{\mu}\right)+\frac{5}{72\pi^2}p_0p_i+\frac{1}{24\pi^2}p_0p_i
\ln\left(\frac{{\cal{A}}_1^8}{{\cal{A}}_2}\right),}
\eqn\PifourfourRD{\frac{1}{\lambda}\Pi_{00}^{(RD)}=-\frac{11}{24\pi^2}p^2
\ln\left(\frac{M}{\mu}\right)-\frac{p^2}{3}F_2-\frac{1}{72\pi^2}p^2+
2M^2\left(C_1-C_2\right)
+\frac{1}{24\pi^2}p^2\ln\left(\frac{{\cal{A}}_2^5}{{\cal{A}}_1^{16}}\right).}

From the discussion above, the required quadratic counterterm
Lagrangian must be chosen to precisely cancel these regulator
dependent contributions, in order to give agreement with dimensional
regularization. Thus, we must have
\eqn\ctlag{{\cal{L}}_{ct}=-\tr(A^{\mu}A^{\nu})\Pi^{(RD)}_{\mu\nu}.}
In the next subsection, we perform two consistency checks on these results.

\subsec{The Slavnov-Taylor identity and $SO(4)$ invariance}

In this subsection, we first use the fact that our result for the
self-energy must be consistent with gauge invariance and $SO(4)$
symmetry at short distances to demonstrate the consistency of our
results for the logarithmic divergences in \PiikRD -
\PifourfourRD.  In particular, since the logarithmically divergent
and finite contributions must satisfy various Slavnov-Taylor
identities for these symmetries independently, we can determine
the structure of the former without any knowledge of the latter.

We begin by considering the Slavnov-Taylor identity relevant for
the gauge symmetry in Coulomb gauge.  As usual, we start with the
Euclidean gauge-fixed action
\eqn\stact{S=\int
d^4x\left\{{\cal{L}}_{YM}+\frac{1}{2\epsilon}(\nabla_iA^i)^2-
\bar{c}\nabla_iD_ic\right\}}
and take $\epsilon\rightarrow 0$ to get the Coulomb gauge.  The
BRST charge $Q$ satisfies
\eqn\stbrst{\eqalign{[Q,A_{\mu}]&=D_{\mu}c,\cr \{Q,c\}&=ic^2,\cr
\{Q,\bar{c}\}&=\frac{1}{\epsilon}\nabla_iA_i.}}
To obtain the Slavnov-Taylor identity, we study the partition
function with sources added for the operators generated by the
BRST transformation
\eqn\stpart{Z[J_{\mu},\xi,K_{\mu},L]=\int\,\exp\left\{-S+\int
d^4x\left(J_{\mu}A_{\mu}+\bar{\xi}c+\bar{c}\xi+
K_{\mu}[Q,A_{\mu}]-L\{Q,c\}\right)\right\}.}
Performing the change of variables $A\rightarrow
A+[\bar{\epsilon}Q,A]$, $c\rightarrow c+[\bar{\epsilon}Q,c]$,
$\bar{c}\rightarrow\bar{c}+[\bar{\epsilon}Q,\bar{c}]$ in the path
integral, we eventually obtain the standard identity
\eqn\stopi{\frac{\delta\hat{\Gamma}}{\delta
A_{\mu}}\frac{\delta\hat{\Gamma}}{\delta
K_{\mu}}+\frac{\delta\hat{\Gamma}}{\delta
c}\frac{\delta\hat{\Gamma}}{\delta L}=0,}
where $\hat{\Gamma}$ is the 1PI effective action less the gauge
fixing term.  From this, we may easily derive a 1-loop
Slavnov-Taylor identity relating the self-energy, $\Pi_{\mu\nu}$,
and the coefficient $\Phi_{\mu}$ of the $K_{\mu}c$ term of
$\hat{\Gamma}$:
\eqn\stid{\partial_{\mu}\Pi_{\mu\nu}+
(-\partial^2g_{\mu\nu}+\partial_{\mu}\partial_{\nu})\Phi_{\mu}=0.}
An analogous relation arising from $SO(4)$ invariance is difficult
to obtain since we are working in a noncovariant gauge.
Fortunately, a simple restriction that arises by requiring
$SO(4)$-invariance of the S-matrix will be enough for our
purposes.  The specific condition that we will impose is the
existence of a double pole in the full propagator at zero
momentum.  This requirement, combined with the weaker
Slavnov-Taylor identity
$\partial_{\mu}\partial_{\nu}\Pi_{\mu\nu}=0$, restricts the local
part of $\Pi_{\mu\nu}$ to be of the form
\eqn\serest{\eqalign{\Pi_{ij}&=C\left([p^2+p_0^2]g_{ij}-p_ip_j\right)\cr
\Pi_{i0}&=Dp_ip_0\cr \Pi_{00}&=-(C+2D)p^2}}
where $C$ and $D$ are dimensionless constants.  It is easy to
demonstrate, using the definition of $\Phi_{\mu}$, that the full
Slavnov-Taylor identity \stid\  fixes $C+D$, leaving one degree of
freedom that is in principle determined by a wave function
renormalization condition.  The logarithmic divergences must have
this structure independent of the finite contribution to the local
part of $\Pi_{\mu\nu}$.  That our result \PiikRD - \PifourfourRD\
is consistent with these conditions is easy to verify.

As a second, and less formal consistency check, we have used our
regularization scheme, together with the counterterms \ctlag, to
compute a physical quantity; the two-loop free energy of Yang-Mills
theory at infinite volume\foot{As far as we are aware, this is the
first time that this computation has been done in Coulomb gauge using any
regulating scheme.}. The counterterms computed in the previous
subsection play a crucial role in our calculation, which we
present in detail in appendix A.3. Our final answer,
$F_{2-loop}={V \lambda T^4/ 72}$, agrees with the previously
computed result (using dimensional regularization in Feynman gauge)
\ArnoldPS. We regard this as a rather nontrivial check of our
regularization scheme.

\subsec{The $\tr(A_iA_j)\tr(A_kA_l)$ counterterms}

We have seen in section 3.6 that the counterterms required to evaluate
$b$ take the form of double-trace terms quartic in the spatial
components of the gauge field. We will now follow the method of the
previous subsection (requiring the order $\lambda^2/N^2$ contribution
to the four point function $\langle A_i A_j A_k A_l\rangle$ to agree
with the result obtained by (split) dimensional regularization) to
evaluate the coefficients of the two possible counterterms of this
form, given in \countertermform.

The one loop diagrams contributing to the nonplanar part of the
four-point correlator are depicted in figure 6 in appendix A.5. It
follows from power counting that the leading divergence in each of
these diagrams is logarithmic. As a consequence, the regulator
dependent part of each of these diagrams may be evaluated with all
external momenta set to zero and has a divergent part proportional to
the single integral
\eqn\divint{\int\frac{d^3q}{4\pi\sqrt{q^2}(q^2+a^2)}.}
In Appendix A.5 we list the coefficient of this divergent integral
computed for each of the diagrams with a particular index
structure\foot{We keep the number of spatial dimensions, $d$, explicit
(as opposed to setting $d=3$), in order to determine the finite
counterterm needed to bring our result into agreement with that of
(split) dimensional regularization.}.  To obtain the full expression,
we must also sum over distinct permutations of indices.  We list the
results $R^{(*)}_{ijkl}$ (for the coefficient of the integral
\divint), as well as the corresponding contribution to the
counterterm, diagram by diagram in Appendix A.5.

Summing over the expressions in Appendix A.5 we find that the sum
of the diagrams depicted in Appendix A evaluates to
\eqn\logdiv{\frac{d-3}{2\pi^2}\left[1-\frac{4d-1}{d(d+2)}\right]
\int\frac{d^3q}{4\pi\sqrt{q^2}(q^2+a^2)}+{\rm finite}.}
Notice that the first term in \logdiv\ is simply zero in the cut off
regulator scheme (on setting $d=3$). However, in dimensional
regularization this term evaluates to
\eqn\logdivexp{\left(-\frac{2\epsilon}{15\pi^2}+{\cal{O}}(\epsilon^2)
\right) \times {1 \over  \epsilon}=-{2\over 15 \pi^2}.}
It follows
that perturbative computations in the damping function scheme must
be accompanied by the counterterm
\eqn\nonplanarcterm{{\cal{L}}_{CT}=\left(\frac{\lambda^2}{N^2}\right)
\left(\frac{1}{120\pi^2}\ln\left(\frac{{\cal{A}}_1^4{\cal{A}}_2^3
{\cal{A}}_4^{15}}{{\cal{A}}_3^{22}}\right)+
\frac{1}{60\pi^2}\right)\left[\tr(A_i A_i) \tr(A_j
A_j)+2\tr(A_iA_j)\tr(A_iA_j)\right],}
where the $\cal{A}$s are regulator dependent constants defined in
appendix A.1. For the sharp cutoff used in section 3, all ${\cal A}$'s
give the same result, so the term involving ${\cal A}$'s evaluates to
zero.

\newsec{Validity of Perturbation Theory}

In this section we will determine the precise regime of validity
of perturbation theory for pure Yang Mills theory on a sphere of
radius $R$, in order to make sure that the computation we
described above is valid. Naively, perturbation theory is good
whenever $\Lambda_{QCD} R \ll 1$, since the running coupling
constant is then small at all scales above the scale $1/R$ of the
classical mass gap. In thermal Yang Mills theory it turns out that
this expectation is modified by IR divergences; as we explain
below, perturbation theory is valid at small $\Lambda_{QCD} R$
only for $TR \ll {1\over \lambda(T)}$, where $\lambda(T)$ is the
running 't Hooft coupling at the energy scale $T$. It follows in
particular that for $\Lambda_{QCD}R \ll 1$ perturbation theory is
good at $TR \sim 1$, which is the regime of interest for this
paper.

\subsec{Review of IR divergences in flat space}

In this subsection we review the well known effects of IR
divergences on thermal Yang Mills theory in flat space; see
\GrossBR\ and references therein for more details.

Perturbative computations in Yang Mills theory on $\IR^3 \times S^1$
(where the $S^1$ is a thermal circle) are beset by IR divergences,
as is easily seen from power counting. IR divergences arise from
the $\omega=0$ sector ($\omega$ is the Euclidean energy) of the
theory. Consider a Feynman diagram made up entirely of $\omega=0$
modes. Let $q$ be the scale of spatial momenta in such a diagram.
Each additional loop is accompanied by a factor of ${\lambda T
\over q^4}$ (from vertices and propagators) times $ q^3$ (from
phase space), giving a net factor of ${\lambda T \over q}$.
Consequently, higher loop graphs are increasingly infrared
divergent.

These infrared divergences are cured by the fact that the gauge
field $A$ is effectively massive. Working in Feynman gauge, the
one loop self energy of the $A_0$ field at zero energy and
momentum, $\Pi_{00}(0, {\vec 0})$, is nonzero and of order
$\lambda T^2$. As a consequence, $A_0$ is effectively massive with
mass of order $m_{el}=\sqrt{\Pi_{00}(0, {\vec 0} )} \sim
\sqrt{\lambda} T$. Consequently, infrared divergences in loops
involving $A_0$ are cut off at this mass; thus, the
effective loop counting parameter for $A_0$ loops with $\omega=0$
is ${\lambda T \over m_{el}} \sim \sqrt{\lambda}$. So, $A_0$ IR
divergences change the perturbative expansion parameter from
$\lambda$ to $\sqrt{\lambda}$. The first fractional power of
$\lambda$ that appears in the expansion of the free energy is
$\lambda^{{3\over 2}}$ (from a one loop graph using a mass
corrected propagator for $A_0$). The next fractional power,
$\lambda^{{5\over 2}}$, follows from the electric mass
regularization of a 3-loop IR divergence.

 IR divergences involving the spatial gauge field $A_i$ are more
 serious.  It turns out that the $A_i$ self energy at zero momentum
 vanishes at one loop, but is nonvanishing at two loops. As a
 consequence, the effective mass for $A_i$ is of order $m_{mag}\sim
 \sqrt{\lambda^2 T^2} = T \lambda$.  IR divergences involving spatial
 $A_i$ fields are cut off by this mass; as a consequence the effective
 loop counting parameter for loops involving spatial $A_i$ is of order
 one. Graphs of low enough order do not suffer from spatial IR
 divergences; however, a detailed investigation reveals that an
 infinite number of graphs contribute to the free energy at order
 $\lambda^3$ and higher. In summary, the free energy may be expanded
 up to order $\lambda^{{5\over 2}}$; all coefficients in the expansion
 of the free energy to this order are perturbatively computable, and
 have been computed (see \gfive\ and references therein).
 Higher order terms are, in principle, inaccessible to perturbative
 analysis.

The generation of an electric mass simply reflects the fact that
the high temperature dynamics of Yang Mills theory deconfines.
Indeed, space is filled with a plasma of charged particles of
density $\sim T^3$. As each of these particles carries a charge
$\sqrt{\lambda}$, the screening length of this plasma is $1 /
\sqrt{\lambda} T$, explaining the magnitude of $m_{el}$ described
above.

The generation of a magnetic mass may be explained from the
observation that Yang Mills theory on $\IR^3 \times S^1$ reduces, at
high temperatures, to a (Euclidean) 3 dimensional Yang Mills
theory with an effective Yang Mills coupling constant $\lambda T$,
coupled to an adjoint scalar field of much larger mass
$\sqrt{\lambda} T$. The low energy dynamics of this theory is
simply that of pure 3 dimensional Yang Mills theory, which
non-perturbatively develops a mass gap of order $\lambda T$.

\subsec{IR behaviour on $S^3$}

We now turn to a study of the IR behaviour of Yang Mills theory on
$S^3$. Yang Mills theory on $S^3$ has a mass gap $1/R$ even
classically. As a consequence, even ignoring the dynamical mass
generation, the power counting arguments of the previous subsection
indicate that (assuming $TR \gg 1$) loops of low energy $A_0$ and
$A_i$ fields are both weighted by the effective coupling
\eqn\effc{\lambda_{eff}\sim \lambda TR \simeq m_{mag} R.}
Perturbation theory is valid when this effective coupling is small.
When $m_{mag}R \sim 1$ this effective coupling is of order one, and
perturbation theory breaks down \foot{Once naive perturbation
theory breaks down, the effects of dynamical mass generation are
important. In this regime the correct way to proceed is to mimic the
flat space analysis, and to shift the bare quadratic action by $
A_{\mu} \Pi_{\mu \nu}(0, 0) A_{\nu}$, where the first index of $\Pi$
refers to the energy and the second to the spherical harmonic number
on $S^3$. At low enough temperatures this effective mass is of order
$1/R$, while at higher temperatures the effective mass crosses over
to its flat space value. For instance, the effective $A_i$ mass is
of order $1/R$ for $\lambda TR \ll 1$, but is given by $m_{mag}$ for
$\lambda TR \gg 1$. Consequently, for $T \gg 1 / \lambda R$, the
flat space analysis of the previous subsection applies, and
perturbation theory breaks down. Were we to ignore the dynamically
generated contribution to the masses, we would be faced with a
paradox. The free energy would receive contributions proportional to
increasingly high powers of $R$, in conflict with extensivity.}.

In summary, finite temperature Yang Mills perturbation theory on
$S^3$ is useful provided both that $\Lambda_{QCD} R \ll 1$ and
that temperatures are low enough so that $\lambda (T) TR \ll 1$.

\newsec{Conclusions}

In this paper we have computed the leading perturbative correction
to the thermal partition function of pure $SU(N)$ Yang-Mills
theory on $S^3$ around the phase transition point, and found that
it leads to a first order deconfinement phase transition. The
analysis, requiring diagrams with up to 3 loops, is quite
complicated, and the interesting result is a single number
\bdefdag\ which governs the order of the phase transition. As
described above, we have subjected our formalism to various
consistency checks, including the cancellation of all divergent
contributions, but we do not have any way to independently verify the
correctness of our final result \bdefdag. It would be 
useful to have an independent computation of \bdefdag\ as a check
of our results.

In the pure Yang-Mills theory on $S^3$ we found that $b$ is negative;
a similar result was found in the corresponding analysis of various
quantum mechanical systems \refs{\HadizadehBF,\toruspaper}. It would
be interesting to compute the sign of $b$ in other $3+1$ dimensional
(or lower dimensional) field theories, and to see how it depends on
the matter content and on the various coupling constants of the
theory. In particular, we are planning to compute the value of $b$ in
the $3+1$ dimensional ${\cal N}=4$ supersymmetric Yang-Mills theory on
$S^3$, to see if the order of the deconfinement phase transition at
weak coupling is the same as the first order behaviour found at strong
coupling \refs{\HawkingDH,\WittenZW}.

Since $b$ is negative in all examples that have been analyzed so
far, one might conjecture that for some reason it always has to be
negative. However, it is easy to see that this is not the case, at
least when one adds additional scalar fields with arbitrary
couplings. For example, let us consider a $0+1$ or $1+1$
dimensional gauge theory with a massive adjoint scalar field
$\Phi$, with a potential of the form \eqn\scalpot{V(\Phi) =
\tr(m^2 \Phi^2 + c_4 g_{YM}^2 \Phi^4 + c_6 g_{YM}^4 \Phi^6 + c_8
g_{YM}^6 \Phi^8),} where the $c_i$ are kept fixed in the 't Hooft
large $N$ limit. By analyzing the vacuum diagrams up to 3-loop
order, it is easy to see that $c_6$ contributes linearly to $b$ at
order $\lambda^2$ (through a diagram similar to 3n of figure
\diagrams), while $c_8$ only contributes to $b$ at higher orders.
Thus, for given values of $m^2$ and $c_4$, we can achieve any sign
for the leading perturbative contribution to $b$ just by varying
$c_6$. We may need to choose $c_6$ to be negative for this, but we
can always choose $c_8$ to be large enough so that $\Phi=0$ is
still the unique minimum of \scalpot. In higher dimensions we have
not yet been able to find a similar example involving purely
single-trace interactions, but a potential term of the form $c
g_{YM}^2 \tr(\Phi^2)^2 / N$ may be shown by similar arguments to
lead to arbitrary values for $b$ (already at order $\lambda$).
Thus, large $N$ weakly coupled deconfinement phase transitions may
generally be of either first order or second order.

\bigskip

\centerline{\bf Acknowledgements}

We would like to thank N. Arkani-Hamed, A. Dabholkar, R.  Gopakumar,
D. Gross, C. Korthals Altes, H. Liu, J. Maldacena, G. Mandal, L. Motl,
A. Neitzke, R. Pisarski, J. Polchinski, E. Rabinovici, A. Sen, G.  Semenoff,
S. Shenker, A. Strominger, M. Teper, S. Trivedi, S. Wadia, L. Yaffe, and X. Yin,
as well as everybody else that encouraged us to go through with this
computation. OA would like to thank Harvard University, UBC, KITP, the
Aspen Center for Physics, the Fields Institute, the Perimeter Institute,
and the second Crete regional meeting on
String Theory for hospitality during the course of this long
project. JM would like to thank UBC, IAS, and KITP for hospitality
during various stages of this work. SM would like to thank the second
Crete regional meeting on String Theory, the Indian Institute for the
Cultivation of Sciences, and KITP for hospitality while this work was
in progress. KP would like to thank KITP for hospitality while this
work was in progress. The work of OA was supported in part by the
Israel-U.S. Binational Science Foundation, by the Israel Science
Foundation (grant number 1399/04), by the Braun-Roger-Siegl
foundation, by the European network HPRN-CT-2000-00122, by a grant from
the G.I.F., the German-Israeli Foundation for Scientific Research and
Development, and by
Minerva.  The work of JM was supported in part by an NSF Graduate
Research Fellowship. The work of SM was supported in part by DOE grant
DE-FG01-91ER40654, NSF career grant PHY-0239626, a Sloan fellowship,
and a Harvard Junior Fellowship. The work of KP was
supported in part by DOE grant DE-FG01-91ER40654.  The work of MVR was
supported in part by NSF grant PHY-9870115, by funds from the Stanford
Institute for Theoretical Physics, by NSERC grant 22R81136 and by the
Canada Research Chairs programme.

\appendix{A}{Details Related to Regularization}

In this appendix, we provide many of the details behind
calculations discussed in sections 3 and 4.

\subsec{Definitions of useful regulator-dependent constants and
functions}

We start by presenting definitions for the various constants which
encode the dependence of the results in section 4 on the
regularization functions $R$, $\tilde{R}$, $\bar{R}$ and $\tilde{\bar{R}}$.  
We also define three
functions of external momentum that will arise in the expressions
for individual diagrams presented later in this appendix.

\eqn\rmomsdefdone{
\ln\left(\frac{{\cal{A}}_nM}{\mu}\right)=\int_0^{\infty}\,dq\,
\frac{\sqrt{q^2}R^n(q)}{q^2+\frac{\mu^2}{M^2}}}
\eqn\rmomsdeftwo{\eqalign{
B_1&=\frac{1}{2\pi}\int_{-\infty}^{\infty}dq_0\,\tilde{R}(q_0)\cr
\bar{B}_1&=\frac{1}{2\pi}\int_{-\infty}^{\infty}dq_0\,\tilde{\bar{R}}(q_0)\cr
\bar{B}_2&=\frac{1}{2\pi}\int_{-\infty}^{\infty}dq_0\,\tilde{\bar{R}}(q_0)^2\cr
C_1&=\frac{1}{4\pi^2}\int_0^{\infty}\,dq\,qR(q)\cr
\bar{C}_1&=\frac{1}{4\pi^2}\int_0^{\infty}\,dq\,q{\bar{R}}(q)\cr
C_2&=\frac{1}{4\pi^2}\int_0^{\infty}\,dq\,qR(q)^2\cr
F_2&=\frac{1}{4\pi^2}\int_0^{\infty}\,dq\,q\,R(q)R^{\prime\prime}(q)\cr
}}
\eqn\rmomsdefthree{\eqalign{
(H_1)_{ij}(p)&=\int\frac{d^3q}{(2\pi)^3}\frac{q_iq_j}
{q^2\left(\frac{p}{M}-q\right)^2}R(q)\cr
(\bar{H}_2)_{ij}(p)&=\int\frac{d^3q}{(2\pi)^3}\frac{q_iq_j}
{q^2\left(\frac{p}{\bar{M}}-
q\right)^2}
\bar{R}(\sqrt{q^2})\bar{R}\left(\sqrt{\left(\frac{p}
{\bar{M}}-q\right)^2}\right)\cr
(\bar{J}_2)_i(p)&=\int\frac{d^3q}{(2\pi)^3}\frac{q_i}
{q^2\left(\frac{p}{\bar{M}}-q\right)^2}\bar{R}(\sqrt{q^2})\bar{R}
\left(\sqrt{\left(\frac{p}{\bar{M}}-q\right)^2}\right)}}

Note the parameter $\mu$ appearing in the definition of the
constants ${\cal A}_i$. This parameter is needed when splitting
logarithmically divergent integrals in order to avoid introducing
artificial IR divergences into the ``regularized'' pieces. Moreover,
$\mu$ is identified with the scale associated to (split)
dimensional regularization when comparing results obtained in that
scheme with those obtained in ours.

\subsec{Diagram by diagram contribution to the regulator dependent
piece of $\Pi_{\mu \nu}$}

\fig{Diagrams contributing to the gauge boson self-energy at
1-loop. Solid lines denote $A_i$ propagators, dashed lines denote
$A_0$ propagators, and arrowed lines denote ghost
propagators.}{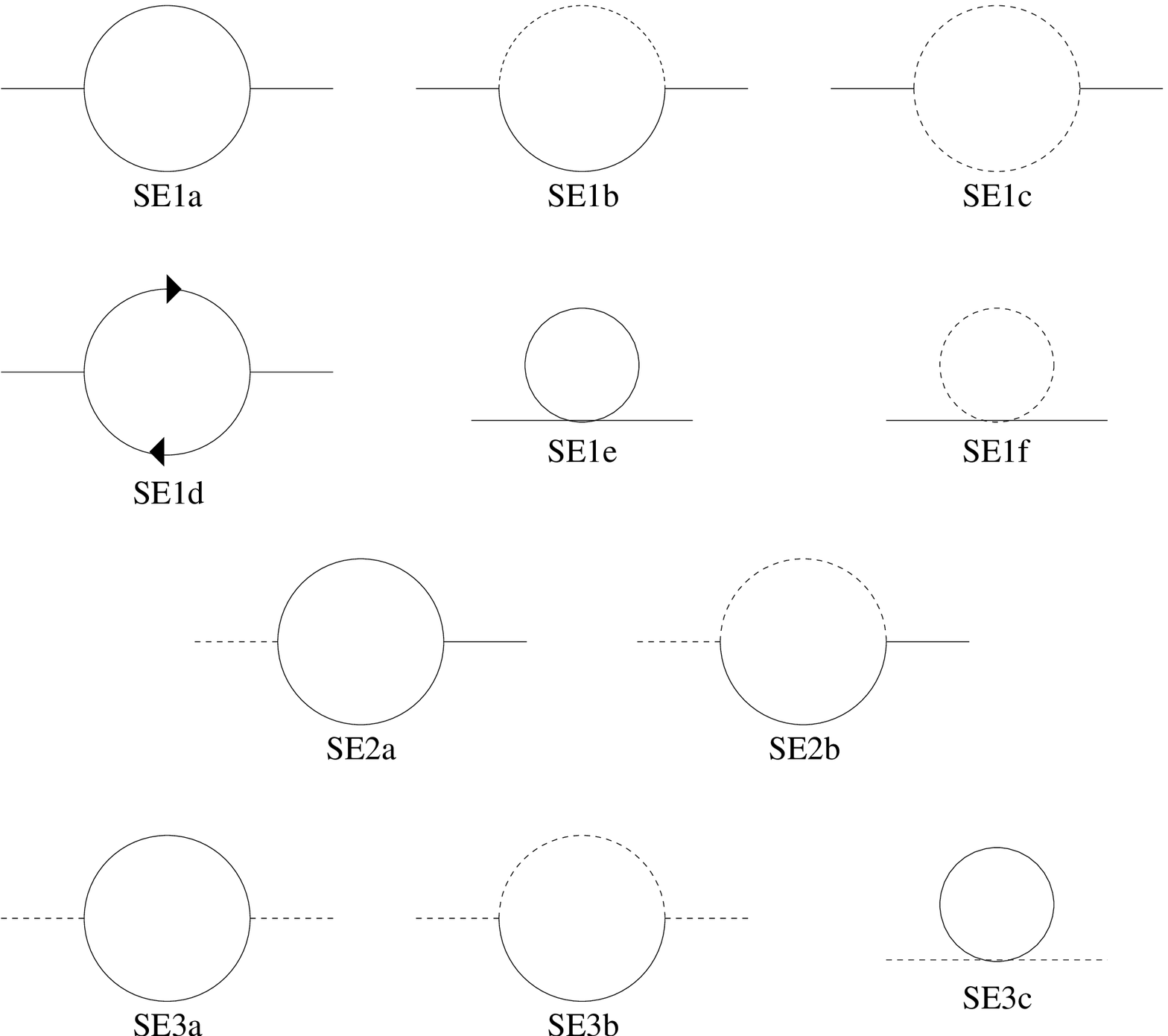}{6truein} \figlabel{\SEdiags}

We now list the regulator dependent contribution,
$\Pi^{(RD)}_{*}$, of each diagram in figure \SEdiags\ to the
self-energy $\Pi_{\mu \nu}=-\frac{1}{2}\langle A_{\mu} A_{\nu}
\rangle$ at momentum $p$ :
\eqn\SEonea{\eqalign{
\frac{1}{\lambda}\left(\Pi^{(RD)}_{SE1a}\right)_{ij}&=-\frac{2}{3}M^2C_2
g_{ij}+
\left(\frac{1}{24\pi^2}p_0^2g_{ij}-\frac{9}{40\pi^2}p^2g_{ij}+\frac{31}{120\pi^2}p_ip_j\right)
\ln\left(\frac{{\cal{A}}_2M}{\mu}\right)\cr
&-\frac{1}{15}\left(p^2g_{ij}+4p_ip_j\right)F_2+\frac{269}{3600\pi^2}p_ip_j-
\frac{17}{400\pi^2}p^2g_{ij}+\frac{1}{144\pi^2}p_0^2g_{ij},}}
\eqn\SEoneb{\eqalign{
\frac{1}{\lambda}\left(\Pi^{(RD)}_{SE1b}\right)_{ij}&=\frac{2}{3}M^2C_1g_{ij}+
\left(-\frac{1}{6\pi^2}p_0^2g_{ij}+\frac{1}{10\pi^2}p^2g_{ij}-\frac{2}{15\pi^2}p_ip_j\right)
\ln\left(\frac{{\cal{A}}_1M}{\mu}\right)\cr
&-\frac{1}{36\pi^2}p_0^2g_{ij}+\frac{16}{225\pi^2}p_ip_j-\frac{31}{300\pi^2}p^2g_{ij}
+\Lambda M
B_1\left(\delta^m_i\delta^n_j-g^{mn}g_{ij}\right)(H_1)_{mn},}}
%
%
\eqn\SEonec{
\frac{1}{\lambda}\left(\Pi^{(RD)}_{SE1c}\right)_{ij}=\bar{\Lambda}
\bar{B}_2\left(p_i[\bar{J}_2]_j-2 \bar{M} (\bar{H}_2)_{ij}\right),}
\eqn\SEoned{
\frac{1}{\lambda}\left(\Pi^{(RD)}_{SE1d}\right)_{ij}=\bar{\Lambda}
\bar{B}_2\left(\bar{M} [\bar{H}_2]_{ij}-p_i[\bar{J}_2]_j\right),}
\eqn\SEonee{
\frac{1}{\lambda}\left(\Pi_{SE1e}^{(RD)}\right)_{ij}=\frac{4}{3}M^2C_1
g_{ij},}
\eqn\SEonef{
\frac{1}{\lambda}\left(\Pi_{SE1f}^{(RD)}\right)_{ij}=2\bar{\Lambda}
\bar{M} \bar{B}_1 \bar{C}_1 g_{ij},}
\eqn\SEtwoa{\eqalign{
\frac{1}{\lambda}\left(\Pi^{(RD)}_{SE2a}\right)_{0i}&=-\frac{p_0p_i}{24\pi^2}
\ln\left(\frac{{\cal{A}}_2M}{\mu}\right)+\frac{p_0p_i}{72\pi^2},}}
\eqn\SEtwob{\eqalign{
\frac{1}{\lambda}\left(\Pi^{(RD)}_{SE2b}\right)_{0i}&=\frac{p_0p_i}{3\pi^2}
\ln\left(\frac{{\cal{A}}_1M}{\mu}\right)+\frac{p_0p_i}{18\pi^2},}}
\eqn\SEthreea{\eqalign{
\frac{1}{\lambda}\left(\Pi_{SE3a}^{(RD)}\right)_{00}&=-2C_2M^2+\frac{5p^2}{24\pi^2}
\ln\left(\frac{{\cal{A}}_2M}{\mu}\right)-\frac{p^2}{3}F_2+\frac{7}{72\pi^2}p^2},}
\eqn\SEthreeb{\eqalign{
\frac{1}{\lambda}\left(\Pi_{SE3b}^{(RD)}\right)_{00}&=-\frac{2p^2}{3\pi^2}
\ln\left(\frac{{\cal{A}}_1M}{\mu}\right)-\frac{1}{9\pi^2}p^2},}
\eqn\SEthreec{\eqalign{
\frac{1}{\lambda}\left(\Pi_{SE3c}^{(RD)}\right)_{00}&=2M^2C_1}.}

\subsec{A check : the 2-loop free energy}

We now proceed to compute a physical quantity, the free energy of
Yang Mills theory at order $\lambda$, using our regularization
scheme. We will find a result that agrees with the standard result
obtained by traditional methods (utilizing dimensional
regularization in Feynman gauge). We view this agreement as a
significant check on the consistency of our regularization scheme.

\fig{Diagrams contributing to the 2-loop free
energy.}{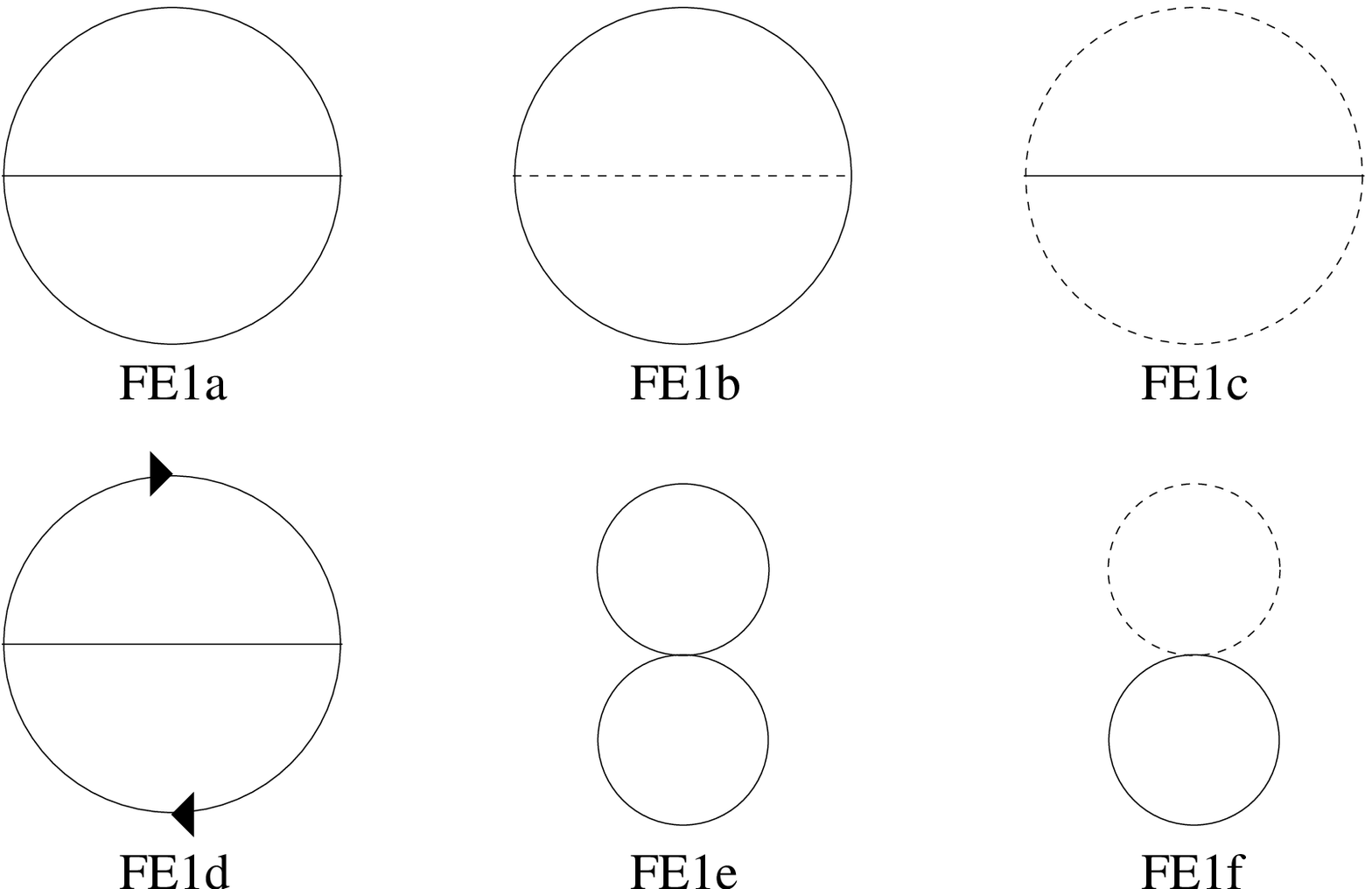}{4truein} \figlabel{\FEdiags}

The two-loop free energy of Yang Mills theory, at order $\lambda$,
receives contributions from six graphs depicted in figure
\FEdiags. A set of one-loop counterterm graphs also contribute to
the same order. We will find it useful to group each two-loop
graph with a set of one-loop counterterm graphs, in the manner
(and for the reasons) that we now explain.

Consider the counterterm contributions to the free energy at order
$\lambda$. For the purpose of this discussion, it will be useful
to regard counterterms that cancel different one-loop
contributions to the self energy (see above) as
distinct\foot{Furthermore, it is easy to convince oneself that
only the counterterm with external vector lines yields a
temperature-dependent contribution to the free energy.}. Any
one-loop contribution to the free energy with a counterterm
insertion may uniquely be associated with one of the two-loop free
energy graphs by ``blowing up'' the counterterm into its parent
self-energy diagram.  It is plausible (and true) that the sum of
any particular 2-loop free energy graph with all its associated
1-loop counterterm graphs, is finite\foot{It may also seem natural
to guess that this sum will equal the corresponding diagram
evaluated in split dimensional regularization, but this is not
precisely the case, as we explain in detail in the next
subsection. } and independent of the damping function $R({q \over
\Lambda})$. This fact suggests a natural grouping of graphs that
we will use.

Next, we list the results related to the diagrams, depicted in
figure \FEdiags, that contribute to the 2-loop free energy. For
each such diagram, we list the result of evaluating it in our
damping function scheme, $F_{*}$, the contribution arising from
its associated one-loop self-energy diagram (see section 4 and the
previous subsection),
$F_*^{(CT)}$, and the result obtained by direct evaluation in
(split) dimensional regularization, $F_*^{(DR)}$ (all divided by
$\lambda V$, where $V$ is the volume of space) :
\eqn\FEonea{\eqalign{F_{FE1a}&=-\frac{1}{9}C_2M^2T^2-
\frac{4}{225}T^4\ln\left(\frac{{\cal{A}}_2M}{a}\right)-
\frac{\pi^2}{225}T^4F_2+
\frac{4}{15\pi^4}\int\frac{dp}{e^{p/T}-1}p^3\ln\left(\frac{p}{a}\right)\cr
&-\frac{4393}{216000}T^4
+\frac{1}{128\pi^4}\int_0^{\infty}\frac{dq}{q^2(e^{q/T}-1)}
\int_0^{\infty}\frac{dp}{p^2(e^{p/T}-1)}\cr
&\qquad\qquad\qquad\qquad\qquad
\times\left\{4q^5p+4p^5q+(p^2+q^2)(p^4+6p^2q^2+q^4)
\ln\left[\frac{(p-q)^2}{(p+q)^2}\right]\right\}\cr
F^{(CT)}_{FE1a}&=\frac{1}{9}C_2M^2T^2+\frac{4}{225}T^4
\ln\left(\frac{{\cal{A}}_2M}{\mu}\right)+\frac{\pi^2}{225}T^4F_2+
\frac{89}{27000}T^4\cr
F^{(DR)}_{FE1a}&=\frac{4}{15\pi^4}\int\frac{dp}{e^{p/T}-1}p^3
\ln\left(\frac{p}{a}\right)+
\frac{4}{225}T^4\ln\left(\frac{a}{\mu}\right)-\frac{587}{72000}T^4\cr
&+\frac{1}{128\pi^4}\int_0^{\infty}\frac{dq}{q^2(e^{q/T}-1)}
\int_0^{\infty}\frac{dp}{p^2(e^{p/T}-1)}\cr
&\qquad\qquad\qquad\times\left\{4q^5p+4p^5q+(p^2+q^2)(p^4+6p^2q^2+q^4)
\ln\left[\frac{(p-q)^2}{(p+q)^2}\right]\right\}
}}
\eqn\FEoneb{\eqalign{F_{FE1b}&=\frac{1}{9}M^2T^2C_1+
\frac{4}{225}T^4\ln\left(\frac{{\cal{A}}_1M}{\mu}\right)
-\frac{4}{15\pi^4}\int_0^{\infty}\frac{dp}{e^{p/T}-1}p^3
\ln\left(\frac{p}{a}\right)+\frac{3593}{216000}T^4\cr
&-\Lambda M B_1\int\frac{d^3q}{(2\pi)^3}\frac{q^2g_{ij}-q_iq_j}
{q^3\left(e^{-q/T}-1\right)}H_1^{ij}(q)\cr
&-\frac{1}{128\pi^4}\int_0^{\infty}\frac{dq}{q^2(e^{q/T}-1)}
\int_0^{\infty}\frac{dp}{p^2(e^{p/T}-1)}\cr
&\qquad\qquad\qquad\times\left\{4(p^5q+q^5p)+(p^2+q^2)(p^4+6p^2q^2+q^4)
\ln\left[\frac{(p-q)^2}{(p+q)^2}\right]\right\}\cr
F^{(CT)}_{FE1b}&=-\frac{1}{9}M^2T^2C_1-
\frac{4}{225}T^4\ln\left(\frac{{\cal{A}}_1M}{\mu}\right)+\frac{17}{3375}T^4\cr
&+\Lambda M B_1\int\frac{d^3q}{(2\pi)^3}\frac{q^2g_{ij}-q_iq_j}
{q^3\left(e^{-q/T}-1\right)}H_1^{ij}(q)\cr
F^{(DR)}_{FE1b}&=-\frac{4}{15\pi^4}\int\frac{dp}{e^{p/T}-1}p^3
\ln\left(\frac{p}{a}\right)-
\frac{4}{225}T^4\ln\left(\frac{a}{\mu}\right)+\frac{2761}{216000}T^4\cr
&-\frac{1}{128\pi^4}\int_0^{\infty}\frac{dq}{q^2(e^{q/T}-1)}
\int_0^{\infty}\frac{dp}{p^2(e^{p/T}-1)}\cr
&\qquad\qquad\qquad\times\left\{4\left(p^5q+q^5p\right)+
(p^2+q^2)(p^4+6p^2q^2+q^4)\ln\left[\frac{(p-q)^2}{(p+q)^2}\right]\right\}\cr
}}
%
%
\eqn\FEonec{\eqalign{ F_{FE1c}&=-2\bar{\Lambda}\bar{M} \bar{B}_2
T\sum_{p_0}\int\frac{d^3p}{(2\pi)^3}\frac{\left(p^2g^{ij}-p^ip^j\right)(\bar{H}_2)_{ij}(p)}{p^2(p^2+p_0^2)}\cr
F^{(CT)}_{FE1c}&=2\bar{\Lambda}\bar{M} \bar{B}_2
T\sum_{p_0}\int\frac{d^3p}{(2\pi)^3}\frac{\left(p^2g^{ij}-p^ip^j\right)(\bar{H}_2)_{ij}(p)}{p^2(p^2+p_0^2)}\cr
F^{(DR)}_{FE1c}&=0}}
\eqn\FEoned{\eqalign{ F_{FE1d}&=\bar{\Lambda}\bar{M} \bar{B}_2
T\sum_{p_0}\int\frac{d^3p}{(2\pi)^3}\frac{\left(p^2g^{ij}-p^ip^j\right)(\bar{H}_2)_{ij}(p)}{p^2(p^2+p_0^2)}\cr
F^{(CT)}_{FE1d}&=-\bar{\Lambda}\bar{M} \bar{B}_2
T\sum_{p_0}\int\frac{d^3p}{(2\pi)^3}\frac{\left(p^2g^{ij}-p^ip^j\right)(\bar{H}_2)_{ij}(p)}{p^2(p^2+p_0^2)}\cr
F^{(DR)}_{FE1d}&=0}}
\eqn\FEonee{\eqalign{F_{FE1e}&=\frac{2}{9}M^2T^2C_1+\frac{1}{108}T^4\cr
F^{(CT)}_{FE1e}&=-\frac{2}{9}M^2T^2C_1\cr
F^{(DR)}_{FE1e}&=\frac{1}{108}T^4}}
\eqn\FEonef{\eqalign{F_{FE1f}&=\frac{1}{3}\bar{\Lambda} \bar{M} \bar{C}_1
\bar{B}_1 T^2\cr
F^{(CT)}_{FE1f}&=-\frac{1}{3}\bar{\Lambda} \bar{M} \bar{C}_1 \bar{B}_1 T^2\cr
F^{(DR)}_{FE1f}&=0}}

Summing the contributions of the individual diagrams, we find that
the 2-loop free energy is
\eqn\twoloopFE{\frac{F_{2-loop}}{V}=\frac{\lambda}{72} T^4,}
in agreement with results previously computed using dimensional
regularization in Feynman gauge \ArnoldPS \foot{The result often
seen in the literature is $\frac{\lambda}{144}T^4$, which differs
from this by a factor of 2. The reason for this apparent
difference arises from a choice of convention.  Our definition of
the propagators in \props\ corresponds to a normalization of the
basis matrices $(t^A)^{ab}$ such that
$(t^A)^{ab}(t^A)^{cd}=\delta^{ad}\delta^{bc}$, which is equivalent
to taking the quadratic Casimir of the fundamental representation
as $C(N)=1$.  The convention most prevalent in the literature is
$C(N)=\frac{1}{2}$.  This difference amounts to an effective
difference in the definition of $\lambda$ which, when properly
accounted for, gives an extra factor of 2.} \foot{We also find the
same result for the free energy evaluated directly in Coulomb
gauge using split dimensional regularization.}.

We may explicitly verify the diagram by diagram cancellation of
divergences discussed above.  In addition, we note that adding the
contributions from a free energy diagram in our damping function
scheme and its associated self-energy diagram does not lead to
full diagram by diagram agreement with the result of (split)
dimensional regularization.  We discuss this in greater detail in
the next subsection.

\subsec{Diagram by diagram comparison with split dimensional
regularization}

In this subsection, we will explain in detail how the sum of any
particular two loop self energy graph plus all its associated 1-loop
counterterm graphs differs from the same graph evaluated in
dimensional regularization.

 If we consider summing
a free energy diagram computed in our scheme with its corresponding
counterterm, this yields a result equivalent to evaluating the
integral over the internal loop momentum from which the divergence
arises {\it{before}} contracting the legs of this loop with the
$A_i$ propagator.  After the evaluation, we perform this contraction
but, at that point, the number of spatial dimensions is fixed at
$d=3$.  On the other hand, when we compute the free energy diagram
in pure (split) dimensional regularization, contraction of the
divergent momentum integral with the final propagator is done at
unspecified $d$, leading to additional dependence on $\epsilon=3-d$
and changing the finite result.

As a check on our calculations, we now compute the difference
between the finite parts of diagrams evaluated with our scheme and
with (split) dimensional regularization.  Any factors of $d$ that
arise only affect the evaluation of the logarithmic divergences and
thus we restrict attention to these.  We write a generic
logarithmically divergent integral that arises in the asymptotic
expansion of a given self-energy diagram as
\eqn\genlog{[f(p^2,\epsilon)g_{ij}+g(p^2,\epsilon)p_ip_j]
\int\frac{d^3q}{(2\pi)^3}\frac{1}{\sqrt{q^2}(q^2+a^2)}.}
Contracting this with an $A_i$ propagator at momentum $p$, and
evaluating the integral in dimensional regularization, we obtain
\eqn\genlogcon{(2-\epsilon)f(p^2,\epsilon)\int\frac{d^3q}
{(2\pi)^3}\frac{1}{\sqrt{q^2}(q^2+a^2)}=(2-\epsilon)
\left[f(p^2,0)+{\cal{O}}(\epsilon)\right]\left[\frac{1}{\epsilon}-
\ln\left(\frac{a}{\mu}\right)\right].}

We see that the new finite contribution that arises due to the
$\epsilon$ in the $(2-\epsilon)$ factor is precisely $-\frac{1}{2}$
times the coefficient of $\ln \mu$ in the final result.  This
implies that, to get the (split) dimensional regularization result
for diagram $FE1a$ ($FE1b$), we should add $\frac{2}{225}T^4$
($-\frac{2}{225}T^4$).  It is easy to check that this is consistent
with equations \FEonea\ and \FEoneb.  Moreover, we see why this
diagram by diagram discrepancy didn't ruin agreement in the final
result for the free energy; the difference between our scheme and
(split) dimensional regularization is proportional to the
logarithmic divergence of a given diagram and all logarithmic
divergences cancel when the diagrams are summed.

\subsec{Coefficients of divergences in non-planar one-loop 4-point
graphs}

\fig{The diagrams contributing to $\langle
\tr(A_iA_j)\tr(A_kA_l)\rangle$ at order $\lambda^2/N^2$, whose
corresponding counterterms are relevant to the calculation of
$b$.}{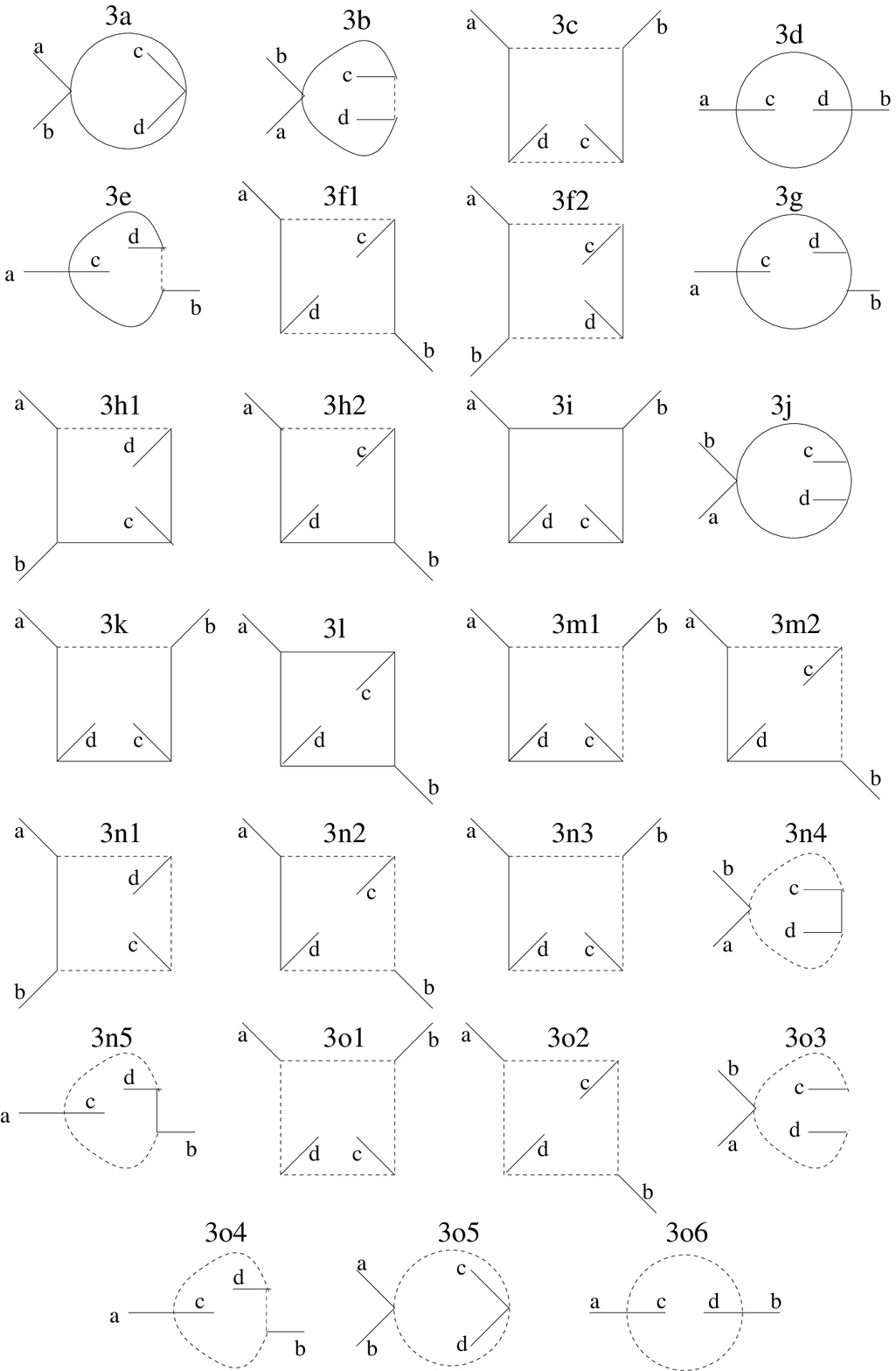}{5.5truein}
\figlabel{\nonplanardiags}

In this subsection, we consider details related to the computation
of the $\tr(A_iA_j)\tr(A_kA_l)$ counterterm that is relevant for
the computation of $b$ in section 3.  The relevant diagrams are depicted in
figure \nonplanardiags.  The seemingly chaotic ordering and
labeling of diagrams was chosen for easy identification with
3-loop free energy diagrams via the correspondence discussed in
the main text.

We compute the counter-term by computing four-point functions of
gauge fields, with the color index structure $\vev{\tr(A_i A_j)
\tr(A_k A_l)}$ which arises from one-loop non-planar diagrams.
These diagrams generally have a logarithmic divergence
proportional to
\eqn\logdivv{\int\frac{d^3q}{4\pi\sqrt{q^2}(q^2+a^2)}.}
We begin by listing the coefficient, $M^*_{abcd}$, of this
logarithmically divergent integral for each diagram in figure
\nonplanardiags\ with the indices fixed as in the figure, computed
in dimensional regularization with $d=3-\epsilon$ :
\eqn\mslistedone{\eqalign{
M^{(3a)}_{abcd}&=\frac{1}{8\pi^2}\left[\frac{1}{d(d+2)}\left[(5+2d)g_{ab}g_{cd}+(5d^-19)g_{ac}g_{bd}+(17-4d^2)g_{ad}g_{bc}\right]+(d-3)g_{ab}g_{cd}\right]\cr
M^{(3b)}_{abcd}&=\frac{1}{8\pi^2}\left(\frac{1}{d(d+2)}\right)\left[(3-d-d^2)g_{ab}g_{cd}+(2d^2-7)g_{ac}g_{bd}+(5-d^2)g_{ad}g_{bc}\right]\cr
M^{(3c)}_{abcd}&=-\frac{3}{8\pi^2}\left(\frac{1}{d(d+2)}\right)\left[g_{ab}g_{cd}+g_{ac}g_{bd}+(d^2-3)g_{ad}g_{bc}\right]\cr
}}
\eqn\mslistedonen{\eqalign{
M^{(3d)}_{abcd}&=\frac{1}{4\pi^2}\left[\frac{1}{d(d+2)}\left[(d^2-2)g_{ab}g_{cd}+(d^2-2)g_{ad}g_{bc}+(4d+10)g_{ac}g_{bd}\right]+2(d-3)g_{ac}g_{bd}\right]\cr
M^{(3e)}_{abcd}&=\frac{1}{8\pi^2}\left(\frac{1}{d(d+2)}\right)\left[(d^2-2)g_{ab}g_{cd}+(d^2-2)g_{ad}g_{bc}+2(3-d-d^2)g_{ac}g_{bd}\right]\cr
M^{(3f1)}_{abcd}&=-\frac{3}{8\pi^2}\left(\frac{1}{d(d+2)}\right)\left[g_{ab}g_{cd}+g_{ac}g_{bd}+(d^2-3)g_{ad}g_{bc}\right]\cr
M^{(3f2)}_{abcd}&=-\frac{3}{8\pi^2}\left(\frac{1}{d(d+2)}\right)\left[g_{ac}g_{bd}+g_{ad}g_{bc}+(d^2-3)g_{ab}g_{cd}\right]\cr
}}
\eqn\mslistedtwo{\eqalign{
M^{(3g)}_{abcd}&=-\frac{3}{4\pi^2}\left(\frac{1}{d(d+2)}\right)\left[(d^2-3)g_{ac}g_{bd}+g_{ab}g_{cd}+g_{ad}g_{bc}\right]\cr
M^{(3h1)}_{abcd}&=\frac{1}{8\pi^2}\left(\frac{1}{d(d+2)}\right)\left[(d+1)g_{ad}g_{bc}-g_{ac}g_{bd}-g_{ab}g_{cd}\right]\cr
M^{(3h2)}_{abcd}&=\frac{1}{8\pi^2}\left(\frac{1}{d(d+2)}\right)\left[(d+1)g_{ac}g_{bd}-g_{ab}g_{cd}-g_{ad}g_{bc}\right]\cr
M^{(3i)}_{abcd}&=\frac{5}{4\pi^2}\left(\frac{d-1}{d(d+2)}\right)\left[g_{ab}g_{cd}+g_{ac}g_{bd}+g_{ad}g_{bc}\right]\cr
M^{(3j)}_{abcd}&=-\frac{3}{8\pi^2}\left(\frac{1}{d(d+2)}\right)\left[g_{ac}g_{bd}+g_{ad}g_{bc}+(d^2-3)g_{ab}g_{cd}\right]\cr
}}
\eqn\mslistedmore{\eqalign{
M^{(3k)}_{abcd}&=\frac{1}{8\pi^2}\left(\frac{1}{d(d+2)}\right)\left[(d+1)g_{ab}g_{cd}-g_{ac}g_{bd}-g_{ad}g_{bc}\right]\cr
M^{(3l)}_{abcd}&=\frac{5}{4\pi^2}\left(\frac{d-1}{d(d+2)}\right)\left[g_{ab}g_{cd}+g_{ac}g_{bd}+g_{ad}g_{bc}\right]\cr
M^{(3m1)}_{abcd}&=\frac{1}{2\pi^2}\left(\frac{1}{d(d+2)}\right)\left[(d+1)g_{ac}g_{bd}-g_{ab}g_{cd}-g_{ad}g_{bc}\right]\cr
M^{(3m2)}_{abcd}&=\frac{1}{2\pi^2}\left(\frac{1}{d(d+2)}\right)\left[(d+1)g_{ab}g_{cd}-g_{ac}g_{bd}-g_{ad}g_{bc}\right]\cr
}}
\eqn\mslistedmoren{\eqalign{
M^{(3n1)}_{abcd}&=\frac{1}{\pi^2}\left(\frac{1}{d(d+2)}\right)\left[g_{ad}g_{cb}+g_{ac}g_{bd}-(d+1)g_{ab}g_{cd}\right]\cr
M^{(3n2)}_{abcd}&=\frac{1}{\pi^2}\left(\frac{1}{d(d+2)}\right)\left[g_{ac}g_{bd}+g_{ab}g_{cd}-(d+1)g_{ad}g_{bc}\right]\cr
M^{(3n3)}_{abcd}&=\frac{1}{\pi^2}\left(\frac{1}{d(d+2)}\right)\left[g_{ab}g_{cd}+g_{ac}g_{bd}-(d+1)g_{ad}g_{bc}\right]\cr
M^{(3n4)}_{abcd}&=\frac{1}{4\pi^2}\left(\frac{d-1}{d}\right)g_{ab}g_{cd}\cr
M^{(3n5)}_{abcd}&=\frac{1}{2\pi^2}\left(\frac{d-1}{d}\right)g_{ac}g_{bd}\cr
M^{(3o1)}_{abcd}&=M^{(3o2)}_{abcd}=M^{(3o3)}_{abcd}=M^{(3o4)}_{abcd}=
M^{(3o5)}_{abcd}=M^{(3o6)}_{abcd}=0. }}

We now sum over the appropriate permutations of indices to obtain
the coefficient, $R^{(*)}_{ijkl}$, of the logarithmic divergence
\logdivv\ due to diagrams of each type.  We list these
coefficients, as well as the contribution to the
$\tr(A_iA_j)\tr(A_kA_l)$ counterterm $R_{CT}^{(*)}$ :
\eqn\Ra{\eqalign{R^{(3a)}_{ijkl}&=\frac{1}{4\pi^2d(d+2)}
\left[(10+4d)g_{ij}g_{kl}+(d^2-2)g_{ik}g_{jl}+(d^2-2)g_{il}g_{jk}\right]+
\frac{d-3}{2\pi^2}g_{ij}g_{kl}\cr
&=\left[\frac{11}{30\pi^2}-\frac{167\epsilon}{450\pi^2}+
{\cal{O}}(\epsilon^2)\right]g_{ij}g_{kl}+
\left[\frac{7}{60\pi^2}-\frac{17\epsilon}{450\pi^2}+
{\cal{O}}(\epsilon^2)\right]\left[g_{ik}g_{jl}+g_{il}g_{jk}\right]\cr
R^{(3a)}_{CT}&=\left[\frac{11}{240\pi^2}\ln\left(\frac{{\cal{A}}_2M}{\mu}\right)+
\frac{167}{3600\pi^2}\right]\tr(A_i A_i) \tr(A_j A_j)\cr &\qquad
+2\left[\frac{7}{480\pi^2}\ln\left(\frac{{\cal{A}}_2M}{\mu}\right)+
\frac{17}{3600\pi^2}\right]\tr(A_iA_j)\tr(A_iA_j) }}
\eqn\Rb{\eqalign{R^{(3b)}_{ijkl}&=\frac{1}{2\pi^2d(d+2)}
\left[2(3-d-d^2)g_{ij}g_{kl}+(d^2-2)g_{ik}g_{jl}+(d^2-2)g_{il}g_{jk}\right]\cr
&=\left[-\frac{3}{5\pi^2}+\frac{11\epsilon}{75\pi^2}+{\cal{O}}(\epsilon^2)\right]g_{ij}g_{kl}
+\left[\frac{7}{30\pi^2}-\frac{17\epsilon}{225\pi^2}+
{\cal{O}}(\epsilon^2)\right]\left[g_{ik}g_{jl}+g_{il}g_{jk}\right]\cr
R^{(3b)}_{CT}&=\left[-\frac{3}{40\pi^2}\ln\left(\frac{{\cal{A}}_2M}{\mu}\right)-
\frac{11}{600\pi^2}\right]\tr(A_i A_i) \tr(A_j A_j)\cr &\qquad
+2\left[\frac{7}{240\pi^2}\ln\left(\frac{{\cal{A}}_2M}{\mu}\right)+
\frac{17}{1800\pi^2}\right]\tr(A_iA_j)\tr(A_iA_j) }}
\eqn\Rc{\eqalign{R^{(3c)}_{ijkl}&=-\frac{3}{4\pi^2 d(d+2)}
\left[2g_{ij}g_{kl}+(d^2-2)\left(g_{ik}g_{jl}+g_{il}g_{jk}\right)\right]\cr
&=\left[-\frac{1}{10\pi^2}-\frac{4\epsilon}{75\pi^2}+
{\cal{O}}(\epsilon^2)\right]g_{ij}g_{kl}+\left[-\frac{7}{20\pi^2}+
\frac{17\epsilon}{150\pi^2}+{\cal{O}}(\epsilon^2)\right]
\left[g_{ik}g_{jl}+g_{il}g_{jk}\right]\cr
R^{(3c)}_{CT}&=\left[-\frac{1}{80\pi^2}\ln\left(\frac{{\cal{A}}_2M}{\mu}\right)+
\frac{1}{150\pi^2}\right]\tr(A_i A_i) \tr(A_j A_j)\cr &\qquad
+2\left[-\frac{7}{160\pi^2}\ln\left(\frac{{\cal{A}}_2M}{\mu}\right)-
\frac{17}{1200\pi^2}\right]\tr(A_iA_j)\tr(A_iA_j) }}
\eqn\Rd{\eqalign{R^{(3d)}_{ijkl}&=\frac{1}{4\pi^2d(d+2)}
\left[2(d^2-2)g_{ij}g_{kl}+(d^2+4d+8)g_{ik}g_{jl}+(d^2+4d+8)g_{il}g_{jk}\right]\cr
&\qquad\qquad
+\frac{d-3}{2\pi^2}\left[g_{ik}g_{jl}+g_{il}g_{jk}\right]\cr
&=\left[\frac{7}{30\pi^2}-\frac{17\epsilon}{225\pi^2}+
{\cal{O}}(\epsilon^2)\right]g_{ij}g_{kl}+\left[\frac{29}{60\pi^2}-
\frac{92\epsilon}{225\pi^2}+
{\cal{O}}(\epsilon^2)\right]\left[g_{ik}g_{jl}+g_{il}g_{jk}\right]\cr
R^{(3d)}_{CT}&=\left[\frac{7}{240\pi^2}\ln\left(\frac{{\cal{A}}_2M}{\mu}\right)+
\frac{17}{1800\pi^2}\right]\tr(A_i A_i) \tr(A_j A_j)\cr &\qquad
+2\left[\frac{29}{480\pi^2}\ln\left(\frac{{\cal{A}}_2M}{\mu}\right)+
\frac{92}{1800\pi^2}\right]\tr(A_iA_j)\tr(A_iA_j) }}
\eqn\Re{\eqalign{R^{(3e)}_{ijkl}&=\frac{1}{2\pi^2d(d+2)}
\left[2(d^2-2)g_{ij}g_{kl}+(4-2d-d^2)g_{ik}g_{jl}+(4-2d-d^2)g_{il}g_{jk}\right]\cr
&=\left[\frac{7}{15\pi^2}-\frac{34\epsilon}{225\pi^2}+
{\cal{O}}(\epsilon^2)\right]g_{ij}g_{kl}
+\left[-\frac{11}{30\pi^2}+\frac{16\epsilon}{225\pi^2}+
{\cal{O}}(\epsilon^2)\right]\left[g_{ik}g_{jl}+g_{il}g_{jk}\right]\cr
R^{(3e)}_{CT}&=\left[\frac{7}{120\pi^2}\ln\left(\frac{{\cal{A}}_2M}{\mu}\right)+
\frac{17}{900\pi^2}\right]\tr(A_i A_i) \tr(A_j A_j)\cr &\qquad
+2\left[-\frac{11}{240\pi^2}\ln\left(\frac{{\cal{A}}_2M}{\mu}\right)-
\frac{2}{225\pi^2}\right]\tr(A_iA_j)\tr(A_iA_j) }}
\eqn\Rfone{\eqalign{R^{(3f1)}_{ijkl}&=R^{(3c)}_{ijkl}\cr
R^{(3f1)}_{CT}&=R^{(3c)}_{CT}
}}
%
\eqn\Rftwo{\eqalign{R^{(3f2)}_{ijkl}&=-\frac{3}{2\pi^2d(d+2)}
\left[(d^2-3)g_{ij}g_{kl}+g_{ik}g_{jl}+g_{il}g_{jk}\right]\cr
&=\left[-\frac{3}{5\pi^2}+\frac{7\epsilon}{25\pi^2}+{\cal{O}}(\epsilon^2)\right]g_{ij}g_{kl}
+\left[-\frac{1}{10\pi^2}-\frac{4\epsilon}{75\pi^2}+
{\cal{O}}(\epsilon^2)\right]\left[g_{ik}g_{jl}+g_{il}g_{jk}\right]\cr
R^{(3f2)}_{CT}&=\left[-\frac{3}{40\pi^2}
\ln\left(\frac{{\cal{A}}_2M}{\mu}\right)-\frac{7}{200\pi^2}\right]\tr(A_i
A_i) \tr(A_j A_j)\cr &\qquad
+2\left[-\frac{1}{80\pi^2}\ln\left(\frac{{\cal{A}}_2M}{\mu}\right)+
\frac{1}{150\pi^2}\right]\tr(A_iA_j)\tr(A_iA_j) }}
\eqn\Rg{\eqalign{R^{(3g)}_{ijkl}&=-\frac{3}{\pi^2d(d+2)}
\left[2g_{ij}g_{kl}+(d^2-2)g_{ik}g_{jl}+(d^2-2)g_{il}g_{jk}\right]\cr
&=\left[-\frac{2}{5\pi^2}-\frac{16\epsilon}{75\pi^2}+{\cal{O}}(\epsilon^2)\right]g_{ij}g_{kl}
+\left[-\frac{7}{5\pi^2}+\frac{34\epsilon}{75\pi^2}+
{\cal{O}}(\epsilon^2)\right]\left[g_{ik}g_{jl}+g_{il}g_{jk}\right]\cr
R^{(3g)}_{CT}&=\left[-\frac{1}{20\pi^2}\ln\left(\frac{{\cal{A}}_3M}{\mu}\right)+
\frac{2}{75\pi^2}\right]\tr(A_i A_i) \tr(A_j A_j)\cr &\qquad
+2\left[-\frac{7}{40\pi^2}\ln\left(\frac{{\cal{A}}_3M}{\mu}\right)-
\frac{17}{300\pi^2}\right]\tr(A_iA_j)\tr(A_iA_j) }}
\eqn\Rhone{\eqalign{R^{(3h1)}_{ijkl}&=\frac{1}{2\pi^2 d(d+2)}
\left[-2g_{ij}g_{kl}+d\left(g_{ik}g_{jl}+g_{il}g_{jk}\right)\right]\cr
&=\left[-\frac{1}{15\pi^2}-\frac{8\epsilon}{225\pi^2}+
{\cal{O}}(\epsilon^2)\right]g_{ij}g_{kl}
+\left[\frac{1}{10\pi^2}+\frac{\epsilon}{50\pi^2}+
{\cal{O}}(\epsilon^2)\right]\left[g_{ik}g_{jl}+g_{il}g_{jk}\right]\cr
R^{(3h1)}_{CT}&=\left[-\frac{1}{120\pi^2}\ln\left(\frac{{\cal{A}}_3M}{\mu}\right)+
\frac{1}{225\pi^2}\right]\tr(A_i A_i) \tr(A_j A_j)\cr &\qquad
+2\left[\frac{1}{80\pi^2}\ln\left(\frac{{\cal{A}}_3M}{\mu}\right)-
\frac{1}{400\pi^2}\right]\tr(A_iA_j)\tr(A_iA_j) }}
\eqn\Rhtwo{\eqalign{R^{(3h2)}_{ijkl}&=R^{(3h1)}_{ijkl}\cr
R^{(3h2)}_{CT}&=R^{(3h1)}_{CT}
}}
\eqn\Ri{\eqalign{R^{(3i)}_{ijkl}&=\frac{5}{\pi^2}\left(\frac{d-1}{d(d+2)}\right)
\left[g_{ij}g_{kl}+g_{ik}g_{jl}+g_{il}g_{jk}\right]\cr
&=\left[\frac{2}{3\pi^2}+\frac{\epsilon}{45\pi^2}+{\cal{O}}(\epsilon^2)\right]
\left[g_{ij}g_{kl}+g_{ik}g_{jl}+g_{il}g_{jk}\right]\cr
R^{(3i)}_{CT}&=\left[\frac{1}{12\pi^2}\ln\left(\frac{{\cal{A}}_4M}{\mu}\right)-
\frac{1}{360\pi^2}\right]\left[\tr(A_i A_i) \tr(A_j
A_j)+2\tr(A_iA_j)\tr(A_iA_j)\right] }}
\eqn\Rj{\eqalign{R^{(3j)}_{ijkl}&=-\frac{3}{\pi^2d(d+2)}
\left[(d^2-3)g_{ij}g_{kl}+g_{ik}g_{jl}+g_{il}g_{jk}\right]\cr
&=\left[-\frac{6}{5\pi^2}+\frac{14\epsilon}{25\pi^2}+
{\cal{O}}(\epsilon^2)\right]g_{ij}g_{kl}
+\left[-\frac{1}{5\pi^2}-\frac{8\epsilon}{75\pi^2}+
{\cal{O}}(\epsilon^2)\right]\left[g_{ik}g_{jl}+g_{il}g_{jk}\right]\cr
R^{(3j)}_{CT}&=\left[-\frac{3}{20\pi^2}\ln\left(\frac{{\cal{A}}_3M}{\mu}\right)-
\frac{7}{100\pi^2}\right]\tr(A_i A_i) \tr(A_j A_j)\cr &\qquad
+2\left[-\frac{1}{40\pi^2}\ln\left(\frac{{\cal{A}}_3M}{\mu}\right)+
\frac{1}{75\pi^2}\right]\tr(A_iA_j)\tr(A_iA_j) }}
\eqn\Rk{\eqalign{R^{(3k)}_{ijkl}&=\frac{1}{\pi^2
d(d+2)}\left[(d+1)g_{ij}g_{kl}-g_{ik}g_{jl}-g_{il}g_{jk}\right]\cr
&=\left[\frac{4}{15\pi^2}+\frac{17\epsilon}{225\pi^2}+{\cal{O}}(\epsilon^2)\right]g_{ij}g_{kl}
+\left[-\frac{1}{15\pi^2}-\frac{8\epsilon}{225\pi^2}+{\cal{O}}(\epsilon^2)\right]\left[g_{ik}g_{jl}+g_{il}g_{jk}\right]\cr
R^{(3k)}_{CT}&=\left[\frac{1}{30\pi^2}\ln\left(\frac{{\cal{A}}_3M}{\mu}\right)-\frac{17}{1800\pi^2}\right]\tr(A_i
A_i) \tr(A_j A_j)\cr &\qquad
+2\left[-\frac{1}{120\pi^2}\ln\left(\frac{{\cal{A}}_3M}{\mu}\right)+\frac{1}{225\pi^2}\right]\tr(A_iA_j)\tr(A_iA_j)
}}
\eqn\Rl{\eqalign{R^{(3l)}_{ijkl}&=\frac{5}{2\pi^2}\left(\frac{d-1}{d(d+2)}\right)\left[g_{ij}g_{kl}+g_{ik}g_{jl}+g_{il}g_{jk}\right]\cr
&=\left[\frac{1}{3\pi^2}+\frac{\epsilon}{90\pi^2}+{\cal{O}}(\epsilon^2)\right]\left[g_{ij}g_{kl}+g_{ik}g_{jl}+g_{il}g_{jk}\right]\cr
R^{(3l)}_{CT}&=\left[\frac{1}{24\pi^2}\ln\left(\frac{{\cal{A}}_4M}{\mu}\right)-\frac{1}{720\pi^2}\right]\left[\tr(A_i
A_i) \tr(A_j A_j)+2\tr(A_iA_j)\tr(A_iA_j)\right] }}
\eqn\Rmone{\eqalign{R^{(3m1)}_{ijkl}&=\frac{4}{\pi^2
d(d+2)}\left[-2g_{ij}g_{kl}+d\left(g_{ik}g_{jl}+g_{il}g_{jk}\right)\right]\cr
&=\left[-\frac{8}{15\pi^2}-\frac{64\epsilon}{225\pi^2}+{\cal{O}}(\epsilon^2)\right]g_{ij}g_{kl}
+\left[\frac{4}{5\pi^2}+\frac{4\epsilon}{25\pi^2}+{\cal{O}}(\epsilon^2)\right]\left[g_{ik}g_{jl}+g_{il}g_{jk}\right]\cr
R^{(3m1)}_{CT}&=\left[-\frac{1}{15\pi^2}\ln\left(\frac{{\cal{A}}_2M}{\mu}\right)+\frac{8}{225\pi^2}\right]\tr(A_i
A_i) \tr(A_j A_j)\cr &\qquad
+2\left[\frac{1}{10\pi^2}\ln\left(\frac{{\cal{A}}_2M}{\mu}\right)-\frac{1}{50\pi^2}\right]\tr(A_iA_j)\tr(A_iA_j)
}}
%
%
\eqn\Rmtwo{\eqalign{R^{(3m2)}_{ijkl}&=\frac{4}{\pi^2
d(d+2)}\left[(d+1)g_{ij}g_{kl}-g_{ik}g_{jl}-g_{il}g_{jk}\right]\cr
&=\left[\frac{16}{15\pi^2}+\frac{68\epsilon}{225\pi^2}+{\cal{O}}(\epsilon^2)\right]g_{ij}g_{kl}
+\left[-\frac{4}{15\pi^2}-\frac{32\epsilon}{225\pi^2}+{\cal{O}}(\epsilon^2)\right]\left[g_{ik}g_{jl}+g_{il}g_{jk}\right]\cr
R^{(3m2)}_{CT}&=\left[\frac{2}{15\pi^2}\ln\left(\frac{{\cal{A}}_2M}{\mu}\right)-\frac{17}{450\pi^2}\right]\tr(A_i
A_i) \tr(A_j A_j)\cr &\qquad
+2\left[-\frac{1}{30\pi^2}\ln\left(\frac{{\cal{A}}_2M}{\mu}\right)+\frac{4}{225\pi^2}\right]\tr(A_iA_j)\tr(A_iA_j)
}}
\eqn\Rnone{\eqalign{R^{(3n1)}_{ijkl}&=\frac{8}{\pi^2
d(d+2)}\left[-(d+1)g_{ij}g_{kl}+\left(g_{ik}g_{jl}+g_{il}g_{jk}\right)\right]\cr
&=\left[\frac{-32}{15\pi^2}-\frac{136\epsilon}{225\pi^2}+{\cal{O}}(\epsilon^2)\right]g_{ij}g_{kl}
+\left[\frac{8}{15\pi^2}+\frac{64\epsilon}{225\pi^2}+{\cal{O}}(\epsilon^2)\right]\left[g_{ik}g_{jl}+g_{il}g_{jk}\right]\cr
R^{(3n1)}_{CT}&=\left[-\frac{4}{15\pi^2}\ln\left(\frac{{\cal{A}}_1M}{\mu}\right)+\frac{17}{225\pi^2}\right]\tr(A_i
A_i) \tr(A_j A_j)\cr &\qquad
+2\left[\frac{1}{15\pi^2}\ln\left(\frac{{\cal{A}}_1M}{\mu}\right)-\frac{8}{225\pi^2}\right]\tr(A_iA_j)\tr(A_iA_j)
}}
\eqn\Rntwo{\eqalign{R^{(3n2)}_{ijkl}&=\frac{4}{\pi^2
d(d+2)}\left[2g_{ij}g_{kl}-d\left(g_{ik}g_{jl}+g_{il}g_{jk}\right)\right]\cr
&=\left[\frac{8}{15\pi^2}+\frac{64\epsilon}{225\pi^2}+{\cal{O}}(\epsilon^2)\right]g_{ij}g_{kl}
+\left[-\frac{4}{5\pi^2}-\frac{4\epsilon}{25\pi^2}+{\cal{O}}(\epsilon^2)\right]\left[g_{ik}g_{jl}+g_{il}g_{jk}\right]\cr
R^{(3n2)}_{CT}&=\left[\frac{1}{15\pi^2}\ln\left(\frac{{\cal{A}}_1M}{\mu}\right)-\frac{8}{225\pi^2}\right]\tr(A_i
A_i) \tr(A_j A_j)\cr &\qquad
+2\left[-\frac{1}{10\pi^2}\ln\left(\frac{{\cal{A}}_1M}{\mu}\right)+\frac{1}{50\pi^2}\right]\tr(A_iA_j)\tr(A_iA_j)
}}
\eqn\Rnthree{\eqalign{R^{(3n3)}_{ijkl}&=R^{(3n2)}_{ijkl}\cr
R^{(3n3)}_{CT}&=R^{(3n2)}_{CT}
}}
\eqn\Rnfour{\eqalign{R^{(3n4)}_{ijkl}&=\frac{2}{\pi^2}\left(\frac{d-1}{d}\right)g_{ij}g_{kl}\cr
&=\left[\frac{4}{3\pi^2}-\frac{2\epsilon}{9\pi^2}+
{\cal{O}}(\epsilon^2)\right]g_{ij}g_{kl}\cr
R^{(3n4)}_{CT}&=\left[\frac{1}{6\pi^2}\ln\left(\frac{{\cal{A}}_1M}{\mu}\right)+\frac{1}{36\pi^2}\right]\tr(A_i
A_i) \tr(A_j A_j) }}
\eqn\Rnfive{\eqalign{R^{(3n5)}_{ijkl}&=\frac{2(d-1)}{\pi^2d}\left[g_{ik}g_{jl}+g_{il}g_{jk}\right]\cr
&=\left[\frac{4}{3\pi^2}-\frac{2\epsilon}{9\pi^2}+{\cal{O}}(\epsilon^2)\right]\left[g_{ik}g_{jl}+g_{il}g_{jk}\right]\cr
R^{(3n5)}_{CT}&=2\left[\frac{1}{6\pi^2}\ln\left(\frac{{\cal{A}}_1M}{\mu}\right)+\frac{1}{36\pi^2}\right]\tr(A_iA_j)\tr(A_iA_j)
}}

Summing these yields the following result for the order
$\lambda^2/N^2$ double-trace counterterm which is needed in equation
\countertermres\ :
\eqn\logdivct{{\cal{L}}_{CT}=\left(\frac{\lambda^2}{N^2}\right)
\left({1\over {120 \pi^2}}
\ln\left(\frac{{\cal{A}}_1^4{\cal{A}}_2^3{\cal{A}}_4^{15}}
{{\cal{A}}_3^{22}}\right)
+\frac{1}{60\pi^2}\right)\left[\tr(A_i A_i) \tr(A_j
A_j)+2\tr(A_iA_j)\tr(A_iA_j)\right].}

\appendix{B}{Useful Spherical Harmonic Identities}

In this appendix, we collect various properties of the $S^3$ spherical
harmonics required to study field theory on $S^3$. Many of the basic
results were derived in \Cutkosky.

\subsec{Basic properties of spherical harmonics}

Scalar functions on the sphere may be expanded in a complete set of
spherical harmonics $S_j^{m \; m'}$ transforming in the $(j/2,j/2)$
representation of $SU(2) \times SU(2) \equiv SO(4)$, where $j$ is any
nonnegative integer, and $-j/2 \le m,m' \le j/2$. It is convenient to
denote the full set of indices $(j,m,m')$ by $\alpha$.  These obey an
orthonormality condition (we take the radius of the $S^3$ to be one)
\eqn\sortho{
\int_{S^3} S^\alpha S^\beta = \delta^{\alpha \bar{\beta}},
}
where $S^{\bar{\alpha}}$ denotes the complex conjugate of $S^\alpha$,
\eqn\scomcon{
(S_j^{m \; m'})^* = (-1)^{m+ m'} S_j^{-m \; -m'}.
}
The spherical harmonics are eigenfunctions of the Laplace operator on
the sphere,
\eqn\lapeq{
\nabla^2 S^\alpha = -j_\alpha (j_\alpha + 2) S^\alpha,
}
and under a parity operation transform with eigenvalue $(-1)^{j_\alpha}$.

A general vector field on the sphere may be expanded as a combination
of gradients of the scalar spherical harmonics plus a set of vector
spherical harmonics $\vec{V}_{j \pm}^{m \; m'}$.  These transform in
the $({j \pm 1 \over 2}, {j \mp 1 \over 2})$ representation of
$SO(4)$, where $j$ is a positive integer. Again, it is convenient to
denote the full set of indices $(j,m,m',\epsilon)$ by a single index
$\alpha$. These obey orthonormality relations
\eqn\vorth{\eqalign{
\int_{S^3} V^\alpha \cdot V^\beta &= \delta^{\alpha \bar{\beta}}, \cr
\int_{S^3} V^\alpha \cdot \nabla S^\beta &= 0 \; .
}}
Again $V^{\bar{\alpha}}$ indicates the complex conjugate of
$V^\alpha$, given by
\eqn\vcomcon{
({\bf V}_{j \pm}^{m \; m'})^* = (-1)^{m+ m'+1} {\bf V}_{j \pm}^{-m \; -m'} \; .
}
The vector spherical harmonics are eigenfunctions of parity with
eigenvalue $(-1)^{j+1}$, and satisfy
\eqn\lapcurl{\eqalign{
\nabla^2 V^\alpha &= -(j_\alpha + 1)^2 V^\alpha, \cr
\nabla \times V^\alpha &= -\epsilon_\alpha (j_\alpha + 1) V^\alpha, \cr
\nabla \cdot V^\alpha &= 0.}}

Explicit expressions for the scalar and vector spherical harmonics may
be found in \Cutkosky.

\subsec{Spherical harmonic integrals}

In expanding the action in modes, we require the integrals over the
sphere of products of various numbers of spherical harmonics. For two
spherical harmonics, the integrals are given by the orthonormality
relations. For products of three spherical harmonics, we require the
set of integrals given in \sphconv. These were calculated in
\Cutkosky, and the results may be expressed in terms of the functions
\rtwoo\ and \Rcombo\ as\foot{The expression for $C$
below differs by a factor of two from the expression in \Cutkosky,
but we believe that this expression is correct.}
\eqn\threeints{\eqalign{
C^{\alpha \beta \gamma} & = \left(\matrix{{j_\alpha \over 2} &{j_\beta
+ \epsilon_\beta \over 2} & {j_\gamma \over 2} \cr
m_\alpha&m_\beta&m_\gamma }\right) \left(\matrix{{j_\alpha \over 2}
&{j_\beta - \epsilon_\beta \over 2} & {j_\gamma \over 2} \cr
m'_\alpha&m'_\beta&m'_\gamma}\right) R_2(j_\alpha, j_\beta, j_\gamma),
\cr D^{\alpha \beta \gamma} & = \left(\matrix{{j_\alpha +
\epsilon_\alpha \over 2} & {j_\beta + \epsilon_\beta \over 2} &
{j_\gamma \over 2} & \cr m_\alpha&m_\beta&m_\gamma }\right)
\left(\matrix{{j_\alpha - \epsilon_\alpha \over 2} & {j_\beta -
\epsilon_\beta \over 2} & {j_\gamma \over 2} & \cr
m'_\alpha&m'_\beta&m'_\gamma}\right) R_{3 \epsilon_\alpha
\epsilon_\beta}(j_\alpha, j_\beta, j_\gamma), \cr E^{\alpha \beta
\gamma} & = \left(\matrix{{j_\alpha + \epsilon_\alpha \over 2} &
{j_\beta + \epsilon_\beta \over 2} & {j_\gamma + \epsilon_\gamma \over
2} & \cr m_\alpha&m_\beta&m_\gamma}\right) \left(\matrix{{j_\alpha -
\epsilon_\alpha \over 2} & {j_\beta - \epsilon_\beta \over 2} &
{j_\gamma - \epsilon_\gamma \over 2} & \cr
m'_\alpha&m'_\beta&m'_\gamma }\right) R_{4 \epsilon_\alpha
\epsilon_\beta \epsilon_\gamma}(j_\alpha, j_\beta, j_\gamma). \cr }}
To evaluate integrals appearing in quartic terms, we do not require
any additional information, since any product of two spherical
harmonics may be expressed as a sum of single spherical harmonics
using the completeness property and the integrals above. For example,
we find
\eqn\vdotprod{
V^\alpha \cdot V^\beta = D^{\alpha \beta \bar{\gamma}} S^\gamma.
}

\subsec{Identities involving 3j symbols}

The expressions for two and three loop vacuum diagrams involve
products of the integrals in the previous subsection, with indices
contracted and summed over in various ways. Since the $m$ and $m'$
indices appear only in 3j-symbols, we can always evaluate the sums
over these using standard 3j-symbol identities.

For basic manipulations, we require the identities
\eqn\threeiidentbb{\left(\matrix{j_1&j_2&j_3\cr
m_1&m_2&m_3}\right)=\left(\matrix{j_3&j_1&j_2\cr
m_3&m_1&m_2}\right)}
\eqn\threeiidenaa{\left(\matrix{j_1&j_2&j_3\cr
m_1&m_2&m_3}\right)=(-1)^{(j_1+j_2+j_3)}\left(\matrix{j_1&j_2&j_3\cr
-m_1&-m_2&-m_3}\right)}
\eqn\threeiidentcc{\left(\matrix{j_1&j_2&j_3\cr
m_1&m_2&m_3}\right)=(-1)^{(j_1+j_2+j_3)}\left(\matrix{j_2&j_1&j_3\cr
m_2&m_1&m_3}\right)}
%
To evaluate two-loop sums, we require
\eqn\threeiident{\sum_{m_1}\sum_{m_2}\left(\matrix{j_1&j_2&j_3\cr
m_1&m_2&m_3}\right)\left(\matrix{j_1&j_2&j_4\cr
m_1&m_2&m_4}\right)=\frac{\delta_{j_3 j_4} \delta_{m_3
m_4}}{2j_3+1}}
Finally, in three-loop computations we require
\eqn\theejone{\eqalign{&\sum_{m's}(-1)^{m_1+m_2+m_3+m_4+m_5+m_6}\cr &
\left(\matrix{j_1&j_1&j_3\cr m_1&-m_1&m_3}\right)
\left(\matrix{j_3&j_4&j_5\cr -m_3&m_4&m_5}\right)
\left(\matrix{j_5&j_4&j_6\cr -m_5&-m_4&m_6}\right)
\left(\matrix{j_6&j_2&j_2\cr -m_6&m_2&-m_2}\right)\cr
&=(-1)^{j_1+j_2}\sqrt{(2j_1+1)(2j_2+1)}\delta_{j_3,0}\delta_{j_6,0}
\delta_{j_4,j_5}}}
\eqn\mercidentone{\eqalign{&\sum_{m's}(-1)^{m_1+m_2+m_3+m_4+m_5+m_6}\cr
& \left(\matrix{j_1&j_1&j_2\cr m_1&-m_2&m_2}\right)
\left(\matrix{j_2&j_3&j_4\cr -m_2&m_3&m_4}\right)
\left(\matrix{j_3&j_5&j_6\cr -m_3&m_5&m_6}\right)
\left(\matrix{j_4&j_6&j_5\cr -m_4&-m_6&-m_5}\right)\cr &=(-1)^{j_1
+ j_3} \sqrt{2j_1 +1 \over 2 j_3 +1} \delta_{j_3, j_4}
\delta_{j_2,0} \delta(j_4,j_5,j_6)}}
\eqn\mercidenttwo{\eqalign{&\sum_{m's}(-1)^{m_1+m_2+m_3+m_4+m_5+m_6}\cr
& \left(\matrix{j_1&j_2&j_3 \cr m_1&m_2&m_3}\right)
\left(\matrix{j_3&j_2&j_4 \cr -m_3&-m_2&m_4}\right)
\left(\matrix{j_4&j_5&j_6 \cr -m_4&m_5&m_6}\right)
\left(\matrix{j_6&j_5&j_1 \cr -m_6&-m_5&-m_1}\right)\cr &={1 \over
2 j_1 + 1} \delta_{j_1,j_4} \delta(j_1,j_2,j_3)
\delta(j_1,j_5,j_6)}}
\eqn\mercidentthree{\eqalign{&\sum_{m's}(-1)^{m_1+m_2+m_3+m_4+m_5+m_6}\cr
& \left(\matrix{j_1&j_2&j_3\cr m_1&m_2&-m_3}\right)
\left(\matrix{j_3&j_4&j_5\cr m_3&m_4&-m_5}\right)
\left(\matrix{j_5&j_6&j_1\cr m_5&m_6&-m_1}\right)
\left(\matrix{j_2&j_6&j_4\cr -m_2&-m_6&-m_4}\right)\cr
&=(-1)^{j_1+j_2+j_3+j_4+j_5+j_6}\{\matrix{j_1&j_2&j_3\cr
j_4&j_5&j_6}\}}}
In equations \mercidentone\ and \mercidenttwo, the delta function
with three arguments is either 1 or 0, depending on whether or not
the triangle relation is satisfied. In equation \mercidentthree,
$\{\matrix{j_1&j_2&j_3\cr j_4&j_5&j_6}\}$ is the 6j symbol.

\subsec{Identities for sums of spherical harmonics}

Using the 3j-identities, it is straightforward to derive expressions
for sums over $m$,$m'$, and $\epsilon$ in various products of the
spherical harmonic integrals. For the two loop diagrams, we require:
\eqn\sphids{\eqalign{ \sum_{m's}D^{\alpha \bar{\alpha} \gamma}
D^{\beta \bar{\beta} \bar{\gamma}} &= {1 \over 2 \pi^2} j_\alpha
(j_\alpha + 2) j_\beta (j_\beta +2) \delta_{\gamma,0}, \cr
\sum_{m's, \epsilon's, j_\gamma} D^{\alpha \bar{\alpha} \gamma}
D^{\beta \bar{\beta} \bar{\gamma}} &= {2 \over \pi^2} j_\alpha
(j_\alpha + 2) j_\beta (j_\beta +2), \cr \sum_{m's, \epsilon's}
D^{\alpha \beta  \gamma} D^{\bar{\alpha} \bar{\beta} \bar{\gamma}}
&= 2 R^2_{3+}(j_\alpha,j_\gamma,j_\beta) + 2 R^2_{3-}(j_\alpha,
j_\gamma,j_\beta), \cr
 \sum_{m's, \epsilon's,  j_\gamma} D^{\alpha \beta \gamma}
D^{\bar{\alpha} \bar{\beta} \bar{\gamma}} &= {2 \over 3\pi^2}
j_\alpha (j_\alpha + 2) j_\beta (j_\beta +2), \cr \sum_{m's}
E^{\alpha \beta \gamma} E^{\bar{\alpha} \bar{\beta} \bar{\gamma}}
&= R^2_{4 \epsilon_\alpha \epsilon_\beta
\epsilon_\gamma}(j_\alpha, j_\beta, j_\gamma). }}

For the three-loop diagrams it is useful first to note the basic
relations (related to those above):
\eqn\Dsum{\eqalign{ \sum_{m's, \epsilon} D^{\a \bal \l} &=
{\sqrt{2} \over \pi} \delta_{\lambda,0} j_\a (j_\a +2), \cr
\sum_{m's, \epsilon's} D^{\a \b \g} D^{\bal \bb \t} &= 2
\delta_{\g \bt} {1 \over (j_\g + 1)^2} (R_{3+}^2(j_\alpha, j_\g,
j_\b) + R_{3-}^2(j_\a, j_\g, j_\b)), \cr \sum_{m's, \epsilon's}
D^{\a \g \t} D^{\b \bg \bt} &=  \delta_{\a \bb} {1 \over j_\a
(j_\a +2)} (R_{3+}^2(j_\alpha, j_\t, j_\g) + R_{3-}^2(j_\a, j_\t,
j_\g)), \cr \sum_{m's, \epsilon's} C^{\a \d \g} C^{\bg \bd \b} &=
-2 \delta_{\a \bb} {1 \over (j_\a + 1)^2} R_2^2(j_\a, j_\d, j_\g),
\cr \sum_{m's} E^{\a \g \d} E^{\bd \bg \b} &=  -\delta_{\a \bb} {1
\over j_\a (j_\a +2)} R_{4 \e_\d \e_\g \e_\a}^2 (j_\d, j_\g,
j_\a), \cr \sum_{m's} D^{\g \t \a} E^{\bt \bg \b} &=  0, \cr
\sum_{m's} C^{\a \g \t} D^{\bg \b \bt} &=  0. \cr }}
In each of these, we are summing over $m$, $m'$, and $\epsilon$
for each of the contracted indices only.

Using these basic relations and the results of the previous
subsection, we find the following results for the non-vanishing
spherical harmonic sums that appear in the three loop calculations:
\eqn\fourDsum{\eqalign{ \sum_{m's, \epsilon's,j_\l, j_\t} D^{\a
\bal \l} D^{\g \d \bl} D^{\bg \bd \bt} D^{\b \bb \t} &=
\delta_{j_\gamma, j_\delta} {2 \over \pi^4} j_\alpha (j_\a +2)
j_\b (j_\b +2) j_\g (j_\g +2), \cr \sum_{m's, \epsilon's, j_\l}
D^{\a \bal \l} D^{\g \d \bl} D^{\b \bg \t} D^{\bb \bd \bt} &=
\delta_{j_\gamma, j_\delta} {2 \over \pi^2} j_\alpha (j_\a +2)
(R_{3+}^2(j_\gamma, j_\t, j_\b) +R_{3-}^2(j_\gamma, j_\t,
j_\b)),\cr \sum_{m's, \epsilon's} D^{\a \g \l} D^{\bal \d \bl}
D^{\b \bd \t} D^{\bb \bg \bt} &= \delta_{j_\gamma, j_\delta}  {2
\over j_\g (j_\g + 2)} (R_{3+}^2 (j_\g, j_\l, j_\a) +
R_{3-}^2(j_\g, j_\l, j_\a)) \cr &\qquad \qquad \qquad \cdot
(R_{3+}^2 (j_\g, j_\t, j_\b) + R_{3-}^2(j_\g, j_\t, j_\b)), \cr
\sum_{m's, \epsilon's} D^{\a \g \l} D^{\bal \bg \t} D^{\d \b \bl}
D^{\bd \bb \bt} &= \delta_{j_\l, j_\t}  {4 \over (j_\l+1)^2}
(R_{3+}^2 (j_\a, j_\l, j_\g) + R_{3-}^2(j_\a, j_\l, j_\g))  \cr
&\qquad \qquad \qquad \cdot (R_{3+}^2 (j_\b, j_\l, j_\d) +
R_{3-}^2(j_\b, j_\l, j_\d)), \cr \sum_{m's} D^{\a \b \l} D^{\g \d
\bl} D^{\bb \bd \t} D^{\bal \bg \bt} &= (-1)^{\sum j's}
\left\{\matrix{{j_\g + \e_\g \over 2} & {j_\l \over 2} & {j_\d +
\e_\d \over 2} \cr {j_\b + \e_\b \over 2} & {j_\t \over 2} & {j_\a
+ \e_\a \over 2}}\right\} \left\{\matrix{{j_\g - \e_\g \over 2} &
{j_\l \over 2} & {j_\d - \e_\d \over 2} \cr {j_\b - \e_\b \over 2}
& {j_\t \over 2} & {j_\a - \e_\a \over 2}}\right\} \cr &
\!\!\!\!\!\!\!\!\!\!\!\!\!\!\!\!\!\!\!\!\!\!\!\!\!\!\! 
\cdot R_{3 \e_\a \e_\b} (j_\a, j_\l,
j_\b) R_{3 \e_\g \e_\d} (j_\g, j_\l, j_\d) R_{3 \e_\b \e_\d}
(j_\b, j_\t, j_\d) R_{3 \e_\a \e_\g} (j_\a, j_\t, j_\g), \cr }}
\eqn\moresumsone{\eqalign{ \sum_{m's, \epsilon's,j_\l} D^{\r \br
\l} D^{\a \b \bl} \hat{E}^{\bal \t \s} \hat{E}^{\bb \bs \bt} &=
-\delta_{j_\a, j_\b} {1 \over \pi^2} j_\r(j_\r+2) \sum_{\e
's}(\hat{R}_{4 \e_\s \e_\t \e_\a}(j_\s, j_\t, j_\a))^2, \cr
\sum_{m's, \epsilon's} C^{\s \g \r} C^{\br \bg \bt} D^{\a \b \bs}
D^{\bal \bb \t} &= -\delta_{j_\s, j_\t}  {4 \over (j_\s+1)^2}
R_2^2 (j_\s, j_\g, j_\r)\cr &\qquad \qquad \qquad \cdot (R_{3+}^2
(j_\a, j_\s, j_\b) + R_{3-}^2 (j_\a, j_\s, j_\b)), \cr \sum_{m's,
\epsilon's} D^{\a \b \l} D^{\bal \g \bl} \hat{E}^{\bb \d \r}
\hat{E}^{\bd \bg \br} &= -\delta_{j_\b, j_\g} {1 \over j_\b (j_\b
+2)} (R_{3+}^2 (j_\a, j_\l, j_\b) + R_{3-}^2 (j_\a, j_\l, j_\b))
\cr &\qquad \qquad \qquad \cdot \sum_{\e's} (\hat{R}_{4 \e_\b
\e_\d \e_\r}(j_\b, j_\d, j_\r))^2, \cr \sum_{m's, \epsilon's}
\hat{E}^{\g \a \r} \hat{E}^{\br \bal \bd} \hat{E}^{\bg \t \b}
\hat{E}^{\bb \bt \d} &= \delta_{j_\g, j_\d} {1 \over j_\g (j_\g
+2)} \sum_{\e_\g} [ \sum_{\e's} (\hat{R}_{4 \e_\a \e_\r
\e_\g}(j_\a, j_\r, j_\g))^2 \cr & \qquad\qquad\qquad\qquad\qquad
\cdot \sum_{\e's} (\hat{R}_{4 \e_\b \e_\t \e'_\g}(j_\b, j_\t,
j_\g))^2 ], \cr }}
\eqn\moresumstwo{\eqalign{
\sum_{m's} D^{\a \g \r} D^{\b \d \br}
\hat{E}^{\bal \bb \t} \hat{E}^{\bd \bg \bt} &= -(-1)^{\sum j's}
\left\{\matrix{{j_\a + \e_\a \over 2} & {j_\b + \e_\b \over 2} &
{j_\t + \e_\t \over 2} \cr {j_\d + \e_\d \over 2} & {j_\g + \e_\g
\over 2} & {j_\r \over 2}} \right\} \left\{\matrix{{j_\a - \e_\a
\over 2} & {j_\b - \e_\b \over 2} & {j_\t - \e_\t \over 2} \cr
{j_\d - \e_\d \over 2} & {j_\g - \e_\g \over 2} & {j_\r \over 2}}
\right\} \cr & \!\!\!\!\!\!\!\!\!\!\!\!\!\!\!\!\!\!\!\!\!\!\!\!\!
\cdot R_{3 \e_\a \e_\g} (j_\a, j_\r, j_\g) R_{3 \e_\b \e_\d}
(j_\b, j_\r, j_\d) \hat{R}_{4 \e_\a \e_\b \e_\t} (j_\a, j_\b,
j_\t) \hat{R}_{4 \e_\d \e_\g \e_\t} (j_\d, j_\g, j_\t), \cr
\sum_{m's} \hat{E}^{\r \s \t} \hat{E}^{\bal \b \bt} \hat{E}^{\bb
\g \br} \hat{E}^{\bg \a \bs} &= (-1)^{\sum j's}
\left\{\matrix{{j_\r + \e_\r \over 2} & {j_\s + \e_\s \over 2} &
{j_\t + \e_\t \over 2} \cr {j_\a + \e_\a \over 2} & {j_\b + \e_\b
\over 2} & {j_\g + \e_\g \over 2}} \right\} \left\{\matrix{{j_\r -
\e_\r \over 2} & {j_\s - \e_\s \over 2} & {j_\t - \e_\t \over 2}
\cr {j_\a - \e_\a \over 2} & {j_\b - \e_\b \over 2} & {j_\g -
\e_\g \over 2}} \right\} \cr & \! \! \! \! \! \! \! \! \! \!
\!\!\!\!\!\!\!\!\!\!\!\!\!\!\! \cdot \hat{R}_{4 \e_\r \e_\s \e_\t}
(j_\r, j_\s, j_\t) \hat{R}_{4 \e_\a \e_\b \e_\t} (j_\a, j_\b,
j_\t) \hat{R}_{4 \e_\b \e_\g \e_\r} (j_\b, j_\g, j_\r) \hat{R}_{4
\e_\g \e_\a \e_\s} (j_\g, j_\a, j_\s), \cr \sum_{m's} D^{\a \g \r}
D^{\b \bg \s} C^{\br \bb \t} C^{\bt \bal \bs} &= -(-1)^{\sum j's}
\left\{\matrix{{j_\r  \over 2} & {j_\t \over 2} & {j_\b + \e_\b
\over 2} \cr {j_\s  \over 2} & {j_\g + \e_\g \over 2} & {j_\a +
\e_\a \over 2}} \right\}
 \left\{\matrix{{j_\r  \over 2}
& {j_\t \over 2} & {j_\b - \e_\b \over 2} \cr {j_\s  \over 2} &
{j_\g - \e_\g \over 2} & {j_\a - \e_\a \over 2}} \right\} \cr &
\cdot R_2 (j_\r, j_\b, j_\t) R_2 (j_\t, j_\a, j_\s) R_{3 \e_\a
\e_\g} (j_\a, j_\r, j_\g) R_{3 \e_\b \e_\g} (j_\b, j_\s, j_\g),
\cr \sum_{m's} \hat{E}^{\a \b \g} C^{\r \bg \s} D^{\bb \t \bs}
D^{\bal \bt \br} &= (-1)^{\sum j's} \left\{\matrix{{j_\a + \e_\a
\over 2} & {j_\b + \e_\b \over 2} & {j_\g + \e_\g \over 2} \cr
{j_\s  \over 2} & {j_\r \over 2} & {j_\t + \e_\t \over 2}}
\right\} \left\{\matrix{{j_\a - \e_\a \over 2} & {j_\b - \e_\b
\over 2} & {j_\g - \e_\g \over 2} \cr {j_\s \over 2} & {j_\r \over
2} & {j_\t - \e_\t \over 2}} \right\} \cr & \!\!\!\!\! \cdot R_2
(j_\r, j_\g, j_\s) \hat{R}_{4 \e_\a \e_\b \e_\g} (j_\a, j_\b,
j_\g) R_{3 \e_\b \e_\t} (j_\b, j_\s, j_\t) R_{3 \e_\a \e_\t}
(j_\a, j_\r, j_\t). \cr }}
Finally, in certain cases, we may simplify expressions further by using
the relations
\eqn\Rsums{\eqalign{ \sum_c R_{3+}^2 (a,c,b) &= \sum_c R_{3-}^2
(a,c,b) = {1 \over 6 \pi^2} a(a+2)b(b+2), \cr \sum_c
(R_2(a,b,c))^2 &= {1 \over 3 \pi^2} b(b+2)a(a+2)(a+1)^2. }}

\listrefs

\end